\documentclass{article}

\usepackage{amssymb, amsmath, amsthm, bm}
\usepackage{bbm}
\usepackage{xspace}
\usepackage{caption}
\usepackage{mathtools}
\usepackage{subcaption}
\usepackage{enumitem}
\usepackage[utf8]{inputenc}
\usepackage{xcolor}
\usepackage{graphicx}
\usepackage[sort&compress]{natbib}
\usepackage{nicefrac}
\usepackage{centernot}
\usepackage{array}
\usepackage[toc,page]{appendix}
\usepackage{mathabx}
\usepackage{hyperref}
\usepackage{subcaption}
\usepackage{titlesec}
\usepackage{soul}
\usepackage[font=itshape,begintext=`,endtext=']{quoting}

\newtheorem{theorem}{Theorem}%[section]
\newtheorem{proposition}{Proposition}%[section]
\newtheorem{corollary}{Corollary}
\newtheorem{lemma}[theorem]{Lemma}

\theoremstyle{definition}
\newtheorem{definition}{Definition}
\newtheorem{example}{Example}

\theoremstyle{remark}

\allowdisplaybreaks

\newcommand{\naturals}{\mathbb{N}}

\newcommand{\Scal}{\mathcal{S}}

\newcommand{\Bcal}{\mathcal{B}}

\newcommand{\Xcal}{\mathcal{X}}
\newcommand{\Ycal}{\mathcal{Y}}

\newcommand{\ze}{UMD\xspace}
\newcommand{\expText}[1]{\exp\left(#1\right)}

\newcommand{\abs}[1]{\left\lvert#1\right\rvert}

\newcommand{\indicator}[1]{\mathbbm{1}\curlybrack{#1}}

\newcommand{\roundbrack}[1]{\left( #1 \right)}
\newcommand{\curlybrack}[1]{\left\lbrace #1 \right\rbrace}

\newcommand{\E}{\mathbb{E}}

\newcommand{\Prob}{{\text{Pr}}}
\newcommand{\Exp}[2]{\mathbb{E}_{#1}\left\lbrack#2\right\rbrack}

\newcommand{\independent}{\!\perp\!\!\!\perp}

\newcommand{\Gcal}{\mathcal{G}}
\newcommand{\Dcal}{\mathcal{D}}

\newcommand{\credit}{\texttt{CREDIT}\xspace}
\newcommand{\hoeffding}{t}

\usepackage{algorithm}
\usepackage{algorithmic}
\usepackage[margin=1.2in]{geometry}
\newcommand{\rev}[1]{{#1}}%{\textcolor{blue}{#1}}}

\begin{document}

\title{Distribution-free calibration guarantees for\\ histogram binning without sample splitting}
\author{Chirag Gupta, Aaditya K. Ramdas
\\\\\texttt{chiragg@cmu.edu, aramdas@cmu.edu}
\\\\Carnegie Mellon University
}

\maketitle
\begin{abstract}
    %We prove calibration guarantees for uniform-mass binning without making any distributional assumptions, and without using sample splitting. This calibration approach was first advocated by the popular work of \citet{zadrozny2001obtaining}, but came without theoretical guarantees despite displaying strong practical performance. Calibration guarantees have since been proved for variants that involve sample splitting, at the expense of worse practical performance. Our paper finally proves a guarantee for the original method, thus providing rigorous justification for a commonly used post-hoc calibration algorithm. 
    We prove calibration guarantees for the popular histogram binning (also called uniform-mass binning) method  of \citet{zadrozny2001obtaining}. Histogram binning has displayed strong practical performance, but theoretical guarantees have only been shown for sample split versions that avoid `double dipping' the data. 
    We demonstrate that the statistical cost of sample splitting is practically significant on a credit default dataset. We then prove calibration guarantees for the original method that double dips the data, using a certain Markov property of order statistics. Based on our results, we make practical recommendations for choosing the number of bins in histogram binning. %Thus, our analysis provides rigorous justification for a commonly used post-hoc calibration method. 
    %We prove calibration guarantees for uniform-mass binning without making any distributional assumptions, and without using sample splitting. This calibration approach was first advocated by the popular work of \citet{zadrozny2001obtaining}, but came without theoretical guarantees despite displaying strong practical performance. Calibration guarantees have since been proved for variants that involve sample splitting, at the expense of worse practical performance. Our paper finally proves a guarantee for the original method, thus providing rigorous justification for a commonly used post-hoc calibration algorithm. 
    In our illustrative simulations, we propose a new tool for assessing calibration---validity plots---which provide more information than an ECE estimate. Code for this work will be made publicly available at \url{https://github.com/aigen/df-posthoc-calibration}.%settling a long-standing open problem in the area of post-hoc calibration, making theory and practice match perfectly.
    %thus providing rigorous justification for a popular algorithm
    %As far as we are aware, these are the first theoretical guarantees for the original algorithm without alteration, making theory and practice match perfectly.
\end{abstract}
%\textbf{Keywords:} uniform-mass binning, post-hoc calibration, distribution-free, binary classification, double dipping, sample splitting, Markov property of order statistics.
\tableofcontents 
\newpage 
\section{Introduction}
In classification, the goal is to learn a model that uses observed feature measurements to make a class prediction on the categorical outcome. However, for safety-critical areas such as medicine and finance, a single class prediction might be insufficient and reliable measures of confidence or certainty may be desired. Such uncertainty quantification is often provided by predictors that produce not just a class label, but a probability distribution over the labels. If the predicted probability distribution is consistent with observed empirical frequencies of labels, the predictor is said to be calibrated \citep{dawid1982well}. %For example, when a calibrated predictor predicts a credit default probability of $0.1$ for $1000$ people, it guarantees that approximately $100$ of them indeed default. 
 
In this paper we study the problem of calibration for binary classification; let $\Xcal$ and $\smash{\Ycal=\{0,1\}}$ denote the feature and label spaces. We focus on the recalibration or post-hoc calibration setting, a standard statistical setting where the goal is to recalibrate existing (`pre-learnt') classifiers that are powerful and (statistically) efficient for classification accuracy, but do not satisfy calibration properties out-of-the-box. This setup is popular for recalibrating pre-trained deep nets. For example, \citet[Figure 4]{guo2017nn_calibration} demonstrated that a pre-learnt ResNet is initially miscalibrated, but can be effectively post-hoc calibrated. In the case of binary classification, the pre-learnt model can be an arbitrary predictor function that provides a classification `score' $g \in \Gcal$, where $\Gcal$ is the space of all measurable functions from $\Xcal \to [0, 1]$. Along with $g$, we are given access to a calibration dataset of size $n \in \naturals$, $\Dcal_n = \{(X_i, Y_i)\}_{i \in [n]}$, drawn independently from a distribution $P\equiv P_X \times P_{Y|X}$. 
The goal is to define a calibrator $\smash{H : \Gcal \times (\Xcal \times [0,1])^n \to \Gcal}$, that `recalibrates' $g$ to an approximately calibrated predictor $H(g, \Dcal_n)$ (formally defined shortly). We denote $H(g, \Dcal_n)$ as $h$. \emph{All probabilities in this paper are conditional on $g$ and thus conditional on the data on which $g$ is learnt.} %For notational simplicity, we simply write %$h(\cdot, \cdot)$ for $h(g, \cdot, \cdot)$ and
%$h(x)$ for $h(g, \Dcal_n, x)$.  %We typically use $g$ to refer to any pre-learnt classifier, and $h$ to refer to a potentially recalibrated classifier.

Let $\Exp{}{\cdot}$ denote the expectation operator associated with $P$, interpreted marginally or conditionally depending on the context. 
The predictor $h$ %$\smash{h : \Xcal \to [0, 1]}$ 
is said to be perfectly calibrated if $\smash{\Exp{}{Y \mid h(X)} = h(X)}$ (almost surely). While perfect calibration is impossible in finite samples, we desire a framework to make transparent claims %about their validity. In other words, we require that $h$ comes packaged with a \textit{valid} guarantee 
about how close $h$ is to being perfectly calibrated. The following notion proposed by \citet{gupta2020distribution} defines a calibrator that provides \emph{probably approximate calibration}  %. Suppose the prescribed levels of 
for chosen levels of approximation $\varepsilon \in (0, 1)$ and failure $\alpha \in (0, 1)$. For brevity, we  skip the qualification `probably approximate'.
%$(\varepsilon, \alpha)$-approximately calibrated predictor for some prescribed values of $\varepsilon, \alpha \in (0, 1)$. %In order to establish \citet{gupta2020distribution} proposed an alternative way of providing valid calibration claims. %Instead, we aim for predictors that come packaged with a transparent claim about their approximate validity. 
%Informally, we say that a predictor has a \textit{valid} approximate calibrated guarantee if one can make transparent claims about how close the predictor it is to being perfectly calibrated.  In this work, we are interested in the achievable goal of approximate calibration first defined formally by \citet{gupta2020distribution}. 
\begin{definition}[Marginal calibration\footnote{This definition is unrelated to that of \citet[Definition 1c]{gneiting2007probabilistic}, where marginal calibration refers to an asymptotic notion of calibration in the regression setting.
}]\label{def:marginal-calib}
		A calibrator $H : (g, \Dcal_n) \mapsto h$ is said to be $(\varepsilon, \alpha)$-marginally calibrated 
		%for some $\alpha, \varepsilon \in(0,1)$ 
		if for every predictor $g \in \Gcal$ and distribution $P$ over $\Xcal \times [0, 1]$, 
		\begin{equation}
		\Prob(\abs{\Exp{}{Y | h(X)}-h(X)}\leq \varepsilon) \geq 1- \alpha.
		\label{eq:marginal-calib}
		\end{equation}
	\end{definition}
% since any achievable notion of calibration cannot be deterministic or exact. 
%The formal statistical setup is described shortly, but we briefly note that the probability in \eqref{eq:marginal-calib} is over the calibration as well as test data. 
\noindent The above probability is taken over both $X$ and $\Dcal_n$ since $\smash{h = H(g, \Dcal_n)}$ contains the randomness of $\Dcal_n$. The qualification \emph{marginal} signifies that the inequality $\abs{\Exp{}{Y \mid h(X)} - h(X)} \leq \varepsilon$ may not hold conditioned on $X$ or $h(X)$, but holds only on \emph{average}. We now define a more stringent conditional notion of calibration, %which is calibration conditional on the prediction $h(X)$. 
which requires that approximate calibration hold simultaneously (or conditionally) for every value of the prediction. 
\begin{definition}[Conditional calibration]\label{def:conditional-calib}
		A calibrator $H : (g, \Dcal_n) \mapsto h$ is $(\varepsilon, \alpha)$-conditionally calibrated if for every predictor $g \in \Gcal$ and distribution $P$ over $\Xcal \times [0, 1]$,
        \begin{equation}
		\Prob(\forall r \in \text{Range}(h),  \abs{\Exp{}{Y \mid h(X) = r}-r}\leq \varepsilon) \geq 1- \alpha.
% 		\text{ a.s. $h(X)$,}
		\label{eq:conditional-calib}
\end{equation}
% or equivalently,
% \begin{equation}
% 		\Prob(\abs{\Exp{}{Y \mid h(X)}-h(X)}\leq \varepsilon \text{ for all }X,  P_X\text{-almost surely}) \geq 1- \alpha.
% % 		\text{ a.s. $h(X)$,}
% 		\label{eq:conditional-calib2}
% \end{equation}
% where a.s.\ $h(X)$ means almost surely (with probability one) with respect to the distribution~of~$h(X)$.
	\end{definition}
\noindent In contrast to \eqref{eq:marginal-calib}, the $\Pr$ above is only over $\Dcal_n$. Evidently, if $H$ is conditionally calibrated, it is also marginally calibrated. %The property \eqref{eq:conditional-calib} 
The conditional calibration property \eqref{eq:conditional-calib} has a PAC-style interpretation:  with probability $ 1 - \alpha$ over $\Dcal_n$, $h$ satisfies the following deterministic property:%\vspace{-0.2cm}
\begin{equation}
\forall r \in \text{Range}(h),  \abs{\Exp{}{Y \mid h(X) = r}-r}\leq \varepsilon. \label{eq:PAC-style-conditional-calib}%\vspace{-0.2cm}
\end{equation}
%On the other hand,  marginal calibration has no immediate PAC-style interpretation. 
Marginal calibration does not have such an interpretation; we cannot infer from \eqref{eq:marginal-calib} a statement of the form ``with probability $1 - \gamma$ over $\Dcal_n$, $h$ satisfies $\cdots$ ". 

Marginal and conditional calibration assess the truth of the event $\indicator{\abs{\Exp{}{Y \mid h(X)} - h(X)} \leq \varepsilon}$ for a given $\varepsilon$. Instead we can consider bounding the expected value of $\abs{\Exp{}{Y \mid h(X)} - h(X)}$ for $X \sim P_X$. This quantity is known as the expected calibration error. %This quantity is known as the expected calibration error. 
\begin{definition}[Expected Calibration Error (ECE)]\label{def:ece}
		For $p \in [1, \infty)$, the $\ell_p$-ECE of a predictor $h$ is%\vspace{-0.15cm}% for any $p \in [1, \infty)$ is defined as
    \begin{equation}
	\ell_p\text{-ECE}(h) =\left( \E_X{\abs{\Exp{}{Y \mid h(X)} - h(X)}^p}\right)^{1/p}. 
	\label{eq:ece}
	%\vspace{-0.2cm}
	\end{equation}
\end{definition}
%Further, we can show that the expected calibration error (ECE) \citep{sanders1963subjective} of $h$ is bounded: with probability $\geq 1 - \alpha$, $\ell_1\text{-ECE}(h) \leq \varepsilon$. 
%\noindent The outer expectation above is over $X\sim P_X$. When we recalibrate $g$ to $h$ using $\Dcal_n$, we will use the phrase ``expected ECE of $h$'' to emphasize that we further take the expectation over $\Dcal_n$ as well.
\noindent Note that the expectation above is only over $X\sim P_X$ and not over $\Dcal_n$. We can ask for bounds on the ECE of $h = H(g, \Dcal_n)$ that hold with high-probability or in-expectation over the randomness in $\Dcal_n$. The conditional calibration property  \eqref{eq:PAC-style-conditional-calib} for $h$ implies a bound on the $\ell_p$-ECE for every $p$,
as formalized by the following proposition which also relates $\ell_p$-ECE for different $p$.
\begin{proposition} \label{prop:ECE-holder}
For any predictor $h$ and $1 \leq p \leq q < \infty$, %\vspace{-0.15cm}
\begin{equation}\label{eq:ECE-holder}
\ell_{p}\text{-ECE}(h) \leq \ell_{q}\text{-ECE}(h). %\vspace{-0.15cm}%\leq \sup_{r \in \text{Range}(h), P_X(h(X) = r) > 0}\abs{\Exp{}{Y \mid h(X) = r}-r}.
\end{equation}
Further, if \eqref{eq:PAC-style-conditional-calib} holds, then $\ell_{p}\text{-ECE}(h) \leq \varepsilon, \forall p \in [1, \infty)$. 
\end{proposition}
\noindent The proof (in Appendix~\ref{appsec:adaptive-binning-proofs}) is a straightforward application of H\"{o}lder's inequality. Informally, one can interpret the L.H.S. of \eqref{eq:PAC-style-conditional-calib} as the $\ell_{\infty}\text{-ECE}$ of $h$ so that \eqref{eq:ECE-holder} holds for $1 \leq p \leq q \leq \infty$. Thus conditional calibration is the strictest calibration property we consider: if $H$ is $(\varepsilon, \alpha)$-conditionally calibrated, then (a) $H$ is $(\varepsilon, \alpha)$-marginally calibration and (b) with probability $1-\alpha$, $\ell_p\text{-ECE}(h) \leq \varepsilon$. 
%(this can be formalized but we skip the tedious analytical details ). 
%When we recalibrate $g$ to $h$ using $\Dcal_n$, we will use the phrase ``expected ECE of $h$'' to emphasize that we further take the expectation over $\Dcal_n$ as well. 

\begin{example}
We verify Proposition~\ref{prop:ECE-holder} on a simple example, which also helps build intuition for the various notions of calibration. Suppose $h$ takes just two values: %it takes the value $0.2$ with probability $0.9$ and the value $0.8$ with probability $0.1$. 
$\Prob(h(X) = 0.2) = 0.9$ and $\Prob(h(X) = 0.8) = 0.1$. Let $\Exp{}{Y \mid h(X) = 0.2} = 0.3$ and $\Exp{}{Y \mid h(X) = 0.8} = 0.6$. %Then $h$ satisfies $(0.1,0.1)$-marginal calibration but not $(0.1,0.1)$- conditional calibration. Also,
Then $\ell_1$-ECE$(h) = %0.9\abs{0.2 - 0.3} + 0.1\abs{0.8-0.6} 
0.11 <  \ell_2$-ECE$(h) 
%= \sqrt{0.9(0.2 - 0.3)^2  + 0.1(0.8-0.6)^2} 
\approx 0.114$. Marginal calibration~\eqref{eq:marginal-calib} for $H(\cdot, \cdot) \equiv h$ is satisfied for $(\varepsilon \geq 0.1, \alpha \leq 0.9)$, while the conditional calibration requirement~\eqref{eq:PAC-style-conditional-calib} is only satisfied for $\varepsilon \geq 0.2$. %If we assume $h$ is fixed, then 
 %If we ignore the randomness in $h$ and only consider the marginal calibration over $X \sim P_X$
\end{example}

In this paper, we show that the histogram binning method of \citet{zadrozny2001obtaining}, described shortly, is calibrated in each of the above senses (marginal and conditional calibration; high-probability and in-expectation bounds on ECE), if the number of bins is chosen appropriately. 

Some safety-critical domains may require calibration methods that are robust to the data-generating distribution. We refer to Definitions~\ref{def:marginal-calib} and \ref{def:conditional-calib} as distribution-free (DF) guarantees since they are required to hold for all distributions over $(X,Y)$ without restriction. This paper is in the DF setting: the only assumption we make is that the calibration data $\Dcal_n$ and $(X, Y)$ are independent and identically distributed (i.i.d.). \citet[Theorem 3]{gupta2020distribution} showed that if $H$ is DF marginally calibrated with a meaningful value of $\varepsilon$ (formally, $\varepsilon$ can be driven to zero as sample size grows to infinity), then $H$ must necessarily produce only discretized predictions (formally, $\text{Range}(h)$ must be at most countable).  We refer to such $H$ as `binning methods' --- this emphasizes that $H$ essentially partitions the sample-space into a discrete number of `bins' and provides one prediction per bin (see Proposition~1 \citep{gupta2020distribution}). Since our goal is DF calibration, we focus on binning methods. 

\subsection{Prior work on binning}
\label{sec:prior-work}
Binning was initially introduced in the calibration literature for assessing calibration.
%In order to assess the calibration properties of a continuous scoring function $h$ using 
Given a continuous scoring function $h$, if we wish to plot a reliability diagram \citep{sanders1963subjective, Niculescu2005predicting} or compute an ECE estimate \citep{miller1962statistical, sanders1963subjective, naeini2015obtaining},  then $h$ must first be discretized using binning. A common binning scheme used for this purpose is `fixed-width binning', where $[0, 1]$ is partitioned into $B \in \naturals$ intervals (called bins) of width $1/B$ each and a single prediction is assumed for every bin. For example, if $B = 10$, then the width of each bin is $0.1$, and if (say) $h(x) \in [0.6, 0.7)$ then the prediction is assumed to be $0.65$. %test points are `binned' depending on which of these intervals the prediction belongs to, and the prediction is replaced by the average prediction in the corresponding interval on the test set. %or binned, so that in effect we are assessing the calibration properties of the binned version of $h$. 

%In this paper, we focus on the usage of binning for achieving calibration. The first proposal to this end was made by \citet{zadrozny2001obtaining}, who used binning to recalibrate a naive Bayes classifier. 
\citet[Theorem 3]{gupta2020distribution} showed that some kind of binning is in fact necessary to \emph{achieve} DF calibration. The first binning method for calibration was proposed by \citet{zadrozny2001obtaining} to calibrate a naive Bayes classifier. 
%They used uniform-mass binning (referred to as `histogram binning' in their paper) to recalibrate a naive Bayes classifier. In uniform-mass binning, 
%They used histogram binning to recalibrate a naive Bayes classifier. 
Their procedure is as follows. 
First, the interval $[0, 1]$ is partitioned into $\smash{B \in \naturals}$ bins using the histogram of the $g(X_i)$ values, to ensure that each bin has the same number of calibration points (plus/minus one). Thus the bins have nearly `uniform (probability) mass'. Then, the calibration points are assigned to bins depending on the interval to which the score $g(X_i)$ belongs to, %. This binning is done using the histogram of the $g(X_i)$ values to ensure that each bin has the same number of calibration points (plus/minus one); thus the bins have nearly `uniform (probability) mass'.
and the probability that $\smash{Y = 1}$ is estimated for each bin as the average of the observed $Y_i$-values in that bin. This average estimates the `bias' of the bin. The binning scheme and the bias estimates together define $h$. A slightly modified version of this procedure is formally described in Algorithm~\ref{alg:efficient-uniform-mass-binary}. 

While Algorithm~\ref{alg:efficient-uniform-mass-binary} was originally called histogram binning, it has also been referred to as uniform-mass binning in some works. In the rest of this paper, we use the latter terminology. 
%We refer to Zadrozny and Elkan's procedure as 
Specifically, we refer to it as \ze, short for Uniform-Mass-Double-dipping. This stresses that the same data is used twice, both to determine  inter-bin boundaries and to calculate intra-bin biases. \ze continues to remain a competitive benchmark in empirical work \citep{guo2017nn_calibration, naeini2015obtaining, roelofs2020mitigating}, but no finite-sample calibration guarantees have been shown for it. Some asymptotic consistency results for a histogram regression algorithm closely related to \ze were shown by \citet{parthasarathy1961some} (see also the work by \citet{lugosi1996consistency}). \citet{zadrozny2002transforming} proposed another popular binning method based on isotonic regression, for which some non-DF analyses exist (see \citet{dai2020bias} and references therein). Recently, two recalibration methods closely related to \ze have been proposed, along with some theoretical guarantees that rely on sample-splitting --- scaling-binning \citep{kumar2019calibration} and sample split uniform-mass binning  \citep{gupta2020distribution}. %These are discussed in more detail below. %Other calibration methods have also been %The broader literature on other nonparameteric calibration methods that can be characterized as binning methods --- isotonic regression, Venn prediction, binning for binary regression, %probability estimation trees, and random forests with probabilistic outputs --- is briefly discussed in Section~\ref{sec:discussion}.%, with a focus on the theoretical calibration guarantees shown for these methods.

%In this subsection, we briefly review various nonparametric calibration methods proposed in literature that we characterize as binning methods. The focus is on the theoretical calibration guarantees (particularly DF guarantees) of these methods.

%In this paper, we develop a novel binning method based on uniform-mass histogram binning, called \ze (short for uniform-mass-double-dipping). We show a distribution-free calibration guarantee for \ze. 

In the scaling-binning method, the binning is performed on the output of another continuous recalibration method (such as Platt scaling~\citep{Platt99probabilisticoutputs}), and the bias for each bin is computed as the average of the output of the scaling procedure in that bin. This is unlike %the algorithm of \citet{zadrozny2001obtaining}, 
other binning methods, where the bias of each bin is computed as the average of the true outputs $Y_i$ in that bin. 
 \citet[Theorem~4.1]{kumar2019calibration} showed that under some assumptions on the scaling class (which includes injectivity), the ECE of the sample split scaling-binning procedure is $\varepsilon$-close to $\sqrt{2}\ \ell_2$-ECE of the scaling procedure if, roughly, $n = \Omega(\log B/\varepsilon^2)$. However, the results of \citet[Section 3.3]{gupta2020distribution} imply that 
%  for any injective scaling procedure, %such as Platt scaling, 
    there exist data distributions on which any injective scaling procedure itself has trivial ECE. %Since conditional calibration implies an ECE bound (as mentioned after Definition~\ref{def:conditional-calib}), 
    %Thus, a DF calibration guarantee cannot be inferred from their theorem. 
    % Thus their theorem does not appear to directly imply a DF calibration guarantee. 
    %Further, experimental evidence in Section~\ref{sec:simulations} indicates that averaging the true outputs leads to better performance (as is done in \ze). 
    
%\paragraph{Uniform-mass binning with sample splitting (UMB). } 
In sample split uniform-mass binning, the first split of the data is used to define the bin boundaries so that the bins are balanced. The second split of the data is used for estimating the bin biases, using the average of the $Y_i$-values in the bin.  We refer to this version as UMS, for Uniform-Mass-Sample-splitting.  \citet[Theorem~5]{gupta2020distribution} showed that UMS is $(\varepsilon, \alpha)$-marginally calibrated if (roughly) $n = \Omega(B\log (B/\alpha)/\varepsilon^2)$. To the best of our knowledge, this is the only known DF guarantee for a calibration method. However, in Section~\ref{sec:ums-inefficient} we demonstrate that the constants in this guarantee are quite conservative, and the loss in performance due to sample splitting is practically significant on a real dataset.
%They allude to the conditional calibration guarantee of UMB, but informally. 
%Note that $(\varepsilon, \alpha)$-marginal calibration also leads to a bound on ECE: w.p. $\geq 1 - \alpha$, $\ell_1$-ECE $\leq \varepsilon + \alpha$. % for any $p \in [1, \infty)$. %This bound is DF and holds irrespective of the underlying parametric algorithm. 
%They also showed that UMB satisfies a notion of

% \begin{figure*}
% \begin{subfigure}[t]{0.49\linewidth}
% \centering
% \includegraphics[width=0.6\textwidth]{umd/uniform_mass.pdf}
% %\caption{UMS with $B = 10$.}
% \end{subfigure}
% \begin{subfigure}[t]{0.49\linewidth}
% \centering
% \includegraphics[width=0.6\textwidth]{umd/umd.pdf}
% %\caption{\ze with $B = 10$.}
% \end{subfigure}
% \caption{Validity plots on the \credit dataset. Validity plots are descibed in Section~\ref{subsec:experiments-conservative}. The plots show that \ze has higher confidence score ($1-\alpha$) for the same values of $n, \varepsilon$, and thus lower $\ell_1$-ECE. For example, for $n=1000$ and $\varepsilon=0.05$, UMS has $1-\alpha \approx 0.63$, while \ze has  $1-\alpha \approx 0.77$.}
% \label{fig:credit-sample-size}
% \end{figure*}

\subsection{Our contribution}
    We show tight DF calibration guarantees for the original method proposed by \citet{zadrozny2001obtaining}, \ze.  %Additionally, the calibration bound of \ze is conditional on the prediction, per Definition~\ref{def:conditional-calib}. 
    While the existing theoretical analyses rely on sample splitting \citep{kumar2019calibration, gupta2020distribution},  %This is done in  by \citet{kumar2019calibration} as well as \citet{gupta2020distribution}.
    it has been observed in experiments that double dipping to perform both bin formation and bias estimation on the same data leads to excellent practical performance \citep{zadrozny2001obtaining, guo2017nn_calibration, kumar2019calibration, roelofs2020mitigating}.
    Our work fills this gap in theory and practice. %To the best of our knowledge, ours is the first formal calibration guarantee for \ze. 
    %This will reinforce the practitioner's confidence in applying \ze to new problems, and calm the theoretician's nerves that \ze is just a heuristic.%We use some elegant but subtle probabilistic facts about order statistics to avoid paying for any selection bias caused by double dipping.  This will reinforce the practitioner's confidence in applying the method to new problems, and calm the theoretician's nerves that \ze is just a heuristic. %: to determine both inter-bin boundaries and calculate intra-bin biases. 

    We exploit a certain Markov property of order statistics, which are a set of classical, elegant results that are not well known outside of certain subfields of statistics (for one exposition of the Markov property, see \citet[Chapter 2.4]{arnold2008first}). The strength of these probabilistic results is not widely appreciated --- judging by their non-appearance in the ML literature --- nor have they had implications for any modern AI applications that we are aware of. Thus, we consider it a central contribution of this work to have recognized that these mathematical tools can be brought to bear in order to shed light on a contemporary ML algorithm. 
    
    A simplified version of the Markov property is as follows: for order statistics $Z_{(1)}, Z_{(2)}, \ldots, Z_{(n)}$ of samples $\{Z_i\}_{i \in [n]}$ drawn i.i.d from any absolutely continuous distribution $Q$, and any indices $\smash{1 < i < j \leq n}$, we have that %\vspace{-0.3cm}
    \[
    Z_{(j)} \perp Z_{(i-1)}, Z_{(i-2)}, \ldots, Z_{(1)} \mid Z_{(i)}.
        %\vspace{-0.1cm}
    \]
    For example, given the empirical median $M$, the points to its left are conditionally independent of the points to its right. Further each of these have a distribution that is identical to that of i.i.d. draws from $Z \sim Q$ when restricted to $Z<M$ (or $Z > M$). The implication is that if we form bins using the order statistics of the scores as the bin boundaries, %eg: the boundaries being $X_{(100)}, X_{(200)}, \ldots X_{(900)}$
    then (a) the points within any bin are independent of the points outside that bin, and (b) conditioned on being in a given bin, say $B_i$, the points in the bin are i.i.d. with distribution $Q_{Z \mid Z \in B_i}$. When we split a calibration sample $\Dcal$ and use one part $\Dcal_1$ for binning and the other $\Dcal\setminus \Dcal_1$ for estimating bin probabilities, the points in $\Dcal \setminus \Dcal_1$ that belong to $B_i$ are also conditionally i.i.d. with distribution $Q_{Z \mid Z \in B_i}$, which is exactly what we accomplished without sample splitting. In short, the Markov property allows us to `double dip' the data, i.e., use the same data for binning and estimating within-bin probabilities.

%\subsection{Organization}
\textbf{Organization. }Section~\ref{sec:ums-inefficient} motivates our research problem by showing that UMS is sample-inefficient both in theory and practice. Empirical evidence is provided through a novel diagnostic tool called validity plots (Section~\ref{sec:validity-plots}). Section~\ref{sec:adaptive-binning} presents \ze formally along with its analysis (main results in Theorems~\ref{thm:umd-binary} and \ref{thm:umd-binary-randomized}). Section~\ref{sec:simulations} contains illustrative simulations. Proofs are in the supplement. %sSection \ref{sec:discussion} concludes the paper with discussion.

%\textbf{Notation. }We denote a general data-sample as $(X, Y) \in \Xcal \times \Ycal$ where $\Xcal$ is some measurable feature space and $\Ycal = \{0, 1\}$. %The distribution of $(X, Y)$ is denoted as $P$. %The probability simplex in $\Real^L$ is denoted as $\Delta_L$. $\Delta_L$ is the prediction space for a calibrated predictor. We also define a vectorized version of $Y$ given by $\Y := \e_Y$ (likewise $\Y_i := \e_{Y_i}$), where $\e_1, \e_2, \ldots, \e_L$ are basis vectors in the canonical basis. 

\section{Sample split uniform-mass binning is inefficient}
\label{sec:ums-inefficient}
The DF framework encourages development of algorithms that are robust to arbitrarily distributed data. At the same time, the hope is that the DF guarantees are adaptive to real data and give meaningful bounds in practice. In this section, we assess if the practical performance of uniform-mass-sample-splitting (UMS) is well explained by its DF calibration guarantee \citep{gupta2020distribution}. As far as we know, this is the only known DF guarantee for a calibration method. However, we demonstrate that the guarantee is quite conservative. Further, we demonstrate that sample splitting leads to a drop in performance on a real dataset. %(Section~\ref{sec:umd-vs-ums}). %This gap between theory and practice has been observed by other works as well, and the problem of proving a (tight) calibration bound for \ze has remained open. %We focus on the comparison to UMS since it is the only procedure for which DF guarantees of the style of Definitions~\ref{def:marginal-calib} and~\ref{def:conditional-calib} have been shown (to the best of our knowledge). 

Suppose we wish to guarantee $\smash{(\varepsilon, \alpha) = (0.1, 0.1)}$-marginal calibration with $B = 10$ bins using UMS. % showed a DF calibration bound based on $B$ and the number of calibration points $n$. 
We unpacked the DF calibration bound for UMS, and computed that to guarantee $\smash{(0.1, 0.1)}$-marginal calibration with $10$ bins, roughly $n \geq 17500$ is required. The detailed calculations can be found in Appendix~\ref{appsec:sample-complexity-gupta20}. This sample complexity seems conservative for a binary classification problem. In Section~\ref{sec:umd-vs-ums}, we use an illustrative experiment to show that the $n$ required to achieve the desired level of calibration is indeed much lower than $17500$. Our experiment uses a novel diagnostic tool called validity plots, introduced next.

\begin{figure*}
\begin{subfigure}[t]{0.34\linewidth}
\centering
\includegraphics[width=0.78\textwidth]{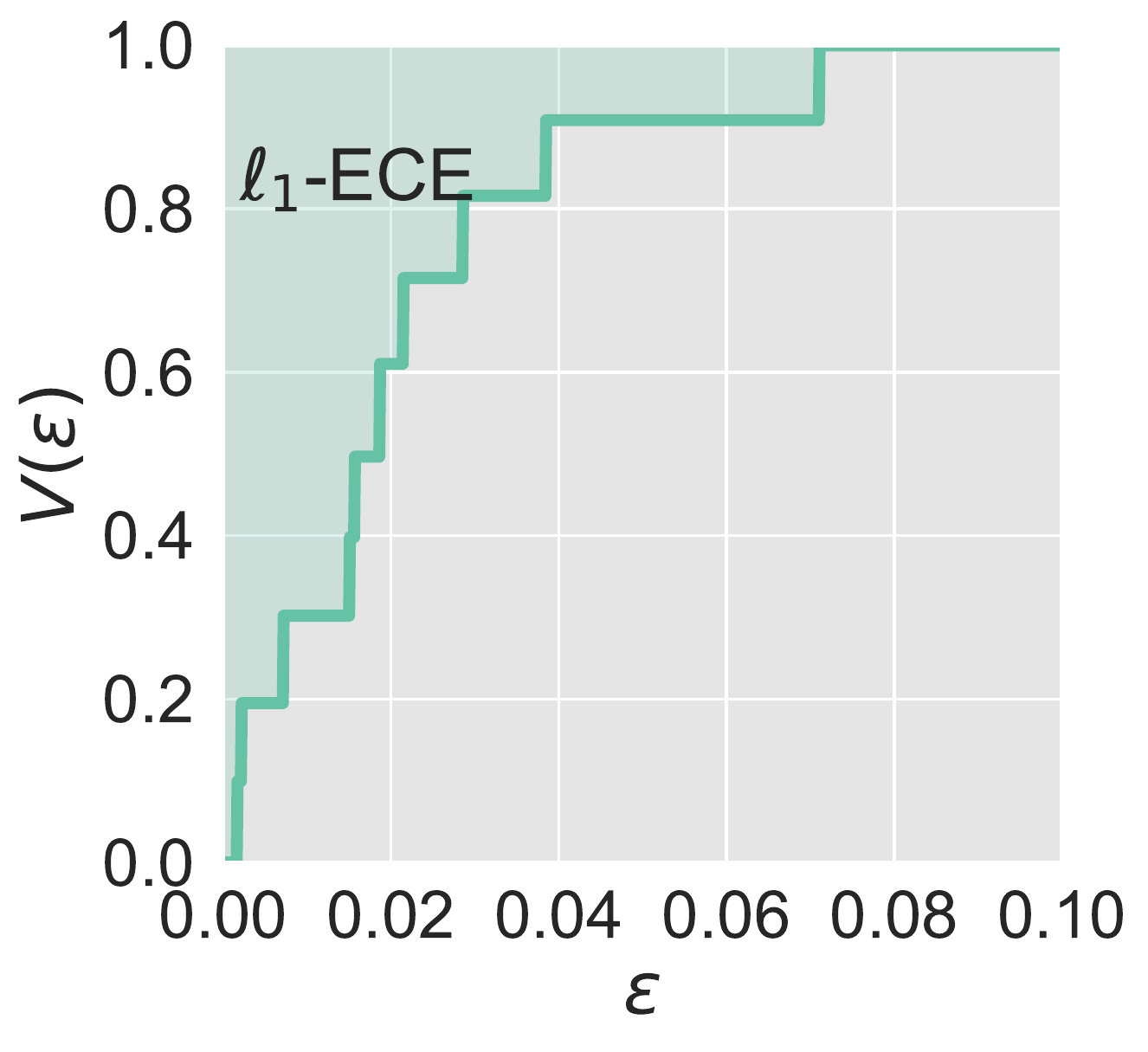}
\caption{An illustrative validity plot. We can read off that marginal calibration is achieved for $(\varepsilon, \alpha) = (0.04, 0.1)$ and $(0.03, 0.2)$. The $\ell_1$-ECE estimate is roughly $0.023$.}
\label{fig:single-validity-plot}
\end{subfigure}
\hspace{0.3cm}
\begin{subfigure}[t]{0.65\linewidth}
\centering
\includegraphics[width=0.41\textwidth]{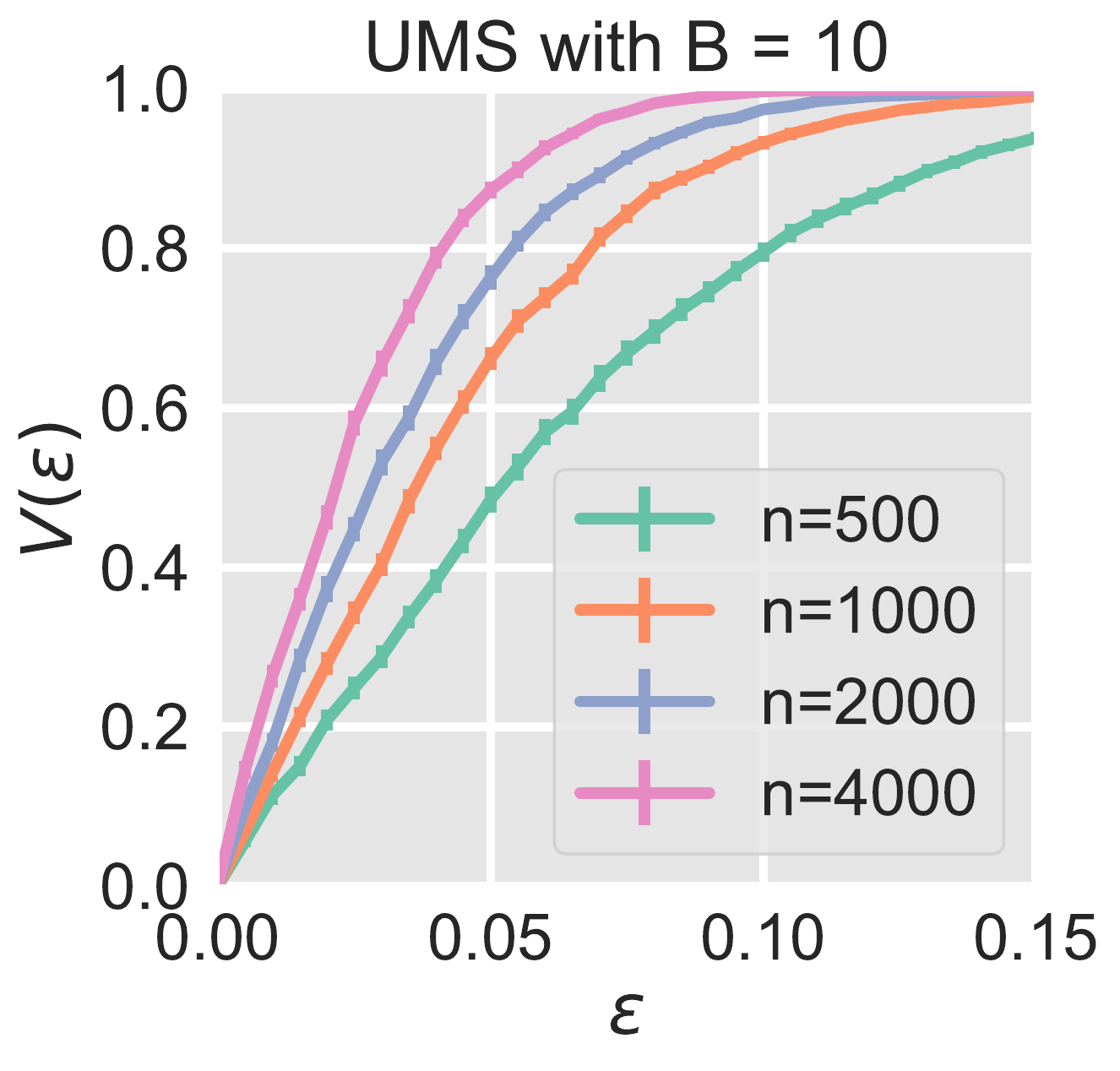}
%\caption{UMS with $B = 10$.}
%\end{subfigure}
%\begin{subfigure}[t]{0.49\linewidth}
%\centering
\includegraphics[width=0.41\textwidth]{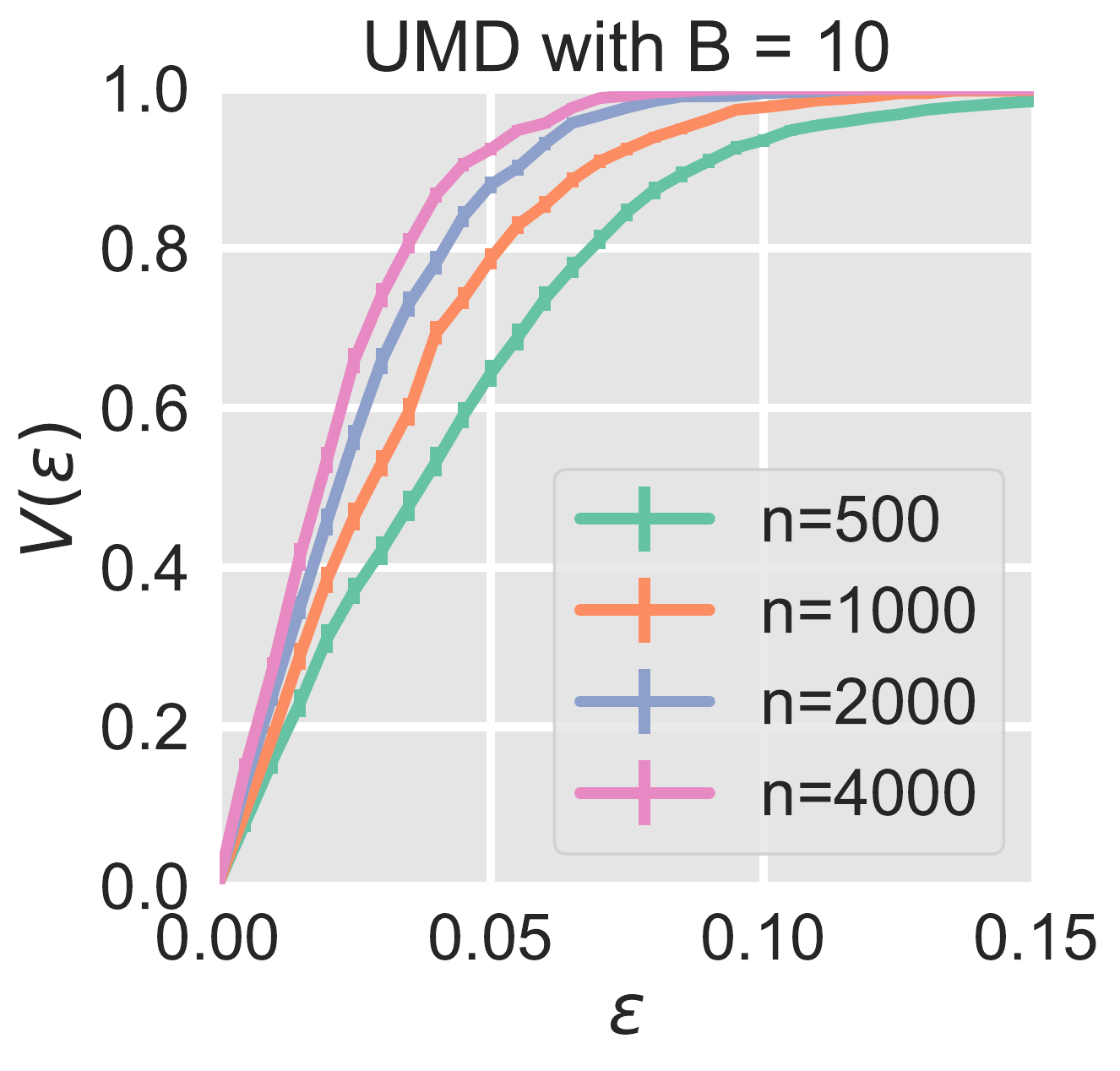}
%\caption{\ze with $B = 10$.}
\caption{Validity plots comparing \ze and UMS on the \credit dataset. %\rev{The experimental setup is described in Section~\ref{subsec:experiments-conservative}.} 
The plots show that \ze has higher validity $V(\varepsilon)$ for the same values of $n, \varepsilon$, and thus lower $\ell_1$-ECE. For example, for $n=1000$ and $\varepsilon=0.05$, UMS has $V(\varepsilon) \approx 0.63$, while \ze has  $V(\varepsilon) \approx 0.79$.}
\label{fig:credit-sample-size}
\end{subfigure}
\caption{Validity plots display estimates of $\smash{V(\varepsilon) = \Prob(\abs{\Exp{}{Y \mid h(X)} - h(X)} \leq \varepsilon)}$ as $\varepsilon$ varies. Validity plots are described in Section~\ref{sec:validity-plots}. The experimental setup for Figure~\ref{fig:credit-sample-size} 
is presented in Section~\ref{sec:umd-vs-ums}.}
\end{figure*}

\subsection{Validity plots}
\label{sec:validity-plots}
Validity plots assess the marginal calibration properties of a calibration method by displaying estimates of the LHS of \eqref{eq:marginal-calib} as $\varepsilon$ varies. Define the function  $V : [0,1] \to [0,1]$ given by $V(\varepsilon) = \Pr(\abs{\Exp{}{Y \mid h(X)} - h(X)} \leq \varepsilon)$. By definition of $V$, $H$ is $(\varepsilon, 1-V(\varepsilon))$-marginally calibrated for every $\varepsilon$. For this reason, we call the graph of $V$, $\{(\varepsilon, V(\varepsilon)) : \varepsilon \in [0,1]\}$, as the `validity curve'. (The term ``curve'' is used informally since  $V$ may have jumps.) %Validity plots are a finite sample estimate of the unknown \emph{validity curve}, described next. %In keeping with the $(\varepsilon, \alpha)$ notation, define 
Note the following neat relationship between the $\ell_1\text{-ECE}$ and the area-under-the-curve (AUC) of the validity curve: 
%\vspace{-0.1cm}
\begin{align*}
    \Exp{}{\ell_1\text{-ECE}(h)}&=\Exp{}{\abs{\Exp{}{Y \mid h(X)} - h(X)}} \\=& \int_{0}^1  \Prob(\abs{\Exp{}{Y \mid h(X)} - h(X)} > \varepsilon)\ d\varepsilon
    \\=& 1-\int_{0}^1  \Prob(\abs{\Exp{}{Y \mid h(X)} - h(X)} \leq \varepsilon)\ d\varepsilon
    \\ =& 1 - \int_{0}^1  V(\varepsilon)\ d\varepsilon = 1 - \text{AUC(validity curve)}.
\end{align*}
%\vspace{-0.3cm}
%(Thus they can be used to judge if a $(\varepsilon,\alpha)$-marginal calibration guarantee is `valid'.) 

A validity plot is a finite sample estimate of the validity curve on a single calibration set $\Dcal_n$ and test set $\Dcal_{\text{test}}$. We now outline the steps for constructing a validity plot. First, $h$ is learned using $\Dcal_n$ and $g$. Next, if $h$ is not a binning method, it must be discretized through binning in order to enable estimation of $\Exp{}{Y \mid h(X)}$. This is identical to the binning step required by plugin ECE estimators and reliability diagrams. For example, one can use fixed-width binning as described in the first paragraph of Section~\ref{sec:prior-work}. In this paper, we empirically assess only binning methods, and so an additional binning step is not necessary. 
%To estimate $\smash{\Pr(\abs{\Exp{}{Y \mid h(X)} - h(X)} \leq \varepsilon)}$ for different values of $\varepsilon$, we first fix a test 
%To produce a validity plot 
%For a given $h$ learned on the calibration data $\Dcal_n$, and a test set $\Dcal_{\text{test}}$, 
Next, the empirical distribution on $\Dcal_{\text{test}}$ is used as a proxy for the true distribution of $(X, Y)$, to estimate $V(\varepsilon)$:
%Given $h$ and $\varepsilon$  we 
%$1 - \alpha(\varepsilon) = \Pr(\abs{\Exp{}{Y \mid h(X)} - h(X)} \leq \varepsilon)$  is estimated as %estimate  $\Pr(\abs{\Exp{}{Y \mid f(X)} - f(X)} \leq \varepsilon)$ for a given $X = x$, 
\begin{align}
%\text{Pr}_{\widehat{P}}(\abs{\Exp{\widehat{P}}{Y \mid h(X)} - h(X)} \leq \varepsilon) \equiv
&\widehat{V}(\varepsilon) =\frac{\sum_{(X_i, Y_i) \in \Dcal_{\text{test}}} \indicator{\abs{\Exp{\widehat{P}}{Y \mid h(X) = h(X_i)} - h(X_i)} \leq \varepsilon} }{\abs{\Dcal_\text{test}}}, \text{ where} \nonumber \\
%\text{ where }
&\quad\Exp{\widehat{P}}{Y \mid h(X) = h(x)} \equiv \frac{\sum_{(X_i, Y_i) \in \Dcal_{\text{test}}} Y_i \rev{\indicator{h(X_i) = h(x)}}}{\sum_{(X_i, Y_i) \in \Dcal_{\text{test}}} \indicator{h(X_i) = h(x)}}. \label{eq:marginal-validity-plot}
\end{align}
For different values of $\varepsilon \in [0,1]$ on the X-axis, the estimate of %$1 - \alpha(\varepsilon)$ 
$V(\varepsilon)$ is plotted on the Y-axis to form the validity plot. %$\Pr(\abs{\Exp{}{Y \mid h(X)} - h(X)} \leq \varepsilon)$ is plotted on the Y-axis to form a validity plot. We denote this estimate as $1 - \alpha$, in keeping with equation~\eqref{eq:marginal-calib}. 
% The area-under-the-curve (AUC) of a validity plot is $1-\ell_1\text{-ECE}$: for any given $h$,
% \begin{align*}
%     \ell_1\text{-ECE}(h)&=\Exp{}{\abs{\Exp{}{Y \mid h(X)} - h(X)}} \\&= \int_{0}^1  \Prob(\abs{\Exp{}{Y \mid h(X)} - h(X)} > \varepsilon)\ d\varepsilon
%     \\&= 1-\int_{0}^1  \Prob(\abs{\Exp{}{Y \mid h(X)} - h(X)} \leq \varepsilon)\ d\varepsilon
%     \\ &= 1 - \text{AUC(validity plot)}.
% \end{align*}
Like the AUC of a validity curve corresponds to $\Exp{}{\ell_1\text{-ECE}}$, the AUC of a validity plot corresponds to the plugin $\ell_1$-ECE estimate \citep{naeini2015obtaining}. 
(There may be small differences in practice since we draw the validity plot for a finite grid of values in $[0, 1]$.) Thus validity plots convey the $\ell_1$-ECE estimate and more. 

Figure~\ref{fig:single-validity-plot} displays an illustrative validity plot for a binning method with $B=10$. $V$ is a right-continuous step function with at most $\abs{\text{Range}(h)} \leq B$ many discontinuities. Each $\varepsilon$ for which there is a discontinuity in $V$
corresponds to a bin that has $\abs{\Exp{}{Y \mid h(X) = r} - r} = \varepsilon$, and the incremental jump in the value of $V$, $V(\varepsilon) - V(\varepsilon^{-})$, corresponds to the fraction of test points in that bin. Figure~\ref{fig:single-validity-plot} was created using UMD, and thus each jump corresponds to roughly a $1/B = 0.1$ fraction of the test points. The $\varepsilon$ values for the bins are approximately $10^{-3} \cdot (1.5, 2, 8, 16, 17, 19, 22, 29, 39, 71)$. 

Unlike reliability diagrams \citep{Niculescu2005predicting}, validity plots do not convey the predictions $h(X)$ to which the $\varepsilon$ values correspond to, or the direction of miscalibration (whether $h(X)$ is higher or lower than $\Exp{}{Y \mid h(X)}$). On the other hand, validity plots convey the bin frequencies for every bin without the need for a separate histogram (such as the top panel in \citet[Figure 1]{Niculescu2005predicting}). In our view, validity plots also `collate' the right entity; we can easily read off from a validity plot practically meaningful statements such as ``for 90\% of the test points, the miscalibration is at most $0.04$".

We can create a smoother validity plot that better estimates $V$ by using multiple runs based on subsampled or bootstrapped data. To do this, for every $\varepsilon \in [0,1]$, $\widehat{V}(\varepsilon)$ is computed separately for each run and the mean value is plotted as the estimate of $V(\varepsilon)$. In our simulations, we always perform multiple runs, and also show $\pm$std-dev-of-mean in the plot. Figure~\ref{fig:credit-sample-size} displays such validity plots (further details presented in the following subsection). 

It is well known that plugin ECE estimators for a binned method are biased towards slightly overestimating the ECE (e.g., see \citet{brocker2012estimating, kumar2019calibration, widmann2019calibration}). %, in the sense that $\abs{\Exp{\widehat{P}}{Y \mid h(X)} - h(X)} > \abs{\Exp{}{Y \mid h(X)} - h(X)}$ is more likely to occur than the other way around. 
%Validity plots have the same kind of bias, which leads to slightly underestimating validity, i.e. curves in Fig 1 are below ... . 
For the same reasons, $\widehat{V}(\varepsilon)$ is a biased underestimate of $V(\varepsilon)$. In other words, the validity plot is on average below the true validity curve.  %  Validity plots have the same kind of bias; the curve in Figure~\ref{fig:single-validity-plot} is 
The reason for this bias is that to estimate ECE as well as to create validity plots, we compute $\abs{\Exp{\widehat{P}}{Y \mid h(X)} - h(X)}$ which can be written as $\abs{\Exp{}{Y \mid h(X)} + \text{mean-zero-noise} - h(X)}$. %This noise could lead to underestimating $\abs{\Exp{}{Y \mid h(X)} - h(X)}$ in some cases, but 
On average, the noise term will lead to overestimating $\abs{\Exp{}{Y \mid h(X)} - h(X)}$.  %In other words, the validity plot for all methods suggests slightly lesser validity than there actually is%, which makes it possible for the Theorem~\ref{thm:umd-binary-randomized} curve to cross the empirical UMD curve for $n=7$K. This phenomenon is exactly the same as the presence of bias in ECE estimates \citep{brocker2012estimating, kumar2019calibration}
However, the noise term is small if there is enough test data (if $n_b$ is the number of test points in bin $b$, then the noise term is $O(\sqrt{1/n_b})$ w.h.p.). Further, it is highly unlikely that the noise will help some methods and hurts others. Thus validity plots can be reliably used to make inferences on the relative performance of different calibration methods. While there exist unbiased estimators for $(\ell_2\text{-ECE})^2$ \citep{brocker2012estimating, widmann2019calibration}, we are not aware of any unbiased $\ell_1$-ECE estimators. If such an estimator is proposed in the future, the same technique will also improve validity plots.

%\subsection{\smash{\ze has better empirical properties}}
\subsection{Comparing UMS and \ze using validity plots}
\label{sec:umd-vs-ums}
Figure~\ref{fig:credit-sample-size} uses validity plots to assess UMS and \ze on \credit, a UCI credit default dataset\footnote{\citet{yeh2009comparisons}; {\scriptsize \url{https://archive.ics.uci.edu/ml/datasets/default+of+credit+card+clients}}}. The task is to accurately predict the probability of default. The experimental protocol is as follows. The entire feature matrix is first normalized\footnote{using Python's \texttt{sklearn.preprocessing.scale}}. \credit has 30K (30,000) samples which are randomly split (once for the entire experiment) into splits (A, B, C) = (10K, 5K, 15K). \rev{First, $g$ is formed by training a logistic regression model on split A and then re-scaling the learnt model using Platt scaling on split B} (Platt scaling before binning was suggested by \citet{kumar2019calibration}; we also observed that this helps in practice). Next, the calibration set $\Dcal_n$ is formed by randomly subsampling $n$ ($\leq$10K) points from split C (without replacement). From the remaining points in split C, a test set of size 5K is subsampled (without replacement). The entire subsampling from split C is repeated 100 times to create 100 different calibration and test sets. For a given subsample, UMS/\ze with $B = 10$ is trained on the calibration set (with 50:50 sample splitting for UMS), %\ze is trained without sample splitting, 
and $\widehat{V}(\varepsilon)$ for every $\varepsilon$ is estimated on the test set. Finally, the (mean$\pm$std-dev-of-mean) of $\widehat{V}(\varepsilon)$ is plotted with respect to $\varepsilon$. This experimental setup assesses marginal calibration for a fixed $g$, in keeping with our post-hoc calibration setting.

The validity plot in Figure~\ref{fig:credit-sample-size} (left) indicates that the desired $(0.1, 0.1)$-marginal calibration is achieved by UMS with just $\smash{n = 1000}$. Contrast this to $\smash{n \geq 17500}$ required by the theoretical bound, as computed in Appendix~\ref{appsec:sample-complexity-gupta20}. In fact,  $\smash{n = 4000}$ nearly achieves $(0.05, 0.1)$-marginal calibration. This gap occurs because the analysis of UMS is complex, with constants stacking up at each step. 

%Figure~\ref{fig:credit-sample-size} (right) presents validity plots for \ze on \credit, with $\smash{B=10}$ and different values of $n$. 
Next, consider the validity plot for \ze in Figure~\ref{fig:credit-sample-size} (right). By avoiding sample splitting, \ze achieves $(0.1, 0.1)$-marginal calibration at $\smash{n = 500}$. In Section~\ref{sec:adaptive-binning} we show that $n \geq 1500$ is provably sufficient for $(0.1, 0.1)$-marginal calibration and $n \geq 2900$ is sufficient for $(0.1, 0.1)$-conditional calibration. Some gap in theory and practice is expected since the theoretical bound is DF, and thus applies no matter how anomalous the data distribution is. However, the gap is much smaller compared to UMS, due to a clean analysis. In Section~\ref{sec:simulations}, we illustrate that the gap nearly vanishes for larger $n$. Section~\ref{sec:simulations} also introduces the related concept of \textit{conditional} validity plots that assess conditional calibration. %, for which we showed that the theoretical guarantee requires (roughly) $n \geq 17500$. 

\section{Distribution-free analysis of uniform-mass binning without sample splitting}
\label{sec:adaptive-binning}
%Later, we connect this setup to top-label binning by imagining that the random variable $S$ corresponds to the score that a scoring function gives for the top-label. 
%In this section we propose a novel uniform-mass binning scheme that does not rely on sample splitting. %This algorithm can be used both for two-class classification and multiclass top-label/label-wise calibration. For simplicity, we start with two-class calibration. 
Define the random variables $S = g(X)$; $S_i = g(X_i)$ for $\smash{i \in [n]}$, called scores. % over $[0, 1]$, to be thought of as the score a scoring function provides for the label $Y = 1$. %(Later, we connect this setup to top-label binning by imagining that the random variable $S$ corresponds to the score that a scoring function gives for the top-label.) %for instance, in top-label binning, this would be given by the $t(X)$. 
%For a specific value of $l$, consider 
Let $\smash{(S, Y)\sim Q}$ and $\smash{S \sim Q_S}$. %Let us denote the joint distribution of the score and the label as $\smash{(S, Y)\sim Q}$ and the marginal distribution of the score as $\smash{S \sim Q_S}$. %Note that $Q$ is over $[0, 1] \times \{0, 1\}$. %We denote a typical draw from $Q$ as $(S, Y)$. 
% In the recalibration setting, the goal is, roughly, to estimate the Bayes function $\Pi(\cdot) :=  \Exp{}{Y \mid S = \cdot}$ using the calibration data:
% $\smash{\{(S_i, Y_i)\}_{i \in [n]} \sim Q^n}$. %$\smash{(S_1, Y_1), (S_2, Y_2), \ldots, (S_n, Y_n) \sim Q^n}$. 
% Let us denote the estimate of $\Pi$ as $\widehat\Pi$. 
In binning, we wish to use the calibration data
$\smash{\{(S_i, Y_i)\}_{i \in [n]} \sim Q^n}$ to (a) define a binning function $\Bcal: [0, 1] \to [B]$ for some number of bins $B \in \naturals$, and (b) estimate the biases in the bins $\smash{\{\Pi_b :=  \Exp{}{Y \mid \Bcal(S) = b}\}_{b \in [B]}}$. We denote the bias estimates as $\widehat{\Pi}_b$. The approximately calibrated function is then defined as $h(\cdot) = \widehat{\Pi}_{\Bcal(\cdot)}$. %$\smash{(S_1, Y_1), (S_2, Y_2), \ldots, (S_n, Y_n) \sim Q^n}$. 
%Let us denote the estimate of $\Pi$ as $\widehat\Pi$.

Suppose the number of recalibration points is $\smash{n \approx 150}$. In the absence of known properties of the data (i.e., in the DF setting), it seems reasonable to have $\smash{B = 1}$ and define $H(g, \Dcal_n)$ as the constant function
$h(\cdot) := n^{-1}\sum_{i=1}^n Y_i.$ Formally, $\smash{n = 150}$ leads to the following Hoeffding-based confidence interval: with probability at least $0.9$, $\abs{n^{-1}\sum_{i=1}^n Y_i - \E Y} \leq \sqrt{\log(2/0.1)/(2\cdot 150)} \approx 0.1$.   %if $n \leq 150$ and we want  
In other words, if $\smash{n=150}$, $H$ satisfies $(0.1, 0.1)$-marginal calibration. Of course, having a single bin completely destroys sharpness of $h$, but it's an instructive special case. 

Suppose now that $n \approx 300$, and we wish to learn a non-constant $h$ using two bins. %It makes sense to split the $S$-values into two bins, and estimate the bias in each bin. 
If $g$ is informative,
%higher values of $g$ indicate higher values of $\Pi$; 
% in other words we hope that $\Pi$ is roughly monotonically increasing %$0 \leq s_1 \leq s_2 \leq 1$, it holds that $\Pi_{s_1} \leq \Pi_{s_2}$ 
we hope that $\Exp{}{Y \mid g(X) = \cdot}$ is roughly a monotonically increasing function. 
%(we don't actually require this to be true for the formal results; it is just a belief that guides us on how to identify the binning partition). 
In light of this belief, it seems reasonable to choose a threshold $t$ and identify the two bins as: $\smash{g(X) \leq t}$ and $\smash{g(X) > t}$. A natural choice for $t$ is $M = \text{Median}(S_1, \ldots, S_n)$ since this ensures that both bins get the same number of points (plus/minus one). This is the motivation for \ze. In this case, $h$ and $\widehat{\Pi}$ are defined as,%\vspace{-0.2cm}%the calibration function $h$ is defined though $\widehat{\Pi}$ as, %Suppose we use the calibration data to estimate the median as well as compute $\widehat{\Pi}$: %What can be said for the CIs given that the median has also been chosen using the data? 
%Formally, let $M = \text{Median}(S_1, \ldots, S_n)$ and consider the predictor given by: 
\begin{equation}
    h(\cdot) :=  \left\{
	\begin{array}{ll}
		\widehat{\Pi}_1 := \text{Average}(\smash{Y_i : S_i \leq M})  \mbox{ if } g(\cdot) \leq M \\
		\widehat{\Pi}_2 := \text{Average}(\smash{Y_i : S_i > M}) \mbox{ if } g(\cdot) > M.
	\end{array}
\right.\label{eq:instructive-two-bins}
\end{equation}
Suppose $M$ were the true median of $Q_S$ instead of the empirical median. Then %the bins defined using this median 
$h$ has a calibration guarantee obtained by applying a Bernoulli concentration inequality separately for both bins and using a union bound (this is done formally by \citet[Theorem~4]{gupta2020distribution}). % who used empirical-Bernstein~\citep{audibert2007tuning}). 
In UMS, we try to emulate the true median case by using one split of the data to estimate the median. $\widehat{\Pi}$ is then computed on the second (independent) split of the data, and concentration inequalities can be used to provide calibration guarantees. % (guarantees for this are shown by \citep[Theorem~5]{gupta2020distribution}). %This $\widehat{\Pi}$ has a Hoeffding based calibration guarantee \citep[Theorem~5]{gupta2020distribution}.
%However, as discussed in Section~\ref{sec:sample-complexity-gupta20}, the practical and theoretical sample complexity of UMS is poor. 
%high due to sample splitting and the theoretical guarantee \citep[Theorem~5]{gupta2020distribution} is loose due to a complex analysis. %split the second half of the data into the two bins, and then apply Hoeffding's inequality within each bin and take a union bound. 

%However, we may want to use the same data to estimate the median, as well as compute the biases in each bin. 
%In the setting described above however, 
\ze does not sample split: in equation~\eqref{eq:instructive-two-bins} above, $M$ is computed using the same data that is later used to estimate $\widehat{\Pi}$. On the face of it, this double dipping eliminates the independence of the $Y_i$ values required to apply a concentration inequality. However, we show that the independence structure can be retained if \ze is slightly modified. This subtle modification is to remove a single point from the bias estimation, namely the $Y_i$ corresponding to the median $M$. (In comparison, in UMS we typically remove a fixed ratio of $n$.) The informal argument is as follows.

For simplicity, suppose $Q_S$ is absolutely continuous (with respect to the Lebesgue measure), so that the $S_i$'s are almost surely distinct, and suppose that the number of samples is odd: $n = 2m+1$. Denote the ordered scores as $\smash{S_{(1)} < S_{(2)} 
<\ldots < S_{(n)}}$ and let $Y_{(i)}$ denote the label corresponding to the score $S_{(i)}$. Thus $\smash{\widehat{\Pi}_1 = m^{-1}\sum_{i = 1}^m Y_{(i)}}$ and $M = S_{(m+1)}$. %Given $\varepsilon \in (0, 1)$, we show that the probability 
% \begin{equation}
% \Prob\roundbrack{|\Exp{}{Y}  - \widehat{\Pi}_1| \geq \varepsilon \mid M =  S_{(m+1)}, S < M},
% \end{equation}
% can in fact be bounded using Hoeffding, which corresponds to a calibration guarantee on the first bin. The informal argument is as follows. %In order to use Hoeffding, we show that in fact $Y_{(i)} \mid S_{(i)} < M \overset{i.i.d.}{\sim} Y \mid S < M$. 
% We can bound this probability if we can bound the following probability: 
% \[
% \Prob\roundbrack{|\Exp{}{Y \mid S \leq M} - m^{-1}\sum_{i = 1}^m Y_{(i)}| \geq \varepsilon \mid S_{(n)} , S_{(n+1)}}.
% \]
%Let us denote the distribution of $(S, Y) \mid S < S_{(m+1)}$ as $Q_{< S_{(m+1)}}$. %Thus $\Exp{}{Y \mid S < M} = \Exp{ P_{< M}}{Y}$.
%In Section~\ref{subsec:order-statistics-lemmas}, we show that the samples $\{(S_{(i)}, Y_{(i)})\}_{i \in [m]}$ can be viewed as $m$ i.i.d. samples drawn from the distribution $Q_{< S_{(m+1)}}$. 
Clearly, $(S_{(i)}, Y_{(i)})$ is not independent of $S_{(m+1)}$ for any $i$. However, it turns out that the following property is true: conditioned on $S_{(m+1)}$, the unordered values $\{(S_{(i)}, Y_{(i)})\}_{i \in [m]}$ can be viewed as $m$ \emph{independent} samples identically distributed as $\smash{(S, Y)}$, given $S < S_{(m+1)}$. (Note that $(S, Y)$
is an unseen and independent random variable.) 
Thus, we can use Hoeffding's inequality to assert:
$\Prob(|\Exp{}{Y \mid M, S < M} - \widehat{\Pi}_1| \geq \varepsilon \mid  M, S < M) %= \Prob\roundbrack{|\Exp{P_{< S_{(m+1)}}}{Y} - m^{-1}\sum_{i = 1}^m Y_{(i)}| \geq \varepsilon \mid S_{(m+1)}} 
%\\ &\qquad\qquad\qquad\qquad\qquad\qquad\leq 
\leq 2\expText{-2m\varepsilon^2}.
$
This can be converted to a calibration guarantee on the first bin. The same bound can be shown if $S > M$, for the estimate $\widehat{\Pi}_2 = m^{-1}\sum_{i=m+1}^{2m+1} Y_{(i)}$. Using a union bound gives a calibration guarantee that holds for both bins simultaneously, which in turn gives conditional calibration. 

% Similarly for the second bin, the samples $(S_{(i)}, Y_{(i)})$, $m+2\leq i \leq 2m+1 $ are $m$ samples from $P_{\geq M}$, which is identical to $P_{> M}$ (since $Q_S$ is assumed to be nonatomic). Thus, 
% \[
% \Prob\roundbrack{|\Exp{P_{> S_{(m+1)}}}{Y} - m^{-1}\sum_{i = m+2}^{2m-1} Y_{(i)}| \geq \varepsilon \mid S_{(m+1)}} \leq 2\expText{-2m\varepsilon^2},
% \]
% which is a calibration guarantee for the second bin.
In the following subsection, we show some key lemmas regarding the order statistics of the $S_i$'s. These lemmas formalize what was argued above: \textit{careful double dipping does not eliminate the independence structure}. In  Section~\ref{subsec:umd-binary}, we formalize the modified \ze algorithm, and prove that it is DF calibrated. %We refer to this modified version as \modifiedze. 
Based on the guarantee for the modified version, Corollary~\ref{cor:ze} finally shows that the original \ze itself is DF calibrated. 

\textbf{Simplifying assumption. }In the following analysis, we assume that $g(X)$ is absolutely continuous with respect to the Lebesgue measure, and thus has a probability density function (pdf). This assumption is made at no loss of generality, for reasons discussed in Appendix~\ref{sec:continuity-assumption}.% (for smoother readability). 

\subsection{Key lemmas on order statistics }
\label{subsec:order-statistics-lemmas}
Consider two indices $i, j \in [n]$. The score $S_i$ is  not independent of the order statistic $S_{(j)}$. However, it turns out that conditioned on $S_{(j)}$, the distribution of $S_i$ given $S_i < S_{(j)}$, is identical to the distribution of an unseen score $S$, given $S < S_{(j)}$. The following lemmas (both proved in Appendix~\ref{appsec:adaptive-binning-proofs}) state versions of this fact that are useful for our analysis of \ze.

We first set up some notation. $S$ is assumed to have a pdf, denoted as $f$. %Conditioned on fixed values of order statistics at some indices $l, u \in [n]$, $l < u$, 
For some $\smash{1 \leq l < u \leq n}$, consider the set of indices $\smash{\{i : S_{(l)} < S_i < S_{(u)}\}}$, and index them arbitrarily as $\{t_1, t_2, \ldots, t_{u-l-1}\}$. This is just an indexing and not an ordering; in particular it is not necessary that $S_{t_1} = S_{(l+1)}$. For $j \in \{l + 1, \ldots, u - 1\}$, define $S_{\{j\}} = S_{t_{j-l}}$. Thus the set $\{S_{\{j\}} : j \in \{l + 1, \ldots, u - 1\} \}$ corresponds to the \textit{unordered} $S_i$ values between $S_{(l)}$ and $S_{(u)}$.

\begin{lemma}
\label{lemma:order-statistics-1}
% Then for any $j \in [n]$,
% the density of $S_j$ conditioned on the event $S_{(k_{i-1})} < S_j <  S_{(k_{i})}$  and the observed values of $S_{(k_{i-1})},  S_{(k_{i})}$
% \[
% f(s_j \mid S_{(k_i)} < S_j <  S_{(k_{i+1})}) %= f(s \mid S_{(k_i)} < S <  S_{(k_{i+1})})
% \]
Fix $l, u \in [n]$ such that $l < u$. The conditional density of the unordered $S_i$ values between the order statistics $S_{(l)}, S_{(u)}$,
$
%f(S_{\{l+1\}} = s_{l+1}, \ldots, S_{\{u-1\}} = s_{u-1}  \mid S_{(l)} =a , S_{(u)} =b)
f(S_{\{l+1\}}, \ldots, S_{\{u-1\}}   \mid S_{(l)}, S_{(u)}),
$
is identical to the  density of independent  $S'_i \sim Q_S$, conditional on lying between $S_{(l)}, S_{(u)}$:
\begin{align*}
%\vspace{-0.5cm}
%&f(S'_1 = s_{l+1}, S'_2 = s_{u+1}, \ldots, S'_{u-l-1} = s_{u-1} \mid S_{(l)} = a,  S_{(u)} = b, \\&\qquad\qquad\qquad\qquad 
%\text{for every $i \in [u-l-1]$, $S_{(l)} < S'_{i} < S_{(u)}$}))
f(S'_1, \ldots, S'_{u-l-1} \mid S_{(l)},  S_{(u)}, 
% \text{for every $i \in [u-l-1], 
S_{(l)} < \{S'_{i}\}_{i \in [u-l-1]} < S_{(u)}).
\end{align*}
 \end{lemma}

%The proof of  %Lemma~\ref{lemma:order-statistics-1} 
%is in Appendix~\ref{appsec:adaptive-binning-proofs}. %In words, Lemma~\ref{lemma:order-statistics-1} shows that each of the $S_i$ values between two given order statistics is identically distributed as an independent random variable $S \sim Q_S$ given that $S$ lies between the order statistics. 
%The above result is conditional on $S_{(l)}$ and $S_{(u)}$. 
In the final analysis, $S_{(l)}$ and $S_{(u)}$ will represent the scores at consecutive bin boundaries, which define the binning scheme. Lemma~\ref{lemma:order-statistics-main} is similar to Lemma~\ref{lemma:order-statistics-1}, but with conditioning on all bin boundaries (order statistics) simultaneously. %However, we will also need to maintain the independence structure conditioned on all bin boundaries simultaneously. The next lemma shows this. 
%is similar to Lemma~\ref{lemma:order-statistics-1}, but conditional on $B-1$ order statistics (corresponding to all bin boundaries) simultaneously. 
To state it concisely, define $S_{(0)} := 0$ and $S_{(n+1)} := 1$ as fixed hypothetical `order statistics'. %with a fixed value at $0$. Similarly we define  as a hypothetical `order statistics' with a fixed value at $1$. 

\begin{lemma}
\label{lemma:order-statistics-main}
% Fix $k_0 = 0, k_B = n+1$ % and fix  $S_{(k_0)} = 0, S_{(k_B)} = 1$.
% and any $B-1$ indices $k_1, k_2, \ldots,  k_{B-1} \in [n]$ such that for every $b \in [B]$, $k_{b-1} < k_b$. 
Fix any $\smash{B-1}$ indices $k_1, k_2, \ldots k_{B-1}$ such that $\smash{0 = k_0 < k_1 < \ldots < k_{B-1} < k_B = n+1}$. % and fix  $S_{(k_0)} = 0, S_{(k_B)} = 1$.
%and any $B-1$ indices $k_1, k_2, \ldots,  k_{B-1} \in [n]$ such that for every $b \in [B]$, $k_{b-1} < k_b$. 
For any $b \in [B]$, the conditional density of the unordered $S_i$ values between the order statistics $S_{(k_{b-1})}, S_{(k_b)}$,
$
%f(S_{\{k_{b-1}+1\}} = s_{k_{b-1}+1}, \ldots, S_{\{k_b-1\}} = s_{k_b-1}  \mid S_{(k_0)} = s_{(k_0)}, \ldots, S_{(k_B)} = s_{(k_B)})
f(S_{\{k_{b-1}+1\}}, \ldots, S_{\{k_b-1\}}  \mid S_{(k_0)} , \ldots, S_{(k_B)})
$,
is identical to the conditional density
\begin{align*}
%\vspace{-0.5cm}
%&f(S'_1 = s_{k_{b-1}+1}, \ldots, S'_{k_b-k_{b-1}-1} = s_{k_b-1} \mid S_{(k_0)} = s_{(k_0)}, \ldots, S_{(k_B)} = s_{(k_B)}, \\&\qquad\qquad\qquad\qquad 
%\text{for every $i \in [k_b-k_{b-1}-1]$, $S_{(k_{b-1})} < S'_{i} < S_{(k_b)}$}))
%f(S'_1, \ldots, S'_{k_b-k_{b-1}-1} \mid S_{(k_0)}, \ldots, S_{(k_B)},
%\text{ for every $i \in [k_b-k_{b-1}-1]$, $S_{(k_{b-1})} < S'_{i} < S_{(k_b)}$}))
&f(S'_1, \ldots, S'_{k_b-k_{b-1}-1} \mid S_{(k_0)}, \ldots, S_{(k_B)}, \\&\qquad
\text{for every $i \in [k_b-k_{b-1}-1]$, $S_{(k_{b-1})} < S'_{i} < S_{(k_b)}$}))
\end{align*}
of independent random variables $S'_i \sim Q_S$.
\end{lemma}
%The proof is in Appendix~\ref{appsec:adaptive-binning-proofs}. %We are now ready to analyze \ze.

\subsection{Main results}
\label{subsec:umd-binary}
%We are now ready to state and analyze the sample-efficient versions of uniform-mass binning, and prove calibration guarantees for them. Our algorithm `double-dips' in the calibration data: the same data is used for quantile estimation as well as mean-estimation per bin. We denote this technique as UMD, short for Uniform-Mass-Double-dipping.

\begin{algorithm}[!t]
\caption{\ze: Uniform-mass binning without sample splitting}
	\label{alg:efficient-uniform-mass-binary}
	\begin{algorithmic}[1]
	%\SetAlgoLined
	\STATE {\bfseries Input:} Scoring function $\smash{g : \Xcal \to [0, 1]}$, \#bins $B$, calibration data $(X_1, Y_1), (X_2, Y_2), \ldots, (X_n, Y_n)$%, test point $(X, Y)$}
	\STATE {\bfseries Output:} {Approximately calibrated  function $h$ }
	\STATE $(S_1, S_2, \ldots, S_n) \gets (g(X_1), g(X_2), \ldots, g(X_n))$\;
	\STATE  $\smash{(S_{(1)}, S_{(2)}, \ldots, S_{(n)}) \gets \text{order-stats}(S_1, S_2, \ldots, S_n)}$\;
	\STATE  $(Y_{(1)}, Y_{(2)}, \ldots, Y_{(n)}) \gets (Y_1, Y_2, \ldots, Y_n)$ ordered as per the ordering of $(S_{(1)}, S_{(2)}, \ldots, S_{(n)})$\;
	\STATE  $\Delta \gets (n+1)/B$\;
    \STATE $\widehat{\Pi} \gets$ empty array of size $B$\;
    \STATE $A \gets 0\text{-indexed array}([0, \lceil \Delta\rceil, \lceil 2 \Delta\rceil, \ldots, n+1])$\;
	\FOR{$b \gets 1$ \textbf{to} $B$}
	    %$l \gets \lceil b\times \Delta\rceil$\\;
        \STATE $l \gets A_{b-1}$\;
        \STATE $u \gets A_{b}$\;
	    \STATE $\widehat{\Pi}_b \gets$ Mean($Y_{(l + 1)}, Y_{(l + 2)}, \ldots, Y_{(u-1)}$)\;
	    \label{line:removed-mean}
	\ENDFOR
	\STATE \rev{ $(S_{(0)}, S_{(n+1)}) \gets (0, 1)$}\;
	\STATE $h(\cdot) \gets \sum_{b=1}^B \indicator{S_{(A_{b-1})} \leq g(\cdot) < S_{(A_{b})}} \widehat{\Pi}_b$\; 
	\end{algorithmic}
\end{algorithm}

\ze is described in Algorithm~\ref{alg:efficient-uniform-mass-binary} (in the description, $\lfloor \cdot \rfloor$ and $\lceil \cdot \rceil$ denote the floor and ceiling operators respectively). \ze takes input $(g, \Dcal_n)$ and outputs $h$. There is a small difference between \ze as stated and the proposal by \citet{zadrozny2001obtaining}. The original version also uses the calibration points that define the bin boundaries for bias estimation --- this corresponds to replacing line~\ref{line:removed-mean} with %\vspace{-0.2cm}
\[\text{line~\ref{line:removed-mean}: } \widehat{\Pi}_b \gets \text{Mean}(Y_{(l+1)}, \ldots, Y_{(u-1)}, \boldsymbol{Y_{(u)}}), %(except if $u = n+1$, which corresponds to the last bin). 
\text{for } b < B. \]
The two algorithms are virtually the same; after stating the calibration guarantee for \ze, we show the result for the original proposal as a corollary. 

By construction, every bin defined by \ze has at least $\smash{\lfloor n/B \rfloor - 1}$ many points for mean estimation. Thus, \ze effectively `uses' only $B-1$ points for bin formulation using quantile estimation. We prove the following calibration guarantee for \ze in Appendix~\ref{appsec:adaptive-binning-proofs}.

% \begin{theorem}
% \label{thm:UMD-binary}
% Suppose the pdf of $g(X)$ is continuous and $\delta \in (0, 1)$. With probability at least $1 - \delta$, simultaneously for every $b \in [B]$,
% \begin{equation}
%     \abs{\Exp{}{Y \mid S_{(k_b)} \leq g(X) < S_{(k_{b+1})}} - \widehat{\Pi}(b)} \leq \sqrt{\frac{\log(2B/\delta)}{2(\lfloor n/B\rfloor -1)}}.
%     \label{eq:epsilon-modifiedze}
% \end{equation}
% As a consequence, we also obtain 
% \[
%     \abs{\Exp{}{Y \mid \widehat{\Pi}(B(g(X)))} - \widehat{\Pi}(B(g(X)))} \leq \sqrt{\frac{\log(2B/\delta)}{2(\lfloor n/B\rfloor -1)}}.
% \]
% % Then the predictor $h$ returned by Algorithm~\ref{alg:efficient-uniform-mass-binary} satisfies with probability $ 1 - \delta$,
% \end{theorem}

\begin{theorem}
\label{thm:umd-binary}
Suppose $g(X)$ is absolutely continuous with respect to the Lebesgue measure and $n \geq 2B$. \ze is $(\varepsilon, \alpha)$-conditionally calibrated for any $\alpha \in (0, 1)$ and%\vspace{-0.2cm}
%For any $\alpha \in (0, 1)$ and 
\begin{equation}
    \varepsilon =\sqrt{\frac{\log(2B/\alpha)}{2(\lfloor n/B\rfloor -1)}}.
    \label{eq:epsilon-modifiedze}%\vspace{-0.2cm}
\end{equation}
Further, for every distribution $P$, w.p. $1-\alpha$ over the calibration data $\Dcal_n$, for all $p \in [1, \infty)$, $\ell_p\text{-ECE}(h) \leq \varepsilon$.
%the function $h$ provided by Algorithm~\ref{alg:efficient-uniform-mass-binary} is $(\varepsilon, \alpha)$-conditionally calibrated.
\end{theorem}

Note that since \ze is $(\varepsilon, \alpha)$-conditionally calibrated, it is also $(\varepsilon', \alpha)$-conditionally calibrated for any $\varepsilon' \in (\varepsilon, 1)$. The absolute continuity requirement for $g(X)$ can be removed with a randomization trick discussed in Section~\ref{sec:continuity-assumption}, to make the result fully DF. The proof sketch is as follows. Given the bin boundaries, the scores in each bin are independent, as shown by Lemma~\ref{lemma:order-statistics-main}. We use this to conclude that the $Y_i$ values in each bin $b$ are independent and distributed as $\text{Bern}(\Exp{}{Y \mid \Bcal(X) = b})$. The average of the $Y_i$ values thus concentrates around $\Exp{}{Y \mid \Bcal(X) = b}$. Since each bin has at least $(\lfloor n/B \rfloor - 1)$ points, Hoeffding's inequality along with a union bound across bins gives conditional calibration for the value of $\varepsilon$ in \eqref{eq:epsilon-modifiedze}. 

The convenient property that every bin has at least $\lfloor n/B\rfloor - 1$ calibration points for mean estimation is not satisfied deterministically even if we used the true quantiles of $g(X)$. In fact, as long as $B = o(n)$, the $\varepsilon$ %guaranteed by Theorem~\ref{thm:umd-binary}
in \eqref{eq:epsilon-modifiedze} approaches the $\varepsilon$ we would get if all the data was used for bias estimation, with at least $\lfloor n/B \rfloor$ points in each bin:%\vspace{-0.2cm}
\[
\text{if $B = o(n)$, }\lim_{n \to \infty} \abs{\sqrt{\frac{\log(2B/\alpha)}{2(\lfloor n/B\rfloor -1)}} - \sqrt{\frac{\log(2B/\alpha)}{2(\lfloor n/B\rfloor)}}} = 0.
\]
In comparison to the clean proof sketch above, UMS requires a tedious multi-step analysis: %Previous methods for uniform-mass binning \citep{gupta2020distribution, kumar2019calibration} have the following drawbacks which UMD eliminates:
\begin{enumerate}
    \item Suppose the sizes of the two splits are $n_1$ and $n_2$. Performing reliable quantile estimation on the first split of the data requires $n_1 = \Omega(B \log(B/\alpha))$ \citep[Lemma 4.3]{kumar2019calibration}). %Previous methods split the calibration data into two (say $\Dcal_{\text{cal}}^1$ and $\Dcal_{\text{cal}}^2$) and use only $\Dcal_{\text{cal}}^1$ for quantile estimation. They require $\abs{\Dcal_{\text{cal}}^1} = \Omega(B \log(B/\alpha))$ samples for reliable quantile estimation \citep[Lemma 4.3]{kumar2019calibration}. 
    \item The estimated quantiles have the guarantee that the \textit{expected} number of points falling into a bin, on the second split is $\geq n_2/2B$. %, for some $k > 1$. The existing analysis uses $k = 2$. % The result of \citet[Lemma 4.3]{kumar2019calibration} proves this for $c \geq 2$ and we are not aware of better results. 
    %Thus, on the second split, the expected number of points in every bin is lower bounded by $n_2/2B$.
    %\item Finally, a high 
    A high probability bound is used to lower bound the actual number of points in each bin. This lower bound is $(n_2/2B) - \sqrt{n_2\log(2B/\alpha)/2}$ \citep[Theorem 5]{gupta2020distribution}.
\end{enumerate} 
\noindent This multi-step analysis leads to a loose bound due to constants stacking up, as discussed in Section~\ref{sec:ums-inefficient}. %In comparison, each of the $B$ bins obtained by \modifiedze have at least $\lfloor (n_1 + n_2)/B \rfloor - 1$ samples.

%As detailed in Section~\ref{sec:sample-complexity-gupta20}, our clean analysis avoids the stacking up of constants in the analysis of 

A guarantee for the original \ze procedure follows as an immediate corollary of Theorem~\ref{thm:umd-binary}. This is because the modification to line~\ref{line:removed-mean} can  %at most increase the approximation error $\varepsilon$ by $1/(\lfloor n/B \rfloor - 1)$
change every estimate $\widehat{\Pi}_b$ by at most $1/(\lfloor n/B \rfloor)$ due to the following fact regarding averages: for any $b \in \naturals, a \in \{0, 1, \ldots, b\}$,% \naturals \cup \{0\}$,
\begin{equation}
\max\roundbrack{\abs{\frac{a}{b+1} - \frac{a}{b}}, \abs{\frac{a+1}{b+1} - \frac{a}{b}}} \leq \frac{1}{b+1}. \label{eq:averages-fact}
\end{equation}
Using \eqref{eq:averages-fact}, we prove the following corollary  in Appendix~\ref{appsec:adaptive-binning-proofs}.
\begin{corollary}\label{cor:ze}
Suppose $g(X)$ is absolutely continuous with respect to the Lebesgue measure and $n \geq 2B$. The original \ze algorithm \citep{zadrozny2001obtaining} is $(\varepsilon, \alpha)$-conditionally calibrated for any $ \alpha \in (0, 1)$ and %\vspace{-0.2cm}
\begin{equation}
    \varepsilon =\sqrt{\frac{\log(2B/\alpha)}{2(\lfloor n/B\rfloor -1)}} + \frac{1}{\lfloor n/B \rfloor}.
    \label{eq:epsilon-ze}
\end{equation}
Further, for every distribution $P$, w.p. $1-\alpha$ over the calibration data $\Dcal_n$, for all $p \in [1, \infty)$, $\ell_p\text{-ECE}(h) \leq \varepsilon$.
\end{corollary}
As claimed in Section~\ref{sec:umd-vs-ums}, if $(n, \alpha, B) = (2900, 0.1, 10)$,  \eqref{eq:epsilon-ze} gives $\smash{\varepsilon< 0.1}$. 
The difference between \eqref{eq:epsilon-ze} and \eqref{eq:epsilon-modifiedze} is small. For example, we computed that if $\smash{\varepsilon \leq 0.1}$, $\smash{\alpha \leq 0.5}$, $B \geq 5$, then \eqref{eq:epsilon-modifiedze} requires $n/B \geq 150$, and thus the additional term in \eqref{eq:epsilon-ze} is at most 0.007. Likewise, in practice, we expect both versions to perform similarly. %However, if the application is sensitive to deviations in $\varepsilon$ at the scale of 0.01, we recommend using Algorithm~\ref{alg:efficient-uniform-mass-binary} under constraint~\eqref{eq:epsilon-modifiedze}. 

 %Thus for simplicity, we recommend using Theorem~\ref{thm:umd-binary} directly. %If the small difference between \eqref{eq:epsilon-ze} and \eqref{eq:epsilon-modifiedze} is a concern for your application, then we recommend using \modifiedze. 
% A practitioner who wishes to use one of these algorithms for calibration 
% may seek an answer to the following question: 
% %In order to build intuition about our calibration guarantee, we discuss answers to the following question:
% ``Given $n$ points for recalibration, how many bins $B$ should be used?" We discuss the answer to this question, guided by the requirement to satisfy \eqref{eq:epsilon-ze} for reasonable values of $\varepsilon, \alpha$. %, building some intuition for the guarantee in the process. 
%We now make sense of this calibration guarantee by showing the relationship between $n, B, \varepsilon, \alpha$ through some indicative plots. 

\begin{figure}
%\vspace{0.5cm}
    \centering
    \includegraphics[trim=0 3cm 0 -3.5cm, clip, width=0.5\columnwidth]{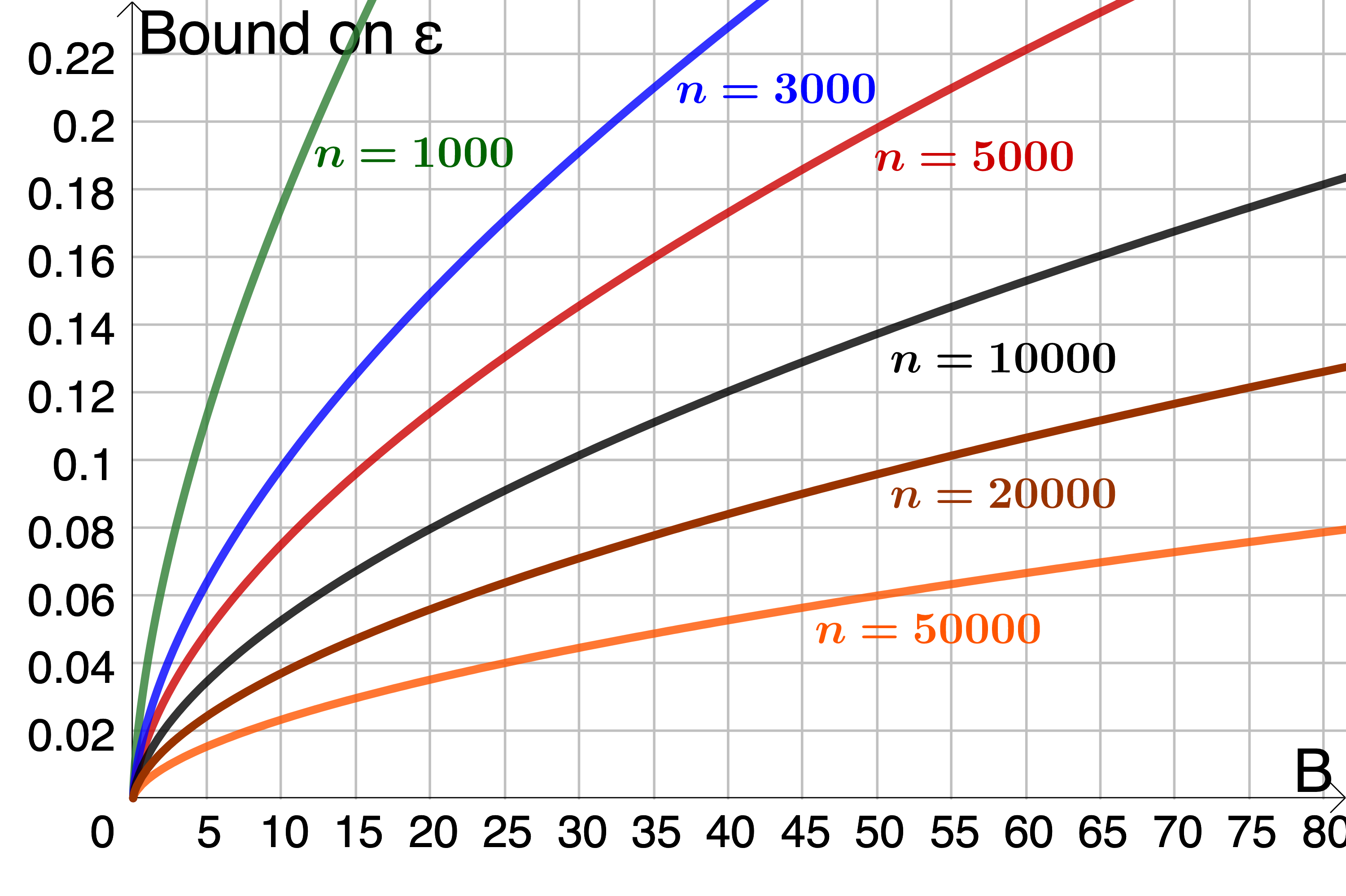}
    \caption{%Values of $\varepsilon$ allowed by Theorem~\ref{thm:umd-binary} as a function of $B$, 
    Plots displaying the relationship \eqref{eq:epsilon-modifiedze} between $\varepsilon$ and $B$ for $\alpha = 0.1$ and different values of $n$. Some indicative suggestions based on the plot: if $n = 1$K, choose $B = 5$ (gives $\varepsilon \leq 0.12$); if $n= 5$K, choose $B =10$ (gives $\varepsilon \leq 0.08$); if $n=20$K, choose $B=22$ (gives $\varepsilon \leq 0.06$).}
    \label{fig:bound-illustration}
    %\vspace{-0.5cm}
\end{figure}

%\subsection{Practical implications of Theorem~\ref{thm:umd-binary}}
At the end of the day, a practitioner may ask: ``Given $n$ points for recalibration, how should I use Theorem~\ref{thm:umd-binary} to decide $B$?" Smaller $B$ gives better bounds on $\varepsilon$, but larger $B$ implicitly means that the $h$ learnt is sharper. As $n$ becomes higher, one may like to have higher sharpness (higher $B$), but at the same time more precise calibration (lower $\varepsilon$ and thus lower $B$). We provide a (subjective) discussion on how to balance these two requirements.

%Constraint~\eqref{eq:epsilon-modifiedze} is a relationship between four parameters. 
First, we suggest fixing a rough domain-dependent probability of failure $\alpha$. Since the dependence of $\varepsilon$ on $\alpha$ in \eqref{eq:epsilon-modifiedze}  is $\log(1/\alpha)$, small changes in $\alpha$ do not affect $\varepsilon$ too much. Typically, 10-20\% failure rate is acceptable, so let us set $\alpha = 0.1$. (For a highly sensitive domain, one can set $\alpha = 0.01$.) %(if needed, a smaller $\alpha$ can also be chosen without affecting the other parameters too much, since the dependence on $\alpha$ is only $\log(1/\alpha)$). 
%For $\smash{\alpha = 0.1}$, 
Then, constraint~\eqref{eq:epsilon-modifiedze} roughly translates to $\varepsilon = \sqrt{B \log(20B)/2n}$. For a fixed $n$, this is a relationship between $\varepsilon$ and $B$, that can be plotted as a curve with $B$ as the independent parameter and $\varepsilon$ as the dependent parameter. Finally, one can eyeball the curve to identify a $B$. We plot such curves in Figure~\ref{fig:bound-illustration} for a range of values of $n$. The caption shows examples of how one can choose $B$ to balance calibration (small $\varepsilon$) and sharpness (high $B$).
% For a given $n$, %
% %To show the dependence between $n/B$ and $\varepsilon$, we consider two cases --- $B \leq 10$  and $B \leq 1000$. 
% For $B \leq 10$  the bound on $\varepsilon$ is plotted as a function of $n/B$ in Figure~\ref{fig:bound-illustration} (green curve). Thus $\smash{n/B \leq 135 \implies \varepsilon \leq 0.15}$, $\smash{n/B \leq 290 \implies \varepsilon \leq 0.1}$, and  $\smash{n/B \leq 1100 \implies \varepsilon \leq 0.05}$.  If lots of data is available (say $n > 10000$), we can use more bins. For $B \leq 100$,  $\smash{n/B \leq 185 \implies \varepsilon \leq 0.15}$, $\smash{n/B \leq 400 \implies \varepsilon \leq 0.1}$, and  $\smash{n/B \leq 1575 \implies \varepsilon \leq 0.05}$. In Figure~\ref{fig:bound-illustration} (orange curve), we show the more extreme case $B \leq 1000$. Here, $\smash{n/B \leq 240 \implies \varepsilon \leq 0.15}$, $\smash{n/B \leq 520 \implies \varepsilon \leq 0.1}$, and  $\smash{n/B \leq 2025 \implies \varepsilon \leq 0.05}$. 

\rev{
While $(\varepsilon, \alpha)$-conditional calibration implies $(\varepsilon, \alpha)$-marginal calibration, we expect to have marginal calibration with smaller $\varepsilon$. Such an improved guarantee can be shown if the bin biases $\widehat{\Pi}_b$ estimated by Algorithm~\ref{alg:efficient-uniform-mass-binary} are distinct. In Appendix~\ref{appsec:randomized-umd}, we propose a randomized version of \ze (Algorithm~\ref{alg:randomized-UMD}) which guarantees uniqueness of the bin biases. %, then %marginal calibration can be shown with $\varepsilon = \sqrt{\log(2/\alpha)/2(\lfloor n/B\rfloor - 1)}$  (this corresponds to replacing improving 
%a better marginal calibration guarantee can be shown.  %is satisfied by $\varepsilon$ from \eqref{eq:epsilon-modifiedze}, but with the numerator reduced to $\log(2/\alpha)$ instead of $\log(2B/\alpha)$. 
%While we expect \ze to have distinct bin biases, this is not guaranteed. However, 
%The requirement of distinct bin biases can be satisfied using randomization. In the following subsection, we describe a randomized version of \ze (Algorithm~\ref{alg:randomized-UMD}) that almost surely has distinct bin biases. 
%The technical details of Algorithm~\ref{alg:randomized-UMD} are discussed in the following subsection, but for continuity of exposition, we state the calibration guarantee satisfied by it here.
Algorithm~\ref{alg:randomized-UMD} satisfies the following calibration guarantee (proved in Appendix~\ref{appsec:adaptive-binning-proofs}).
\begin{theorem}
\label{thm:umd-binary-randomized}
Suppose $n \geq 2B$ and let $\delta > 0$ be an arbitrarily small randomization parameter.  Algorithm~\ref{alg:randomized-UMD} is $(\varepsilon_1, \alpha)$-marginally and $(\varepsilon_2, \alpha)$-conditionally calibrated for any $\alpha \in (0, 1)$,
%For any $\alpha \in (0, 1)$ and 
\begin{equation}
    \varepsilon_1 =\sqrt{\frac{\log(2/\alpha)}{2(\lfloor n/B\rfloor -1)}} + \delta   \text{, }  \varepsilon_2 = \sqrt{\frac{\log(2B/\alpha)}{2(\lfloor n/B\rfloor -1)}} + \delta.   
    \label{eq:marginal-umd-guarantee}
\end{equation}
Further, for every distribution $P$, (a) w.p. $1-\alpha$ over the calibration data $\Dcal_n$, for all $p \in [1, \infty)$, $\ell_p\text{-ECE}(h) \leq \varepsilon_2$, and (b)  $\Exp{\Dcal_n}{\ell_p\text{-ECE}(h)} \leq \sqrt{B/2n} + \delta$ for all $p \in [1, 2]$.% where the expectation is over the calibration data.
%the function $h$ provided by Algorithm~\ref{alg:efficient-uniform-mass-binary} is $(\varepsilon, \alpha)$-conditionally calibrated.
\end{theorem}
%The proof is in Appendix~\ref{appsec:adaptive-binning-proofs}. 
In the proof, we use the law of total expectation to avoid taking a union bound in the marginal calibration result; this gives a $\sqrt{\log(2/\alpha)}$ term in $\varepsilon_1$ instead of the $\sqrt{\log(2B/\alpha)}$ in $\varepsilon_2$. Theorem~\ref{thm:umd-binary-randomized} also does not require absolute continuity of $g(X)$. As claimed in Section~\ref{sec:umd-vs-ums}, if $(n, \alpha, B) = (1500, 0.1, 10)$,  \eqref{eq:marginal-umd-guarantee} gives $\smash{\varepsilon_1 < 0.1}$ (for small enough $\delta$). 
}

\begin{figure*}[t]
\centering
\begin{subfigure}[c]{\linewidth}
\centering
\includegraphics[width=0.8\linewidth]{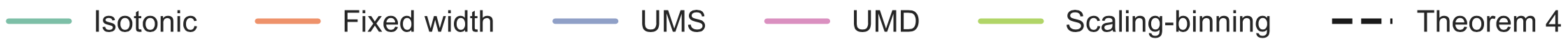}
\end{subfigure}
\begin{subfigure}[c]{0.45\linewidth}
\centering
\includegraphics[width=0.49\linewidth]{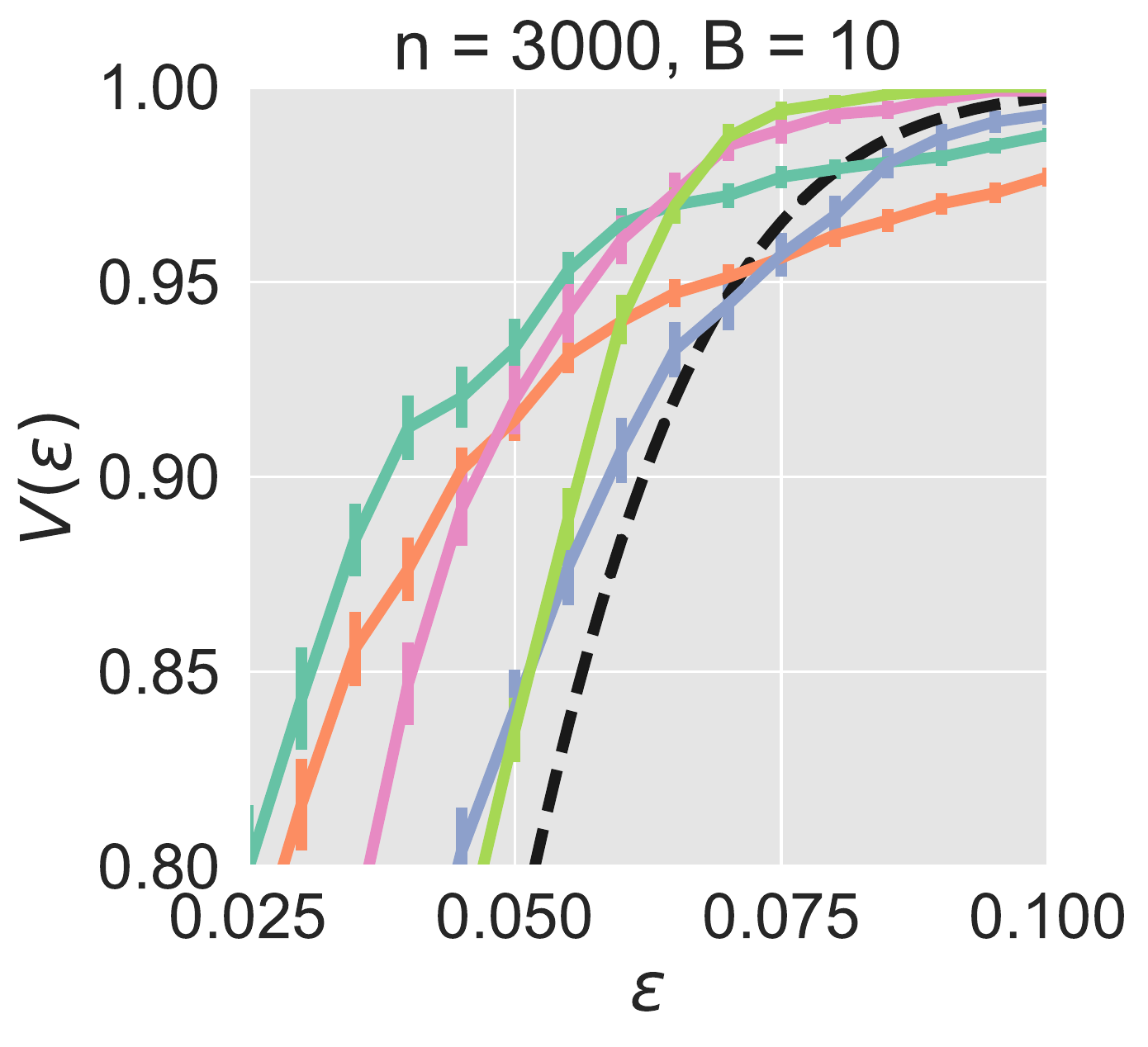}
\includegraphics[width=0.49\linewidth]{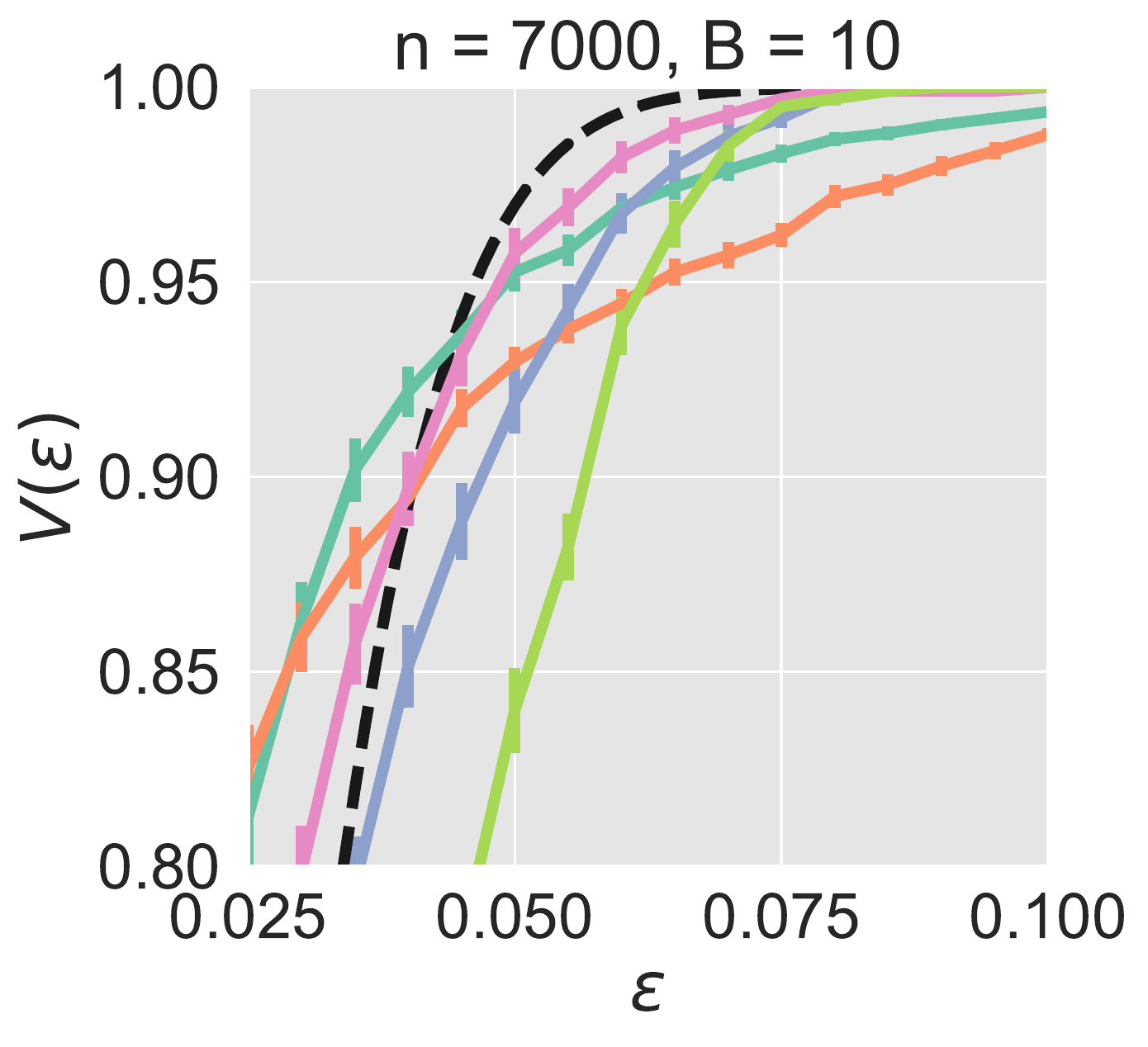}
\caption{Marginal validity plots.}
\label{fig:comparison-across-methods-marginal}
\end{subfigure}
\hspace{1cm}
\begin{subfigure}[c]{0.45\linewidth}
\centering
\includegraphics[width=0.49\linewidth]{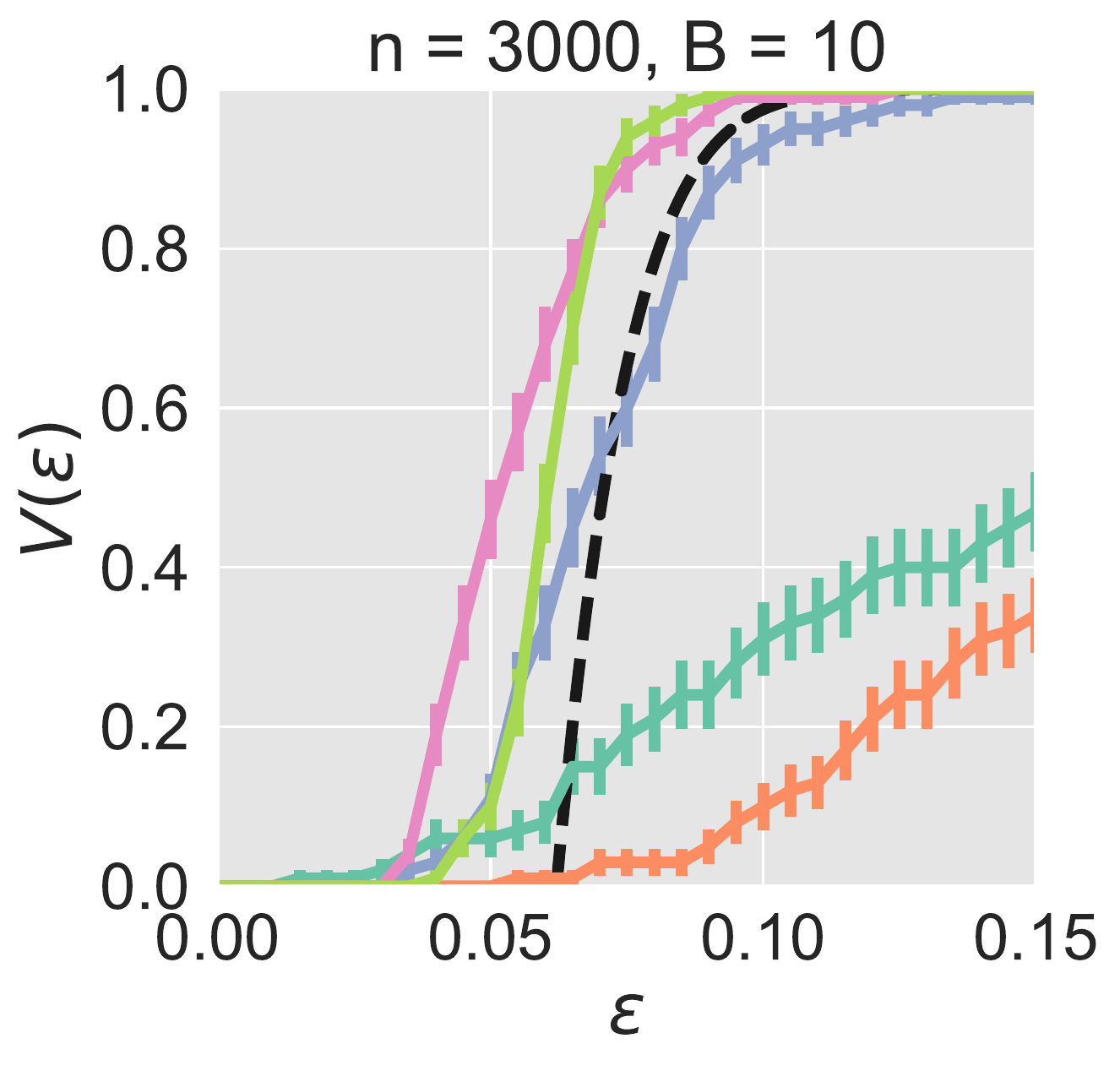}
\includegraphics[width=0.49\linewidth]{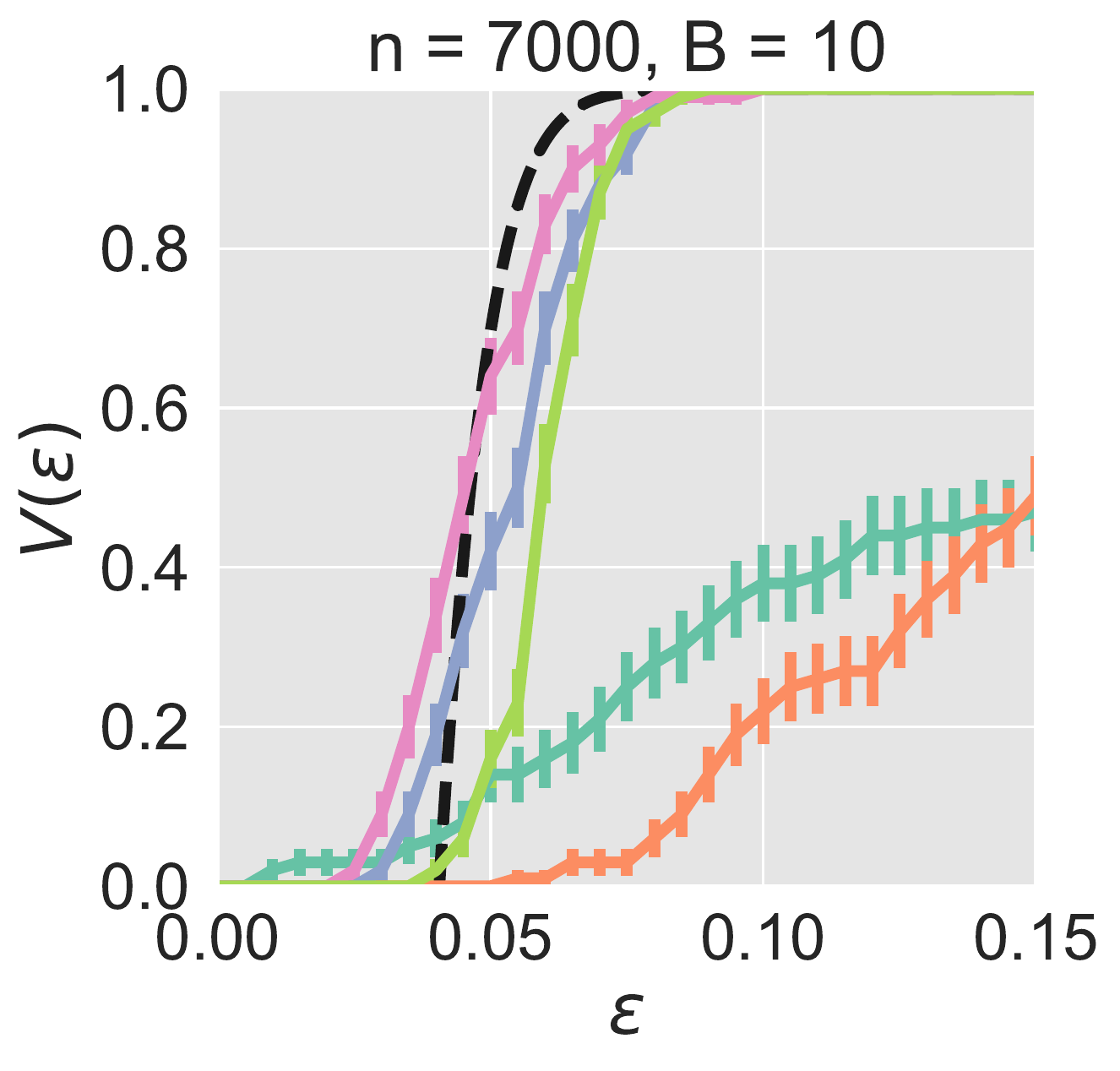}
\caption{Conditional validity plots.}
\label{fig:comparison-across-methods-conditional}
\end{subfigure}
\caption{\ze performs competitively on the \credit dataset. The guarantee of Theorem~\ref{thm:umd-binary-randomized} closely matches empirical behavior. 
}
\label{fig:comparison-across-methods}
\end{figure*}
\section{Simulations}
\label{sec:simulations}
We perform illustrative simulations on the \credit dataset with two goals: (a) to compare the performance of \ze to other binning methods and (b) to show that the guarantees we have shown are reasonably tight, and thus, practically useful.\footnote{Relevant code can be found at \url{https://github.com/aigen/df-posthoc-calibration}} In addition to validity plots, which assess marginal calibration, we use conditional validity plots, that assess conditional calibration. Let $\smash{V : [0,1] \to [0,1]}$ be given by $V(\varepsilon) = \Prob(\forall r \in \text{Range}(h),  \abs{\Exp{}{Y \mid h(X) = r}-r}\leq \varepsilon)$. %To make a conditional validity plot on a given test set 
Given a test set $\Dcal_{\text{test}}$, we first compute $\Exp{\widehat{P}}{Y\mid h(X) = h(x)}$ (defined in \eqref{eq:marginal-validity-plot}), and then estimate $V(\varepsilon)$ as%define the function $\widehat{V} : [0, 1] \to \{0, 1\}$ as
\begin{equation*}
\widehat{V}(\varepsilon) = 
\indicator{\max_{(X_i, Y_i) \in \Dcal_{\text{test}}} \abs{\Exp{\widehat{P}}{Y \mid h(X) = h(X_i)} - h(X_i)} \leq \varepsilon}.
\end{equation*}
%\indicator{\abs{\Exp{\widehat{P}}{Y \mid h(X)} - h(X)}  \leq \varepsilon \mid h(X) \text{ a.s. }\widehat{P}}  \equiv 
%on each resampled test set 
 %This is a Bernoulli random variable for a given recalibration and test set; 
For a single $\Dcal_n$ and $\Dcal_{\text{test}}$, $\widehat{V}(\varepsilon)$ is either $0$ or $1$. Thus to estimate $V(\varepsilon)$, we average $\widehat{V}(\varepsilon)$ across multiple calibration and test sets. The mean$\pm$std-dev-of-mean of the $\widehat{V}(\varepsilon)$ values are plotted as $\varepsilon$ varies. This gives us a conditional validity plot. %$\Prob(\forall r \in \text{Range}(h),  \abs{\Exp{}{Y \mid h(X) = r}-r}\leq \varepsilon)$. %$1 - \alpha$.
It is easy to see that  the conditional validity plot is uniformly dominated by the (marginal) validity plot. %the $1-\alpha$ estimates in a conditional validity plot are always smaller than the $1-\alpha$ estimates in a validity plot. 

% \begin{figure*}[t]
% \begin{subfigure}[c]{0.43\linewidth}
% \includegraphics[width=0.49\linewidth]{marginal_0.3.pdf}
% \includegraphics[width=0.49\linewidth]{marginal_0.7.pdf}
% \caption{Marginal validity plots.}
% \label{fig:comparison-across-methods-marginal}
% \end{subfigure}
% \begin{subfigure}[c]{0.14\linewidth}
% \includegraphics[width=\linewidth,trim=0cm -5cm 0cm 0cm,clip]{legend.png}
% \end{subfigure}
% \begin{subfigure}[c]{0.43\linewidth}
% %\includegraphics[width=0.49\linewidth]{conditional_0.1.pdf}
% \includegraphics[width=0.49\linewidth]{conditional_0.3.pdf}
% %\includegraphics[width=0.49\linewidth]{conditional_0.5.pdf}
% \includegraphics[width=0.49\linewidth]{conditional_0.7.pdf}
% \caption{Conditional validity plots.}
% \label{fig:comparison-across-methods-conditional}
% \end{subfigure}

% \caption{\ze performs competitively on the \credit dataset. The guarantee of Theorem~\ref{thm:umd-binary-randomized} closely matches empirical behavior. 
% }
% \label{fig:comparison-across-methods}
% \end{figure*}

The experimental protocol for \credit is described in Section~\ref{sec:umd-vs-ums}. In our experiments, we used the randomized version of \ze (Algorithm~\ref{alg:randomized-UMD}).  Figure~\ref{fig:comparison-across-methods} presents validity plots for \ze, UMS, fixed-width binning, isotonic regression, scaling-binning, along with the Theorem~\ref{thm:umd-binary-randomized} curve for $\smash{n=3}K$ and $n=7$K. In Appendix~\ref{appsec:additional-exps}, we also present plots for $\smash{n=1}K$ and $n=5$K. Fixed-width binning refers to performing binning with equally spaced bins ($[0, 1/B), \ldots, [1-1/B, 1]$). UMS uses a 50:50 split of the calibration data. We do not rescale in scaling-binning, since it is already done on split B (for all compared procedures) --- instead the comparison is between averaging the predictions of the scaling method (as is done in scaling-binning), against averaging the true outputs in each bin (as is done by all other methods). To have a fair comparison, we use double dipping for scaling-binning (thus scaling-binning and \ze are identical except what is being averaged). 
%As observed in the experiment in Section~\ref{sec:sample-complexity-gupta20},
We make the following observations:% based on Figure~\ref{fig:comparison-across-methods}: 
\begin{itemize}
    %\item %As discussed in Section~\ref{sec:sample-complexity-gupta20}, 
    %, as also discussed in Section~\ref{sec:umd-vs-ums}.
    \item Isotonic regression and fixed-width binning perform well for marginal calibration, but fail for conditional calibration. This is because both these methods tend to have bins with skewed masses, leading to small $\varepsilon$ in bins with many points, and high $\varepsilon$ in bins with few points. 
    \item  Scaling-binning is competitive with \ze for $\smash{n=3}$K, $\smash{\varepsilon > 0.05}$. If $\smash{n =7}$K or $\smash{\varepsilon \leq 0.05}$, \ze outperforms scaling-binning. %The difference is highest in the rightmost plot. 
    In Appendix~\ref{appsec:additional-exps}, we show that for $n=1$K, scaling-binning is nearly the best method. 
    \item \ze always performs better than UMS, and the performance of \ze is almost perfectly explained by the theoretical guarantee.  Paradoxically, for $n=7$K, the theoretical curve \textit{crosses} the validity plot for \ze. This can occur since validity plots are based on a finite sample estimate of $\Exp{}{Y \mid h(X)}$, and the estimation error leads to slight \emph{underestimation} of validity. This phenomenon is the same as the bias of plugin ECE estimators, and is discussed in detail in the last paragraph of Section~\ref{sec:validity-plots}. The curve-crossing shows that Theorem~\ref{thm:umd-binary-randomized} is so precise that 5K test points are insufficient to verify it. %However, in the marginal validity plots, there is a gap between Theorem~\ref{thm:umd-binary} and the performance of \ze. In Appendix~\ref{appsec:marginal-guarantee}, we discuss some conjectures for improving the marginal calibration guarantee to reduce this gap. 
\end{itemize}
%\revicml{We elaborate on the last point a bit more. Due to finite test data, the validity plot displays $\abs{\Exp{\widehat{P}}{Y \mid h(X)} - h(X)}$, which can be written as $\abs{\Exp{}{Y \mid h(X)} + \text{noise} - h(X)}$. This noise could lead to underestimating $\abs{\Exp{}{Y \mid h(X)} - h(X)}$ in some cases, but on average, it will lead to overestimating $\abs{\Exp{}{Y \mid h(X)} - h(X)}$. In other words, the validity plot for all methods suggests slightly lesser validity than there actually is, which makes it possible for the Theorem~\ref{thm:umd-binary-randomized} curve to cross the empirical UMD curve for $n=7$K. This phenomenon is exactly the same as the presence of bias in ECE estimates \citep{brocker2012estimating, kumar2019calibration}. Note that it is highly unlikely that this noise in the estimate of $E[Y\mid h(X)]$ helps some methods and hurt others, and thus we can continue making inferences on the relative behavior of the methods.}

Overall, our experiment indicates that \ze performs competitively in practice and our theoretical guarantee closely explains its performance.
\section{Conclusion}
We used the Markov property of order statistics to prove distribution-free calibration guarantees for the popular uniform-mass binning method of \citet{zadrozny2001obtaining}. %We performed an illustrative experiment to show that this procedure is competitive with other calibration approaches. 
We proposed a novel assessment tool called validity plots, and used this tool to demonstrate that our theoretical bound closely tails empirical performance on a UCI credit default dataset. %Our work has implications beyond the analysis of \ze. 
To the best of our knowledge, we demonstrated for the first time that it is possible to show informative calibration guarantees for binning methods that double dip the data (to both estimate bins and the probability of $\smash{Y=1}$ in a bin). Popular calibration methods such as isotonic regression \citep{zadrozny2002transforming}, probability estimation trees \citep{provost2003tree}, random forests \citep{breiman2001random} and Bayesian binning \citep{naeini2015obtaining} perform exactly this style of double dipping. We thus open up the exciting possibility of providing  DF calibration guarantees for one or more of these methods. 

Another recent line of work for calibration in data-dependent groupings, termed as multicalibration, uses a discretization step similar to fixed-width binning \citep{hebert2018multicalibration}. Our uniform-mass binning techniques can  potentially be extended to multicalibration.  %provable calibration in data-dependent groupings.
A number of non-binned methods for calibrating neural networks have displayed good performance on some tasks \citep{guo2017nn_calibration, kull2017beyond, lakshminarayanan2017simple}.  However, the results of \citet{gupta2020distribution} imply that these methods cannot have DF guarantees. Examining whether they have  guarantees under some (weak) distributional assumptions is also interesting future work.

\section*{Acknowledgments}
We wish to thank Sasha Podkopaev, Anish Sevekari, Saurabh Garg, Elan Rosenfeld, and the anonymous ICML reviewers, for comments on an earlier version of the paper. CG also thanks the instructors and students in the `Writing in Statistics' course at CMU for valuable feedback.

\label{sec:discussion}

\bibliographystyle{plainnat}
\bibliography{references}

\begin{thebibliography}{27}
\providecommand{\natexlab}[1]{#1}
\providecommand{\url}[1]{\texttt{#1}}
\expandafter\ifx\csname urlstyle\endcsname\relax
  \providecommand{\doi}[1]{doi: #1}\else
  \providecommand{\doi}{doi: \begingroup \urlstyle{rm}\Url}\fi

\bibitem[Ahsanullah et~al.(2013)Ahsanullah, Nevzorov, and
  Shakil]{ahsanullah2013introduction}
Mohammad Ahsanullah, Valery~B Nevzorov, and Mohammad Shakil.
\newblock \emph{An introduction to order statistics}, volume~8.
\newblock Springer, 2013.

\bibitem[Arnold et~al.(2008)Arnold, Balakrishnan, and
  Nagaraja]{arnold2008first}
Barry~C Arnold, Narayanaswamy Balakrishnan, and Haikady~Navada Nagaraja.
\newblock \emph{A first course in order statistics}.
\newblock SIAM, 2008.

\bibitem[Breiman(2001)]{breiman2001random}
Leo Breiman.
\newblock Random forests.
\newblock \emph{Machine learning}, 45\penalty0 (1):\penalty0 5--32, 2001.

\bibitem[Br{\"o}cker(2012)]{brocker2012estimating}
Jochen Br{\"o}cker.
\newblock Estimating reliability and resolution of probability forecasts
  through decomposition of the empirical score.
\newblock \emph{Climate dynamics}, 39\penalty0 (3-4):\penalty0 655--667, 2012.

\bibitem[Clopper and Pearson(1934)]{clopper1934use}
Charles~J Clopper and Egon~S Pearson.
\newblock The use of confidence or fiducial limits illustrated in the case of
  the binomial.
\newblock \emph{Biometrika}, 26\penalty0 (4):\penalty0 404--413, 1934.

\bibitem[Dai et~al.(2020)Dai, Song, Barber, and Raskutti]{dai2020bias}
Ran Dai, Hyebin Song, Rina~Foygel Barber, and Garvesh Raskutti.
\newblock The bias of isotonic regression.
\newblock \emph{Electronic journal of statistics}, 14\penalty0 (1):\penalty0
  801, 2020.

\bibitem[Dawid(1982)]{dawid1982well}
A~Philip Dawid.
\newblock The well-calibrated {B}ayesian.
\newblock \emph{Journal of the American Statistical Association}, 77\penalty0
  (379):\penalty0 605--610, 1982.

\bibitem[Gneiting et~al.(2007)Gneiting, Balabdaoui, and
  Raftery]{gneiting2007probabilistic}
Tilmann Gneiting, Fadoua Balabdaoui, and Adrian~E Raftery.
\newblock Probabilistic forecasts, calibration and sharpness.
\newblock \emph{Journal of the Royal Statistical Society: Series B (Statistical
  Methodology)}, 69\penalty0 (2):\penalty0 243--268, 2007.

\bibitem[Guo et~al.(2017)Guo, Pleiss, Sun, and
  Weinberger]{guo2017nn_calibration}
Chuan Guo, Geoff Pleiss, Yu~Sun, and Kilian~Q. Weinberger.
\newblock On calibration of modern neural networks.
\newblock In \emph{International Conference on Machine Learning}, 2017.

\bibitem[Gupta et~al.(2020)Gupta, Podkopaev, and Ramdas]{gupta2020distribution}
Chirag Gupta, Aleksandr Podkopaev, and Aaditya Ramdas.
\newblock Distribution-free binary classification: prediction sets, confidence
  intervals and calibration.
\newblock In \emph{Advances in Neural Information Processing Systems}, 2020.

\bibitem[H{\'e}bert-Johnson et~al.(2018)H{\'e}bert-Johnson, Kim, Reingold, and
  Rothblum]{hebert2018multicalibration}
Ursula H{\'e}bert-Johnson, Michael Kim, Omer Reingold, and Guy Rothblum.
\newblock Multicalibration: Calibration for the (computationally-identifiable)
  masses.
\newblock In \emph{International Conference on Machine Learning}, 2018.

\bibitem[Kull et~al.(2017)Kull, Silva~Filho, and Flach]{kull2017beyond}
Meelis Kull, Telmo~M. Silva~Filho, and Peter Flach.
\newblock Beyond sigmoids: How to obtain well-calibrated probabilities from
  binary classifiers with beta calibration.
\newblock \emph{Electronic Journal of Statistics}, 11\penalty0 (2):\penalty0
  5052--5080, 2017.

\bibitem[Kumar et~al.(2019)Kumar, Liang, and Ma]{kumar2019calibration}
Ananya Kumar, Percy~S Liang, and Tengyu Ma.
\newblock Verified uncertainty calibration.
\newblock In \emph{Advances in Neural Information Processing Systems}, 2019.

\bibitem[Lakshminarayanan et~al.(2017)Lakshminarayanan, Pritzel, and
  Blundell]{lakshminarayanan2017simple}
Balaji Lakshminarayanan, Alexander Pritzel, and Charles Blundell.
\newblock Simple and scalable predictive uncertainty estimation using deep
  ensembles.
\newblock In \emph{Advances in Neural Information Processing Systems}, 2017.

\bibitem[Lugosi and Nobel(1996)]{lugosi1996consistency}
G{\'a}bor Lugosi and Andrew Nobel.
\newblock Consistency of data-driven histogram methods for density estimation
  and classification.
\newblock \emph{Annals of Statistics}, 24\penalty0 (2):\penalty0 687--706,
  1996.

\bibitem[Miller(1962)]{miller1962statistical}
Robert~G Miller.
\newblock Statistical prediction by discriminant analysis.
\newblock In \emph{Statistical Prediction by Discriminant Analysis}, pages
  1--54. Springer, 1962.

\bibitem[Naeini et~al.(2015)Naeini, Cooper, and
  Hauskrecht]{naeini2015obtaining}
Mahdi~Pakdaman Naeini, Gregory Cooper, and Milos Hauskrecht.
\newblock Obtaining well calibrated probabilities using {B}ayesian binning.
\newblock In \emph{AAAI Conference on Artificial Intelligence}, 2015.

\bibitem[Niculescu-Mizil and Caruana(2005)]{Niculescu2005predicting}
Alexandru Niculescu-Mizil and Rich Caruana.
\newblock Predicting good probabilities with supervised learning.
\newblock In \emph{International Conference on Machine Learning}, 2005.

\bibitem[Parthasarathy and Bhattacharya(1961)]{parthasarathy1961some}
KR~Parthasarathy and PK~Bhattacharya.
\newblock Some limit theorems in regression theory.
\newblock \emph{Sankhy{\=a}: The Indian Journal of Statistics, Series A}, pages
  91--102, 1961.

\bibitem[Platt(1999)]{Platt99probabilisticoutputs}
John~C. Platt.
\newblock Probabilistic outputs for support vector machines and comparisons to
  regularized likelihood methods.
\newblock In \emph{Advances in Large Margin Classifiers}, pages 61--74. MIT
  Press, 1999.

\bibitem[Provost and Domingos(2003)]{provost2003tree}
Foster Provost and Pedro Domingos.
\newblock Tree induction for probability-based ranking.
\newblock \emph{Machine learning}, 52\penalty0 (3):\penalty0 199--215, 2003.

\bibitem[Roelofs et~al.(2020)Roelofs, Cain, Shlens, and
  Mozer]{roelofs2020mitigating}
Rebecca Roelofs, Nicholas Cain, Jonathon Shlens, and Michael~C Mozer.
\newblock Mitigating bias in calibration error estimation.
\newblock \emph{arXiv preprint arXiv:2012.08668}, 2020.

\bibitem[Sanders(1963)]{sanders1963subjective}
Frederick Sanders.
\newblock On subjective probability forecasting.
\newblock \emph{Journal of Applied Meteorology}, 2\penalty0 (2):\penalty0
  191--201, 1963.

\bibitem[Widmann et~al.(2019)Widmann, Lindsten, and
  Zachariah]{widmann2019calibration}
David Widmann, Fredrik Lindsten, and Dave Zachariah.
\newblock Calibration tests in multi-class classification: a unifying
  framework.
\newblock In \emph{Advances in Neural Information Processing Systems}, 2019.

\bibitem[Yeh and Lien(2009)]{yeh2009comparisons}
I-Cheng Yeh and Che-hui Lien.
\newblock The comparisons of data mining techniques for the predictive accuracy
  of probability of default of credit card clients.
\newblock \emph{Expert Systems with Applications}, 36\penalty0 (2):\penalty0
  2473--2480, 2009.

\bibitem[Zadrozny and Elkan(2001)]{zadrozny2001obtaining}
Bianca Zadrozny and Charles Elkan.
\newblock Obtaining calibrated probability estimates from decision trees and
  naive {B}ayesian classifiers.
\newblock In \emph{International Conference on Machine Learning}, 2001.

\bibitem[Zadrozny and Elkan(2002)]{zadrozny2002transforming}
Bianca Zadrozny and Charles Elkan.
\newblock Transforming classifier scores into accurate multiclass probability
  estimates.
\newblock In \emph{International Conference on Knowledge Discovery and Data
  Mining}, 2002.

\end{thebibliography}

\newpage
%\onecolumn
\appendix
\addtocontents{toc}{\protect\setcounter{tocdepth}{0}}

%\section{Proofs of results in Section~\ref{sec:adaptive-binning}}
\section{Proofs}
\label{appsec:adaptive-binning-proofs}
\subsection{Proof of Proposition~\ref{prop:ECE-holder}}
Define the random variables $u(X) = \abs{\Exp{}{Y \mid h(X)} - h(X)}^p$ and $v(X) = 1$. Then, by H\"{o}lder's inequality for $r = q/p$ and $s = (1-1/r)^{-1}$,
\begin{align*}
    (\ell_p\text{-ECE}(h))^p &= \Exp{}{u(X)} \\
    &= \Exp{}{\abs{u(X)v(X)}} \\
    &\leq \Exp{}{\abs{u(X)}^r}^{1/r} \Exp{}{\abs{v(X)}^s}^{1/s}
    \\&= \Exp{}{\abs{u(X)}^r}^{1/r}
    \\&= \Exp{}{\abs{\Exp{}{Y \mid h(X)} - h(X)}^q}^{p/q}
    \\ &= (\ell_q\text{-ECE}(h))^p,
\end{align*}
which proves \eqref{eq:ECE-holder}. If $h$ satisfies \eqref{eq:PAC-style-conditional-calib}, then 
$u(X) \leq \varepsilon^p \text{ a.s.}$ Thus $\ell_p\text{-ECE}(h) = \Exp{}{u(X)}^{1/p} \leq \varepsilon$. 
\qed

\subsection{Proof of Lemma~\ref{lemma:order-statistics-1}}
Let $F$ denote the cdf corresponding to $f$. The  structure of the proof is as follows: 
\begin{itemize}
    \item We first compute the conditional density of the order statistics $S_{(l+1)}, S_{(l+2)}, \ldots, S_{(u-1)}$, given $S_{(l)}$ and $S_{(u)}$, in terms of $f$ and $F$ (the expression for this is \eqref{eq:order-statistics-Z}). The basic building block for this computation is a result on the conditional density of order statistics given a single order statistic (equation \eqref{eq:order-statistics-2}).
    \item  Next, we compute the conditional density of the order statistics of the independent random variables $\{S'_i\}_{i \in [u-l-1]}$, given $S_{(l)}$, $S_{(u)}$, and $S_{(l)} < S'_i < S_{(u)}$ for all $i \in [u-l-1]$ (the expression for this is \eqref{eq:order-statistics-Z-2}).
    \item  We verify that \eqref{eq:order-statistics-Z} and \eqref{eq:order-statistics-Z-2} are identical, which shows that the conditional density of the order statistics matches. Finally, we conclude that the unordered random variables must themselves have the same conditional density. This completes the argument.
\end{itemize}

%unordered scores $S_{\{l+1\}}, S_{\{l+2\}}, \ldots, S_{\{u-1\}}$ using the conditional distribution of the order statistics $S_{(l+1)}, S_{(l+2)}, \ldots, S_{(u-1)}$. 

Let $0 \leq s_1 <  \ldots < s_{l-1} < a < s_{l+1} < \ldots < s_n \leq 1$. The conditional density of all the order statistics given $S_{(l)}$ 
\[
f(S_{(1)} = s_1, S_{(2)} = s_2, \ldots , S_{(l-1)
} = s_{l-1}, S_{(l+1)
} = s_{l+1}, \ldots , S_{(n)} = s_n \mid S_{(l)} = a)\]
is given by
\[
\roundbrack{(l-1)!\  \Pi_{i=1}^{l-1}\frac{f(s_i)}{F(a)} } \cdot \roundbrack{(n-l)!\  \Pi_{i=l}^{n}\frac{f(s_i)}{1-F(a)} }.
\]
For one derivation, see \citet[Chapter 5, equation (5.2)]{ahsanullah2013introduction}. This implies that the order statistics larger than $S_{(l)}$ are independent of the order statistics smaller than $S_{(l)}$ given $S_{(l)}$, and %that the conditional density of the order statistics 
\begin{equation}
f(S_{(l+1)} = s_{l+1}, \ldots , S_{(n)} = s_n) \mid S_{(l)} = a) 
=
\roundbrack{(n-l)!\  \Pi_{i=l+1}^{n}\frac{f(s_i)}{1-F(a)} }. \label{eq:order-statistics-2}
\end{equation}
Suppose we draw $n-l$ independent samples $T_1, T_2, \ldots, T_{n-l}$ from the distribution whose density is given by 
\[
g(s) =
\left\{
	\begin{array}{ll}
		\frac{f(s)}{1-F(a)}  & \mbox{if } s \in [a, 1] ~, \\
		0 & \mbox{otherwise.}
	\end{array}
\right. 
\]
(This is the conditional density of  $S$ given $S > S_{(l)} = a$ where $S$ is an independent random variable distributed as $Q_S$.) Consider the order statistics $T_{(1)}, T_{(2)}, \ldots, T_{(n-l)}$ of these $n - l$ samples. It is a standard result --- for example, see \citet[Chapter 2, equation (2.2.3)]{arnold2008first} --- that the density of the order statistics is
\[
g(T_{(1)} = s_{l+1}, T_{(2)} = s_{l+2}, \ldots, T_{(n-l)} = s_{n}) = (n-l)!\ \Pi_{i=1}^{n-l} g(s_{l+1}),
\]
which is identical to \eqref{eq:order-statistics-2}. Thus we can see the following fact:
\begin{equation}
\begin{split}\text{the density of the order statistics larger than $S_{(l)}$, given $S_{(l)} = a$,}\\ \text{is the same as the density of the  order statistics $T_{(1)}, T_{(2)}, \ldots, T_{(n-l)}$.} \label{eq:order-statistics-relationship}
\end{split}
\end{equation}

%  of $n-l$ samples from the distribution whose density $g$ is given by 
% \[
% g(s) =
% \left\{
% 	\begin{array}{ll}
% 		\frac{f(s)}{1-F(a)}  & \mbox{if } s \in [a, 1] \\
% 		0 & \mbox{otherwise.}
% 	\end{array}
% \right. 
% \]
% (This is the conditional density of  $S \mid S > S_{(u)}$ where the marginal density of $S$ is $f$.) 

Now consider the distribution of the order statistics $T_{(1)}, T_{(2)}, \ldots, T_{(u-l-1)}$ given $T_{(u - l)}$. Let $0 < s_{l+1} < \ldots < s_{u-1} < b \leq 1$. Using the same series of steps that led to equation~\eqref{eq:order-statistics-2}, we have
%result used earlier \citep[Chapter 5]{ahsanullah2013introduction}, 
%this density is given by
\begin{align}
&g(T_{(1)} = s_{l+1}, T_{(2)} = s_{l+2}, \ldots, T_{(u-l-1)}= s_{u - 1} \mid T_{(u-l)} = b) \nonumber
\\&\qquad\qquad\qquad =   (u-l-1)!\   \Pi_{i=1}^{u-l-1} \frac{g(s_{l+i})}{G(b)},  \label{eq:order-statistics-Y}
\end{align}
where $G$ is the cdf of $g$:
\[
G(s) =
\left\{
	\begin{array}{ll}
		\frac{F(s)-F(a)}{1-F(a)}  & \mbox{if } s \in [a, 1] ~, \\
		0 & \mbox{if } s \in (-\infty, a) ~, \\
		1 & \mbox{if } s \in (1, \infty) ~.		
	\end{array}
\right. 
\]
Due to fact \eqref{eq:order-statistics-relationship}, the density of $(T_{(1)}, \ldots, T_{(u-l-1)})$ given $T_{(u-l)} = b$ is  the same as the density of $(S_{(l+1)}, \ldots, S_{(u-1)})$ given $S_{(u)} = b$ and $S_{(l)} = a$. Thus, %\eqref{eq:order-statistics-Y} is also equal to 
\[
f(S_{(l+1)} = s_{l+1}, \ldots, S_{(u-1)} = s_{u - 1} \mid S_{(l)} = a, S_{(u)} = b) = (u-l-1)!\   \Pi_{i=1}^{u-l-1} \frac{g(s_{l+i})}{G(b)}. %= g(T_{(1)} = s_{l+1}, T_{(2)} = s_{l+2}, \ldots, T_{(T_{n-l})} = s_{n}).
\]
Writing $g$ and $G$ in terms of $f$ and $F$, we get 
%Theorem 2.4.4 \citep{arnold2008first} shows that the conditional density of the order statistics 
%and using this fact, 
%the fact that the density of $T_{(i)}$Thus we conclude that 
\begin{align}
&f(S_{(l+1)} = s_{l+1}, \ldots, S_{(u-1)} = s_{u-1}  \mid S_{(l)} =a , S_{(u)} =b)  =  (u - l - 1)!\   \Pi_{i=1}^{u-l-1} \frac{f(s_{l+i})}{F(b) - F(a)}.\label{eq:order-statistics-Z}
\end{align}
Now consider the independent random variables $\{Z_i\}_{i = 1}^{u-l-1}$, where the density of each $Z_i$ is the same as the conditional density of $S'_i$, given $ S_{(l)} = a < S'_i < b = S_{(u)} $.

%, \ldots, Z_{u-l-1} = \smash{S'_{u-l-1} \mid S_{(l)} < S'_{u-l-1} < S_{(u)}}$. 
Thus the density $h$ of each $Z_i$ is given by 
\[
h(s) =
\left\{
	\begin{array}{ll}
		\frac{f(s)}{F(b)-F(a)}  & \mbox{if } s \in [a, b] ~, \\
		0 & \mbox{otherwise.}
	\end{array}
\right. 
\]
% and distribution function $H$ given by 
% \[
% H(s) =
% \left\{
% 	\begin{array}{ll}
% 		\frac{F(s) - F(s)}{F(b)-F(a)}  & \mbox{if } s \in [a, b] \\
% 		0 & \mbox{otherwise,}
% 	\end{array}
% \right. 
% \]
%(Thus each $Z_i \sim S'_i \mid S_{(l)} < S'_i < S_{(u)}$.) 
The density of the order statistics $Z_{(1)}, \ldots, Z_{(u-l-1)}$ is given by 
\begin{equation}
h(Z_{(1)} = s_{l+1}, \ldots, Z_{(u-l-1)} = s_{u-1}) = (u - l - 1)!\  \Pi_{i=1}^{u-l-1} h(s_{l+i}),
\label{eq:order-statistics-Z-2}
\end{equation}
which exactly matches the right hand side of \eqref{eq:order-statistics-Z}. Thus, 
\begin{align*}
&~f(S_{(l+1)} = s_{l+1}, \ldots, S_{(u-1)} = s_{u-1}  \mid S_{(l)} =a , S_{(u)} =b) \\&= h(Z_{(1)} = s_{l+1}, \ldots, Z_{(u-l-1)} = s_{u-1} )\\&= f(S'_{(1)} = s_{l+1}, \ldots, S'_{(u-l-1)} = s_{u-1}  \mid S_{(l)} =a , S_{(u)} =b, \text{for every $i \in [u-l-1]$, $S_{(l)} < S'_{i} < S_{(u)}$}).
\end{align*}

Since the conditional densities of the order statistics match, the conditional densities of the unordered random variables must also match. This gives us the claimed result. 

\qed

\subsection{Proof of Lemma~\ref{lemma:order-statistics-main}}
%\citet[Chapter 5, Example 5.1]{ahsanullah2013introduction} show that the order statistics smaller and larger than a given order statistic are independent of each other. 
The sequence of order statistics $S_{(1)}, S_{(2)}, \ldots, S_{(n)}$ form a Markov chain \citep[Theorem 2.4.3]{arnold2008first}. Thus 
\[\roundbrack{S_{(k_{i-1} + 1)}, \ldots, S_{(k_{i} - 1)} \independent S_{(k_0)}, \ldots,  S_{(k_{i-2})}, S_{(k_{i+1})}, \ldots, S_{(k_B)}} \mid S_{(k_{i-1})}, S_{(k_{i})}.\]
%(This is also showed by \citet[Chapter 5, Example 5.1]{ahsanullah2013introduction}.) 
Consequently, for the unordered set of random variables $S_{\{k_{i-1}+1\}}, \ldots, S_{\{k_{i}-1\}}$, we have:%  are also independent:
 \[\roundbrack{S_{\{k_{i-1} + 1\}}, \ldots, S_{\{k_{i} - 1\}} \independent S_{(k_0)}, \ldots,  S_{(k_{i-2})}, S_{(k_{i+1})}, \ldots, S_{(k_B)}} \mid S_{(k_{i-1})}, S_{(k_{i})}.\]
Thus, 
\[
% f(S_{\{k_{i-1}+1\}} = s_{k_{i-1}+1}, \ldots, S_{\{k_i-1\}} = s_{k_i-1}  \mid S_{(k_0)} = s_{(k_0)}, \ldots, S_{(k_B)} = s_{(k_B)})
% \\= f(S_{\{k_{i-1}+1\}} = s_{k_{i-1}+1}, \ldots, S_{\{k_i-1\}} = s_{k_i-1}  \mid S_{(k_{i-1})} = s_{(k_{i-1})}, S_{(k_i)} = s_{(k_i)}).
\smash{f(S_{\{k_{i-1}+1\}}, \ldots, S_{\{k_i-1\}}  \mid S_{(k_0)}, \ldots, S_{(k_B)})
= f(S_{\{k_{i-1}+1\}}, \ldots, S_{\{k_i-1\}}  \mid S_{(k_{i-1})}, S_{(k_i)}).}
\]
Using Lemma~\ref{lemma:order-statistics-1}, the result follows.
\qed

\subsection{Proof of Theorem~\ref{thm:umd-binary}}

For $b \in \{0, 1, \ldots, B\}$, define $k_b = \lceil b(n+1/B)\rceil$. Let $S_{(0)} := 0$ and $S_{(n+1)} := 1$ be fixed hypothetical `order-statistics'. The rest of this proof is conditional on the observed set $\Scal := (S_{(k_1)}, S_{(k_2)}, \ldots, S_{(k_{B-1})})$. (Marginalizing over $\Scal$ gives the theorem result as stated.) Let $\Bcal: \Xcal \to [B]$ be the binning function: for all $x$, $\Bcal(x) = b \iff S_{(k_{b-1})} \leq g(x) < S_{(k_b)}$. Note that given $\Scal$, the binning function $\Bcal$ is deterministic. In particular, this means that for every $b \in [B]$, $\Exp{}{Y \mid \Bcal(X) = b}$ is a fixed number that is not random on the calibration data or $(X, Y)$. 

%Now for a specific $x \in \Xcal$, let $\Bcal(x) = b$,
%and let $b \in [B]$ be the random variable such that $S_{(k_{b-1})} \leq g(x) < S_{(k_b)}$. %(note that $b(X)$ is not random given $\mathcal{S}$ and $X$). 
Let us fix some $b \in [B]$ and denote $l = k_{b-1}, u = k_{b}$. %Conditioned on %these order statistics, 
%$\mathcal{S}$, by
By Lemma~\ref{lemma:order-statistics-main}, the scores $S_{\{l+1\}}, S_{\{l+2\}}, \ldots, S_{\{u-1\}}$ are independent and identically distributed given $\Scal$, and the conditional distribution of each of them equals that of 
$g(X)$ given $\Bcal(X) = b$. 
%$g(X) \mid S_{(u)} \leq g(X) < S_{(l)}$.
Thus $Y_{\{l+1\}}, Y_{\{l+2\}}, \ldots, Y_{\{u-1\}}$ 
are independent and identically distributed given $\Scal$, 
%given only $S_{(k_1)}, S_{(k_2)}, \ldots, S_{(k_{B-1})}$, 
and the conditional distribution of each of them is
$\text{Bernoulli}(\Exp{}{Y 
 \mid \Bcal(X) = b})$.
%\mid S_{(u)} \leq g(X) < S_{(l)}}).\]
Thus for any $\hoeffding \in (0, 1)$, by Hoeffding's inequality, with probability at least $ 1 - \hoeffding$,
\begin{equation}
\abs{\Exp{}{Y \mid 
%S_{(k_b)} \leq g(X) < S_{(k_{b+1})}}
\Bcal(X) = b}- \widehat{\Pi}_b} \leq \sqrt{\frac{\log(2/\hoeffding)}{2\lfloor u - l - 1\rfloor}} \leq \sqrt{\frac{\log(2/\hoeffding)}{2(\lfloor n/B \rfloor - 1) }}. %\varepsilon. %\leq \sqrt{\frac{\log(2B/\alpha)}{2(\lfloor n/B\rfloor - 1)}}.
\label{eq:per-bin-CI}
\end{equation}
The second inequality holds since for any $b$, 
\begin{align*}
    u - l  &= k_b - k_{b-1}
    \\ &= \lfloor (b+1)(n+1)/B\rfloor - \lfloor b(n+1)/B\rfloor
    \\ &= \lfloor U + (n+1)/B\rfloor - \lfloor U\rfloor, \text{ where $U = b(n+1)/B$,}
    \\ &\geq \lfloor(n+1)/B \rfloor  \geq \lfloor n/B \rfloor.
\end{align*}
%Substituting this inequality in \eqref{eq:per-bin-CI}, we obtain \eqref{eq:epsilon-modifiedze} for a specific $b \in B$, with probability at least $ 1 - \alpha/B$. 

Next, we set $\hoeffding = \alpha/B$ in \eqref{eq:per-bin-CI}, and take a union bound over all $b \in B$. Thus, with probability at least $ 1- \alpha$, the event
\begin{equation*}
E : \qquad \text{for every } b \in [B],\ \abs{\Exp{}{Y \mid \Bcal(X) = b} - \widehat{\Pi}_{b}} \leq \varepsilon%  \text{ a.s. }\Bcal(X)
\end{equation*}
occurs. To prove the final calibration guarantee, we need to change the conditioning from $\Bcal(X)$ to $h(X)$. Specifically, we have to be careful about the possibility of multiple bins having the same $\widehat{\Pi}$ values, in which case, conditioning on $\Bcal(X)$ and conditioning on $h(X)$ is not the same. Given that $E$ occurs (which happens with probability at least $ 1- \alpha$),  
% for every $r \in \text{Range}(h)$,
% \begin{align*}
%     &~\abs{\Exp{}{Y \mid h(X) = r} - r } & 
%     \\&= \abs{\Exp{}{\Exp{}{Y \mid \Bcal(X), h(X) = r} \mid h(X) = r} - r } &\text{(applying tower rule)}
%     \\&= \abs{\Exp{}{\Exp{}{Y \mid \Bcal(X)} \mid h(X) = r} - r } &\text{($\Exp{}{Y \mid \Bcal(X), h(X)} = \Exp{}{Y \mid \Bcal(X)}$)}
%     \\ &= \abs{\Exp{}{\Exp{}{Y \mid \Bcal(X)} - r \mid h(X) = r}} %&\text{($h(X)$ is constant since we are conditioning on it)}
%     \\ &= \abs{\Exp{}{\Exp{}{Y \mid \Bcal(X)} - \widehat{\Pi}_{\Bcal(X)} \mid h(X) = r}} &\text{(by definition of $h$)}
%     \\ &\leq \Exp{}{\abs{\Exp{}{Y \mid \Bcal(X)} - \widehat{\Pi}_{\Bcal(X)}  } \mid h(X) = r} &\text{(Jensen's inequality)}
%     \\ &\leq \varepsilon &\text{(since $E$ occurs)}. %$\abs{\Exp{}{Y \mid \Bcal(X) } - \widehat{\Pi}_{\Bcal(X)}} \leq \varepsilon$ a.s.)}. 
% \end{align*}
\begin{align*}
    &~\abs{\Exp{}{Y \mid h(X)} - h(X) } & 
    \\&= \abs{\Exp{}{\Exp{}{Y \mid \Bcal(X), h(X)} \mid h(X)} - h(X) } &\text{(applying tower rule)}
    \\&= \abs{\Exp{}{\Exp{}{Y \mid \Bcal(X)} \mid h(X)} - h(X) } &\text{($\Exp{}{Y \mid \Bcal(X), h(X)} = \Exp{}{Y \mid \Bcal(X)}$)}
    \\ &= \abs{\Exp{}{\Exp{}{Y \mid \Bcal(X)} - h(X) \mid h(X)}} %&\text{($h(X)$ is constant since we are conditioning on it)}
    \\ &= \abs{\Exp{}{\Exp{}{Y \mid \Bcal(X)} - \widehat{\Pi}_{\Bcal(X)} \mid h(X)  }} &\text{(by definition of $h$)}
    \\ &\leq \Exp{}{\abs{\Exp{}{Y \mid \Bcal(X)} - \widehat{\Pi}_{\Bcal(X)}  } \mid h(X)} &\text{(Jensen's inequality)}
    \\ &\leq \varepsilon &\text{(since $E$ occurs)}. %$\abs{\Exp{}{Y \mid \Bcal(X) } - \widehat{\Pi}_{\Bcal(X)}} \leq \varepsilon$ a.s.)}. 
\end{align*}
This completes the proof of the conditional calibration guarantee. The ECE bound follows by Proposition~\ref{prop:ECE-holder}. 
\qed

\subsection{Proof of Corollary~\ref{cor:ze}} Conditioned on $\Scal$ (defined in the proof of Theorem~\ref{thm:umd-binary}), for some $b \in [B]$, $l = k_{b-1}$ and $u = k_b$, we showed in the proof of  Theorem~\ref{thm:umd-binary} that  with  probability at least $1 - \alpha/B$, %From equation~\eqref{eq:per-bin-CI} in the proof of Theorem~\ref{thm:umd-binary},  shows 
\[
\abs{\Exp{}{Y \mid \Bcal(X) = b}- \text{Mean($Y_{(l + 1)}, Y_{(l + 2)}, \ldots, Y_{(u - 1)}$)}} \leq \sqrt{\frac{\log(2B/\alpha)}{2(\lfloor n/B\rfloor - 1)}}.
\]
Thus for $b \in [B-1]$,
\begin{align*}
    \abs{\Exp{}{Y \mid \Bcal(X) = b}- \widehat{\Pi}_b} &\leq \abs{\Exp{}{Y \mid \Bcal(X) = b}- \text{Mean($Y_{(l + 1)}, Y_{(l + 2)}, \ldots, Y_{(u - 1)}$)}} \\ & \qquad + \abs{ \text{Mean($Y_{(l + 1)}, Y_{(l + 2)}, \ldots, Y_{(u - 1)}$)} - \text{Mean($Y_{(l + 1)}, Y_{(l + 2)}, \ldots, Y_{(u)}$)}}
     \\ &\leq  \sqrt{\frac{\log(2B/\alpha)}{2( \lfloor n/B\rfloor - 1)}} +  \frac{1}{\lfloor n/B \rfloor} \qquad \qquad \qquad \qquad \qquad \qquad \text{(by fact \eqref{eq:averages-fact})}
     \\ &\leq \varepsilon. 
\end{align*}
The rest of the argument can be completed exactly as in the proof of Theorem~\eqref{thm:umd-binary} after equation~\eqref{eq:per-bin-CI}. 
\qed

\rev{\subsection{Proof of Theorem~\ref{thm:umd-binary-randomized}}}
Let $\{\widehat{\Pi}_b'\}_{b \in [B]}$ denote the the pre-randomization values of $\widehat{\Pi}_b$ as computed in line~\ref{line:non-diff} of Algorithm~\ref{alg:randomized-UMD}. 
Due to the randomization in line~\eqref{line:diff-4}, no two $\widehat{\Pi}_b$ values are the same. Formally, consider any two indices $1 \leq a \neq b \leq B$. Then, $\widehat{\Pi}_a = \widehat{\Pi}_b$ if and only if $\delta (V_a - V_b) = \widehat{\Pi}_a' - \widehat{\Pi}_b'$, which happens with probability zero. Thus for any $ 1 \leq a\neq b \leq B$, $\widehat{\Pi}_a \neq \widehat{\Pi}_b$  (with probability one).

% Rephrasing \eqref{eq:per-bin-CI} with $\widehat{\Pi}_b'$ instead of $\widehat{\Pi}_b$, we have with probability at least $ 1 - \hoeffding$,
% \begin{equation}
% \abs{\Exp{}{Y \mid 
% \Bcal(X) = b}- \widehat{\Pi}_b'} \leq \sqrt{\frac{\log(2/\hoeffding)}{2\lfloor u - l - 1\rfloor}} \leq \sqrt{\frac{\log(2/\hoeffding)}{2(\lfloor n/B \rfloor - 1) }}.
% \end{equation}

The rest of the proof is conditional on $\Scal$, as defined in the proof of Theorem~\ref{thm:umd-binary}. (Marginalizing over $\Scal$ gives the theorem result as stated.) As noted in that proof, conditioning on $\Scal$ makes the binning function $\Bcal$ deterministic, which simplifies the proof significantly. 

First, we prove a per bin concentration bound for $\widehat{\Pi}_b$ of the form of \eqref{eq:per-bin-CI}. The $\delta$ randomization changes %the the per bin concentration bound  \eqref{eq:per-bin-CI} of Theorem~\ref{thm:umd-binary}
this bound as follows. For any $b\in [B], t \in (0, 1)$, with probability at least $ 1 - t$,
\begin{align}
    \abs{\Exp{}{Y \mid 
    \Bcal(X) = b}- \widehat{\Pi}_b} &\leq \abs{\Exp{}{Y \mid 
    \Bcal(X) = b}- \widehat{\Pi}_b'} + \abs{\widehat{\Pi}_b - \widehat{\Pi}_b'} \nonumber 
    \\ & \leq \sqrt{\frac{\log(2/\hoeffding)}{2(\lfloor n/B \rfloor - 1) }} + \abs{(1 + \delta)^{-1}(\widehat{\Pi}_b' + \delta) - \widehat{\Pi}_b'}  &\text{(Hoeffding's inequaliity \eqref{eq:per-bin-CI})}\nonumber 
    \\ & \leq \sqrt{\frac{\log(2/\hoeffding)}{2(\lfloor n/B \rfloor - 1) }} + \delta. \label{eq:per-bin-CI-2}
\end{align}

Given this concentration bound for every bin, the $(\varepsilon_2, \alpha)$-conditional calibration bound can be shown following the arguments in the proof of Theorem~\ref{thm:umd-binary} after inequality \eqref{eq:per-bin-CI}. We now show the marginal calibration guarantee. Note that since no two $\widehat{\Pi}_b$ values are the same, $\Bcal(X)$ is known given $\widehat{\Pi}_{\Bcal(X)}$, and so $\Exp{}{Y \mid h(X)} = \Exp{}{Y \mid \Bcal(X)}$. Thus, 
\begin{align*}
    &~\Prob(\abs{\Exp{}{Y \mid h(X)} - h(X)} \leq \varepsilon_1) \\&= \sum_{b=1}^B \Prob(\abs{\Exp{}{Y \mid h(X)} - h(X)} \leq \varepsilon_1 \mid \Bcal(X) = b) \ \Prob(\Bcal(X) = b) & \text{(law of total probability)}\\
    &= \sum_{b=1}^B \Prob(\abs{\Exp{}{Y \mid \Bcal(X)} - h(X)} \leq \varepsilon_1 \mid \Bcal(X) = b) \ \Prob(\Bcal(X) = b) &\text{($\Exp{}{Y \mid h(X)} = \Exp{}{Y \mid \Bcal(X)}$)}\\
    &= \sum_{b=1}^B \Prob(\abs{\Exp{}{Y \mid \Bcal(X)} - \widehat{\Pi}_{\Bcal(X)}} \leq \varepsilon_1 \mid \Bcal(X) = b) \ \Prob(\Bcal(X) = b) &\text{(by definition of $h$)}\\
    &\geq \sum_{b=1}^B (1 - \alpha)\ \Prob(\Bcal(X) = b) & \text{($t = \alpha$ in \eqref{eq:per-bin-CI-2})}
    \\ &= 1 - \alpha.
\end{align*}
This proves $(\varepsilon_1, \alpha)$-marginal calibration.

For the ECE bound, note that for every bin $b \in [B]$, $\widehat{\Pi}_b'$ is the average of at least $\lfloor n/B \rfloor - 1$ Bernoulli random variables with bias $\Exp{}{Y \mid \Bcal(X) = b}$. We know the exact form of the variance of averages of Bernoulli random variables with a given bias, giving the following:
\begin{equation}
    \text{Var}(\widehat{\Pi}_b') \leq \frac{\Exp{}{Y \mid \Bcal(X) = b}(1-\Exp{}{Y \mid \Bcal(X) = b})}{\lfloor n/B \rfloor - 1} \leq \frac{1}{4(\lfloor n/B \rfloor - 1)}.
    \label{eq:bound-var}
\end{equation}
We now rewrite the expectation of the square of the $\ell_2$-ECE in terms of $\text{Var}(\widehat{\Pi}_b')$. Recall that all expectations and probabilities in the entire proof are  conditional on $\Scal$, so that $\Bcal$ is known; the same is true for all expectations in the forthcoming panel of equations. To aid readability, when we apply the tower law, we are explicit about the remaining randomness in $\Dcal_n$. %While we do not show the conditioning on $\Scal$ explicitly, we are explicit about the remaining randomness (to aid readability).
% \begin{align*}
%     \Exp{}{(\Exp{}{Y \mid h(X), \Scal} - h(X))^2 \mid \Scal} &= \Exp{}{\Exp{}{(\Exp{}{Y \mid h(X), \Scal} - h(X))^2 \mid \Dcal_n} \mid \Scal} \\ &= 
%     \Exp{}{\sum_{b = 1}^B(\Exp{}{Y \mid \Bcal(X) = b, \Scal} - \widehat{\Pi}_b)^2 \Prob(\Bcal(X) = b \mid \Scal) \mid \Scal}
%     \\ &= 
%     \sum_{b = 1}^B\Exp{}{(\Exp{}{Y \mid \Bcal(X) = b, \Scal} - \widehat{\Pi}_b)^2 \Prob(\Bcal(X) = b \mid \Scal) \mid \Scal}
%     \\ &= 
%     \sum_{b = 1}^B\Exp{}{(\Exp{}{Y \mid \Bcal(X) = b, \Scal} - \widehat{\Pi}_b)^2 \mid \Scal} \Prob(\Bcal(X) = b \mid \Scal) . %&\text{(the outer expectation is the conditional expectation over $\Scal$)}
%     %\\ &= \Exp{}{(\Exp{}{Y \mid \Bcal(X) = b} - \widehat{\Pi}_b(X))^2}
% \end{align*}

\begin{align*}
    \Exp{\Dcal_n}{(\ell_2\text{-ECE}(h))^2} %\Exp{\Dcal_n \cup (X, Y)}{(\Exp{}{Y \mid h(X)} - h(X))^2} 
    &= \Exp{\Dcal_n}{\Exp{(X, Y)}{(\Exp{}{Y \mid h(X)} - h(X))^2 \mid \Dcal_n}} \\ &= 
    \Exp{\Dcal_n}{\sum_{b = 1}^B(\Exp{}{Y \mid \Bcal(X) = b} - \widehat{\Pi}_b)^2 \Prob(\Bcal(X) = b)}
    \\ &= 
    \sum_{b = 1}^B\Exp{\Dcal_n}{(\Exp{}{Y \mid \Bcal(X) = b} - \widehat{\Pi}_b)^2 \Prob(\Bcal(X) = b) }
    \\ &= 
    \sum_{b = 1}^B\Exp{\Dcal_n}{(\Exp{}{Y \mid \Bcal(X) = b} - \widehat{\Pi}_b)^2} \Prob(\Bcal(X) = b) . %&\text{(the outer expectation is the conditional expectation over $\Scal$)}
    %\\ &= \Exp{}{(\Exp{}{Y \mid \Bcal(X) = b} - \widehat{\Pi}_b(X))^2}
\end{align*}

The first equality is by the tower rule. The second equality uses the same simplifications as the panel of equations used to prove the marginal calibration guarantee (law of total probability, using $\Exp{}{Y \mid h(X)} = \Exp{}{Y \mid \Bcal(X)}$, and the definition of $h$). The third equality uses linearity of expectation. The fourth equality follows since $\Bcal$ is deterministic given $\Scal$. Now note that \[\mathbb{E}_{\Dcal_n}(\Exp{}{Y \mid \Bcal(X) = b} - \widehat{\Pi}_b)^2 = \mathbb{E}_{\Dcal_n}(\Exp{}{Y \mid \Bcal(X) = b} - \widehat{\Pi}_b' + \widehat{\Pi}_b' - \widehat{\Pi}_b)^2 \leq \text{Var}(\widehat{\Pi}_b') + \delta^2,\] 
since $\Exp{}{Y \mid \Bcal(X) = b} = \E_{\Dcal_n}(\widehat{\Pi}_b')$ and $\abs{\widehat{\Pi}_b' - \widehat{\Pi}_b} \leq \delta$ deterministically. Thus by bound \eqref{eq:bound-var},
\[
%\Exp{\Dcal_n \cup (X, Y)}{(\Exp{}{Y \mid h(X)} - h(X))^2}
\Exp{\Dcal_n}{(\ell_2\text{-ECE}(h))^2} \leq \sum_{b = 1}^B\roundbrack{\frac{1}{4(\lfloor n/B \rfloor - 1)} + \delta^2 }\Prob(\Bcal(X) = b) = \frac{1}{4(\lfloor n/B \rfloor - 1)} + \delta^2 \leq \frac{B}{2n} + \delta^2.
\]
The last inequality holds since $n \geq 2B$ implies that $\lfloor n/B \rfloor - 1 \geq n/2B$. Jensen's inequality now gives the final result:
\begin{align*}
\Exp{\Dcal_n}{\ell_2\text{-ECE}(h)} &\leq \sqrt{\Exp{\Dcal_n}{(\ell_2\text{-ECE}(h))^2}} &\text{(Jensen's inequality)}
%\Exp{\Dcal_n}{\sqrt{\Exp{(X, Y)}{(\Exp{}{Y \mid h(X)} - h(X))^2 \mid \Dcal_n}}} 
%\\ &= \sqrt{\Exp{\Dcal_n}{\Exp{(X, Y)}{(\Exp{}{Y \mid h(X)} - h(X))^2 \mid \Dcal_n}}}  
%\\ &\leq \sqrt{\frac{1}{4(\lfloor n/B \rfloor - 1)}}
%\\ &=\frac{1}{2\sqrt{\lfloor n/B \rfloor - 1}} 
\\ &\leq \sqrt{\frac{B}{2n} + \delta^2} \leq \sqrt{\frac{B}{2n}} + \delta .
\end{align*}
The bound on $\Exp{\Dcal_n}{\ell_p\text{-ECE}(h)}$ for $p \in [1, 2)$ follows by Proposition~\ref{prop:ECE-holder}. %since by H\"{o}lder's inequality, $\ell_p\text{-ECE}(h)  \leq \ell_2\text{-ECE}(h)$.
\qed
\\

\section{Assessing the theoretical guarantee of UMS}
\label{appsec:sample-complexity-gupta20}
We compute the number of calibration points $n$ required to guarantee $\smash{(\varepsilon, \alpha) = (0.1, 0.1)}$-marginal calibration with $B = 10$ bins using UMS, based on Theorem~5 of \citet{gupta2020distribution}. % Let us unpack their bound to compute the number of calibration points $n$ required if $(\varepsilon, \alpha, B) = (0.1, 0.1, 10)$. %to achieve the desired calibration guarantee. % required to achieve $(\varepsilon, \alpha, B) = (0.1, 0.1, 10)$ and compute an estimate for the number of calibration $n$ needed. %We perform some crude calculations to identify the sample-complexity required to achieve this. 
Following their notation, if the minimum number of calibration points in a bin is denoted as $N_{b^\star}$, then the Hoeffding-based bound on $\varepsilon$, with probablity of failure $\delta$,  is $\sqrt{\log(2B/\delta)/2N_{b^\star}}$. (The original bound is based on empirical-Berstein which is often tighter in practice, but Hoeffding is tighter in the worst case.) Let us set $\delta = \alpha/2 = 0.05$ since the remaining failure budget $\alpha/2$ is for the bin estimation to ensure that $N_{b^\star}$ is lower bounded. Thus, the requirement 
$\sqrt{\log(2\cdot 10/0.05)/2N_{b^\star}} \leq \varepsilon = 0.1$
translates roughly to $N_{b^\star} \geq 300$. 

To ensure $N_{b^\star} \geq 300$, we define the bins to each have roughly $1/B$ fraction of the calibration points in the first split of the data. Lemma 4.3~\citep{kumar2019calibration} shows that w.p. $\geq 1-\delta$, the true mass of the estimated bins is at least $1/2B$, as long as the first split of the data has at least $cB\log(10B/\delta)$ points, for a universal constant $c$. The original proof is for a $c \geq 2000$, but let us suppose that with a tighter analysis it can be improved to (say) $c = 100$. %(The value of $c$ can be improved with a tighter analysis, but that is not the focus of this paper. The message of this subsection remains unchanged even if say $c = 10$.) 
Then for $\delta = \alpha/4 = 0.025$, the first split of the data must have at least $100 \cdot 10 \cdot \log(100/0.025) \geq 8000$ calibration points. Finally, we use Theorem 5~\citep{gupta2020distribution} to bound $N_{b^\star}$. If $n'$ is the cardinality of the second split (denoted as $\abs{\Dcal_{\text{cal}}^2}$ in the original result), then they show that for $\delta = 0.025$,
$N_{b^\star} \geq n'/2B - \sqrt{n'/\log(2B/\delta)/2} %= \frac{n'}{20} - \sqrt{\frac{n'\ln(20/(0.025))}{2}} 
\approx n'/20 - 1.8\sqrt{n'}$. 
% \[
% N_{b^\star} \geq \frac{n'}{2B} - \sqrt{\frac{n'\log(2B/\delta)}{2}} %= \frac{n'}{20} - \sqrt{\frac{n'\ln(20/(0.025))}{2}} 
% \approx \frac{n'}{20} - 1.8\sqrt{n'}.
% \]
Since we require $N_{b^\star} \geq 300$, we must have approximately $n' \geq 9500$. 
Overall, the theoretical guarantee for UMS requires $n \geq 17500$ points to guarantee $(0.1, 0.1)$-marginal calibration with $10$ bins.% This sample complexity seems quite conservative for a binary classification problem. In Section~\ref{sec:umd-vs-ums}, we use an illustrative experiment to show that the $n$ required is indeed much lower. %we demonstrate the conservativeness of the bound of \citet{gupta2020distribution} experimentally. 

%Our experiment uses a novel diagnostic tool called validity plots, introduced next. % (\cg{show this conservativeness experimentally}). 

\begin{algorithm}[t]
\rev{
\caption{Randomized \ze}
	\label{alg:randomized-UMD}
	\begin{algorithmic}[1]
	%\SetAlgoLined
	\STATE {\bfseries Input:} Scoring function $\smash{g : \Xcal \to [0, 1]}$, \#bins $B$, calibration data $(X_1, Y_1), (X_2, Y_2), \ldots, (X_n, Y_n)$, \\ randomization parameter $\delta > 0$ (arbitrarily small)%, test point $(X, Y)$}
	\STATE {\bfseries Output:} {Approximately calibrated  function $h$ }
	%\STATE $(S_1, S_2, \ldots, S_n) \gets (g(X_1), g(X_2), \ldots, g(X_n))$;
	\STATE $(U_1, U_2, \ldots, U_n) \sim \text{Unif}[0,1]^n$\; \label{line:diff-1}
	\STATE $(S_1, S_2, \ldots, S_n) \gets (1+\delta)^{-1}(g(X_1) + \delta U_1, g(X_2) + \delta U_2, \ldots, g(X_n) + \delta U_n)$\; \label{line:diff-2}
	\STATE  $(S_{(1)}, S_{(2)}, \ldots, S_{(n)}) \gets \text{order-stats}(S_1, S_2, \ldots, S_n)$\;
	\STATE  $(Y_{(1)}, Y_{(2)}, \ldots, Y_{(n)}) \gets (Y_1, Y_2, \ldots, Y_n)$ ordered as per the ordering of $(S_{(1)}, S_{(2)}, \ldots, S_{(n)})$\;
	\STATE  $\Delta \gets (n+1)/B$\;
    \STATE $\widehat{\Pi} \gets$ empty array of size $B$\;
    \STATE $A \gets 0\text{-indexed array}([0, \lceil \Delta\rceil, \lceil 2 \Delta\rceil, \ldots, n+1])$\;
	\FOR{$b \gets 1$ \textbf{to} $B$}
	    %$l \gets \lceil b\times \Delta\rceil$\\;
        \STATE $l \gets A_{b-1}$\;
        \STATE $u \gets A_{b}$\;
	    \STATE $\widehat{\Pi}_b \gets$ Mean($Y_{(l + 1)}, Y_{(l + 2)}, \ldots, Y_{(u-1)}$)\;\label{line:non-diff}
	    \STATE $V_b \sim \text{Unif}[0,1]$\; \label{line:diff-3}
	    \STATE$\widehat{\Pi}_b \gets (1+\delta)^{-1}(\widehat{\Pi}_b + \delta V_b)$\; \label{line:diff-4}
	    %\label{line:removed-mean}
	\ENDFOR
 	\STATE $(S_{(0)}, S_{(n+1)}) \gets (0, 1)$\;
	\STATE $h(\cdot) \gets \sum_{b=1}^B \indicator{S_{(A_{b-1})} \leq (1+\delta)^{-1}(g(\cdot) + \delta U) < S_{(A_{b})}} \widehat{\Pi}_b$, for $U \sim \text{Unif}[0, 1]$\;  \label{line:diff-5}
	\end{algorithmic}
}
\end{algorithm}

%\subsection{On the Absolute Continuity of $g(X)$}
\section{Randomized \ze}
\label{appsec:randomized-umd}
We now describe the randomized version of \ze (Algorithm~\ref{alg:randomized-UMD}) that is nearly identical to the non-randomized version in practice, but for which we are able to show better theoretical properties. In this sense, we view randomized \ze as a theoretical tool rather than a novel algorithm (nevertheless, all experimental results in this paper use randomized \ze). Algorithm~\ref{alg:randomized-UMD} takes as input a randomization parameter $\delta > 0$ which can be arbitrarily small, such as $10^{-20}$. The specific lines that induce randomization, in comparison to Algorithm~\ref{alg:efficient-uniform-mass-binary}, are lines \ref{line:diff-1}, \ref{line:diff-2}, \ref{line:diff-3}, \ref{line:diff-4} and \ref{line:diff-5}. %Since $\delta$ can be arbitrarily small, the randomized and non-randomized versions of \ze are nearly identical. However, w
This $\delta$ perturbation leads to a better theoretical result than the non-randomized version --- in comparison to Theorem~\ref{thm:umd-binary}, Theorem~\ref{thm:umd-binary-randomized} does not require absolute continuity of $g(X)$ and provides an improved marginal calibration guarantee. 
\\
\subsection{Absolute continuity of $g(X)$}
\label{sec:continuity-assumption}
In Theorem~\ref{thm:umd-binary}, we assumed that $g(X)$ is absolutely continuous with respect to the Lebesgue measure, or equivalently, it has a pdf. This may not always be the case. For example, $X$ may contain atoms, %in which case $g(X)$ will also contain atoms and not have a pdf. 
or $g$ may have discrete outputs in $[0, 1]$. If $g(X)$ does not have a pdf, a simple randomization trick can be used to ensure that the results hold in full generality (we performed this randomization in our experiments as well).

First, we append the features $X$ with $\text{Unif}[0,1]$ random variables $U$ so that $(X, U) \sim P_X \times \text{Unif}[0, 1]$. Next, for an arbitrarily small value $\delta > 0$, such as $10^{-20}$, we define $\widetilde{g}: \Xcal \times [0, 1] \to [0, 1]$ as $\widetilde{g}(x, u) = (1 + \delta)^{-1}(g(x) + \delta u)$. Thus for every $x$, $\widetilde{g}(x, \cdot)$ is arbitrarily close to $g(x)$, and we do not lose the informativeness of $g$. However, now $\widetilde{g}(X, U)$ is guaranteed to be absolutely continuous with respect to the Lebesgue measure. % and the requirement of Theorem~\ref{thm:UMD-binary} is satisfied. 
The precise implementation details are as follows: (a) to train, draw $\smash{(U_i)_{i\in[n]} \sim \text{Unif}[0, 1]^n}$ and call Algorithm~\ref{alg:efficient-uniform-mass-binary} with $\widetilde{g}, \{((X_i, U_i), Y_i)\}_{i \in [n]}$; (b) to test, draw a new $\text{Unif}[0,1]$ random variable for each test point. \rev{Algorithm~\ref{alg:randomized-UMD} packages this randomization into the pseudocode; see lines \ref{line:diff-1}, \ref{line:diff-2} and \ref{line:diff-5}.} 

The above process is a technical way of describing the following intuitive methodology: ``break ties among the scores arbitrarily but consistently". Lemmas~\ref{lemma:order-statistics-1} and~\ref{lemma:order-statistics-main} fail if two data points have $S_i = S_j$ and one of them is the order statistics we are conditioning on. However, if we fix an arbitrary secondary order through which ties can be broken even if $S_i = S_j$ or $S = S_i$, the lemmas can be made to go through. The noise term $\delta U$ in $\widetilde{g}$ implicitly provides a strict secondary order. \subsection{Improved marginal calibration guarantee}
The marginal calibration guarantee of Theorem~\ref{thm:umd-binary-randomized} hinges on the bin biases $\widehat{\Pi}_b$ being unique. Lines~\ref{line:diff-3} and \ref{line:diff-4} in Algorithm~\ref{alg:randomized-UMD} ensure that this is satisfied almost surely by adding an infinitesimal random perturbation to each $\widehat{\Pi}_b$. This is identical to the technique described in Section~\ref{sec:continuity-assumption}. Due to the perturbation, the $\varepsilon$ required to satisfy calibration as per equation~\eqref{eq:marginal-umd-guarantee} has an additional $\delta$ term. However the $\delta$ can be chosen to be arbitrarily small, and this term is inconsequential. 

We make an informal remark that may be relevant to practitioners. In practice, we expect that the bin biases computed using Algorithm~\ref{alg:efficient-uniform-mass-binary} are unique with high probability without the need for randomization. As long as the bin biases are unique, the marginal calibration and ECE guarantees of Theorem~\ref{thm:umd-binary-randomized} apply to Algorithm~\ref{alg:efficient-uniform-mass-binary} as well. Thus, the $\widehat{\Pi}$-randomization can be skipped if `simplicity' or `interpretability' is desired. Note that the $g(X)$ randomization (Section~\ref{sec:continuity-assumption}) is still crucial since we envision many practical scenarios where $g(X)$ is not absolutely continuous. 
In summary, randomized \ze uses a small random perturbation to ensure that (a) the score values and (b) the bin bias estimates, are unique. The particular randomization strategy we proposed is not special; any other strategy that achieves the aforementioned goals is sufficient (for example, using a (truncated) Gaussian random variable instead of uniform).

\section{Additional experiments}
\label{appsec:additional-exps}
We present additional experiments to supplement those presented in the main paper. 

\begin{figure*}[t]
\begin{subfigure}[c]{0.74\linewidth}
\includegraphics[width=0.49\linewidth]{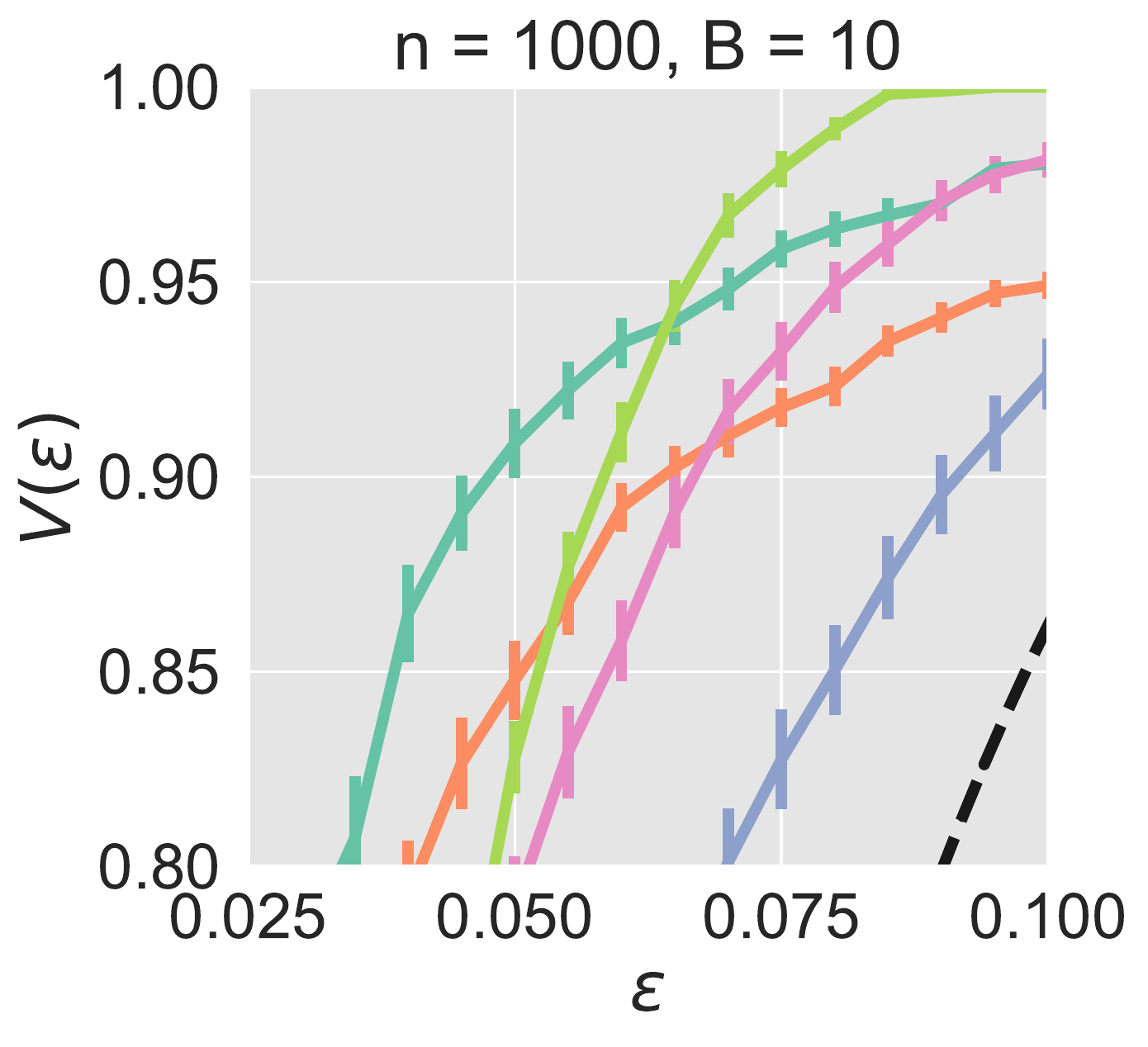}
\includegraphics[width=0.49\linewidth]{marginal_0.3.pdf}
\end{subfigure}
\begin{subfigure}[c]{0.25\linewidth}
\includegraphics[width=\linewidth,trim=0cm -2cm 0cm 0cm,clip]{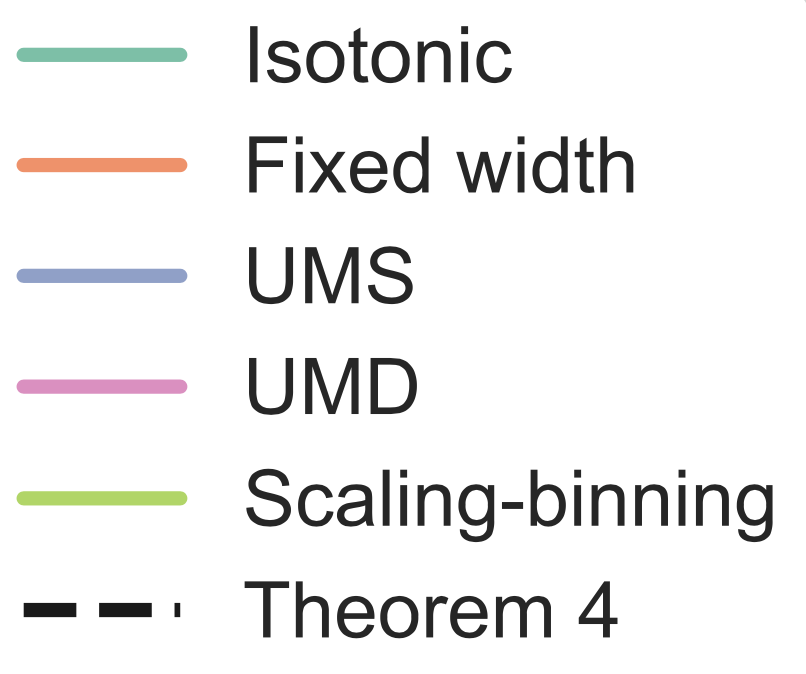}
\end{subfigure}

\begin{subfigure}[c]{0.74\linewidth}
\includegraphics[width=0.49\linewidth]{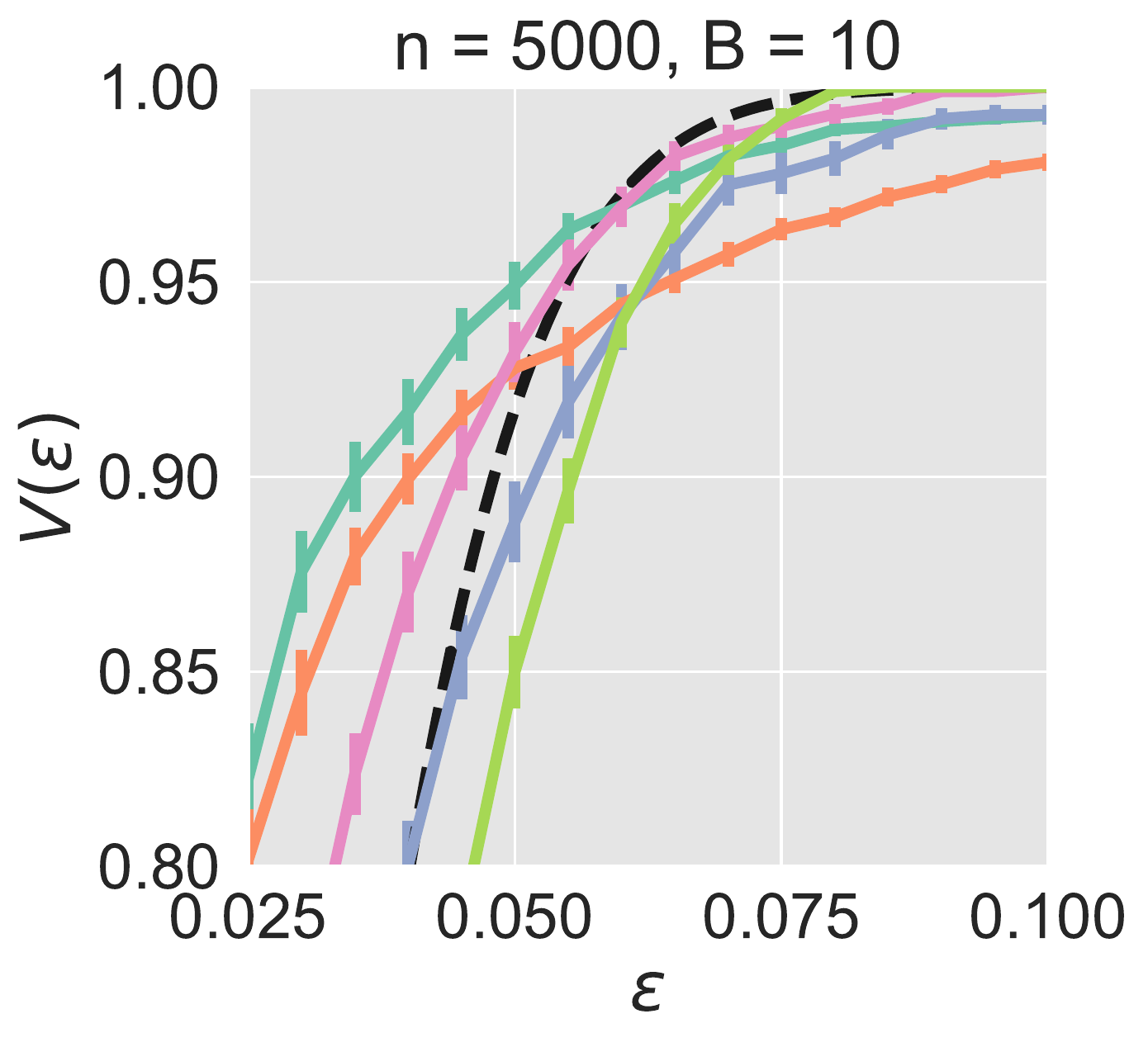}
\includegraphics[width=0.49\linewidth]{marginal_0.7.pdf}
\end{subfigure}
\begin{subfigure}[c]{0.25\linewidth}
\includegraphics[width=\linewidth,trim=0cm -2cm 0cm 0cm,clip]{legend.png}
\end{subfigure}

\caption{Marginal validity plots comparing \ze to other binning methods. The performance of \ze improves at higher values of $n$ and $\varepsilon$, and the performance of \ze is closely explained by its theoretical guarantee. Isotonic regression and fixed-width binning perform well at small values of $\varepsilon$. %The performance of \ze improves as sample size increases. Additional discussion of results is provided in Section~\ref{sec:simulations}. 
}
\label{fig:comparison-across-methods-marginal-all}
\end{figure*}

\begin{figure*}[t]
\begin{subfigure}[c]{0.74\linewidth}
\includegraphics[width=0.49\linewidth]{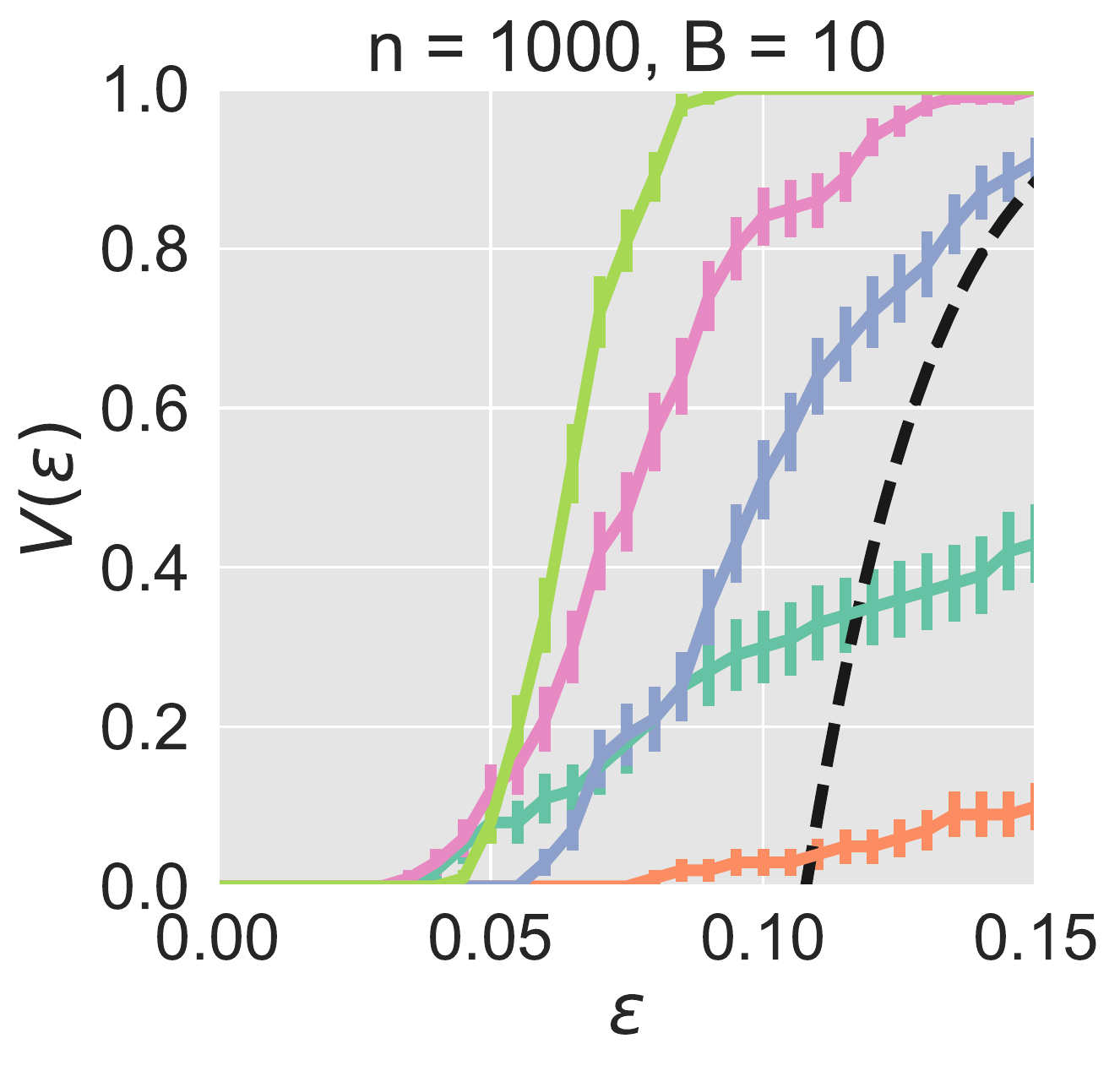}
\includegraphics[width=0.49\linewidth]{conditional_0.3.pdf}
\end{subfigure}
\begin{subfigure}[c]{0.25\linewidth}
\includegraphics[width=\linewidth,trim=0cm -2cm 0cm 0cm,clip]{legend.png}
\end{subfigure}
\begin{subfigure}[c]{0.74\linewidth}
\includegraphics[width=0.49\linewidth]{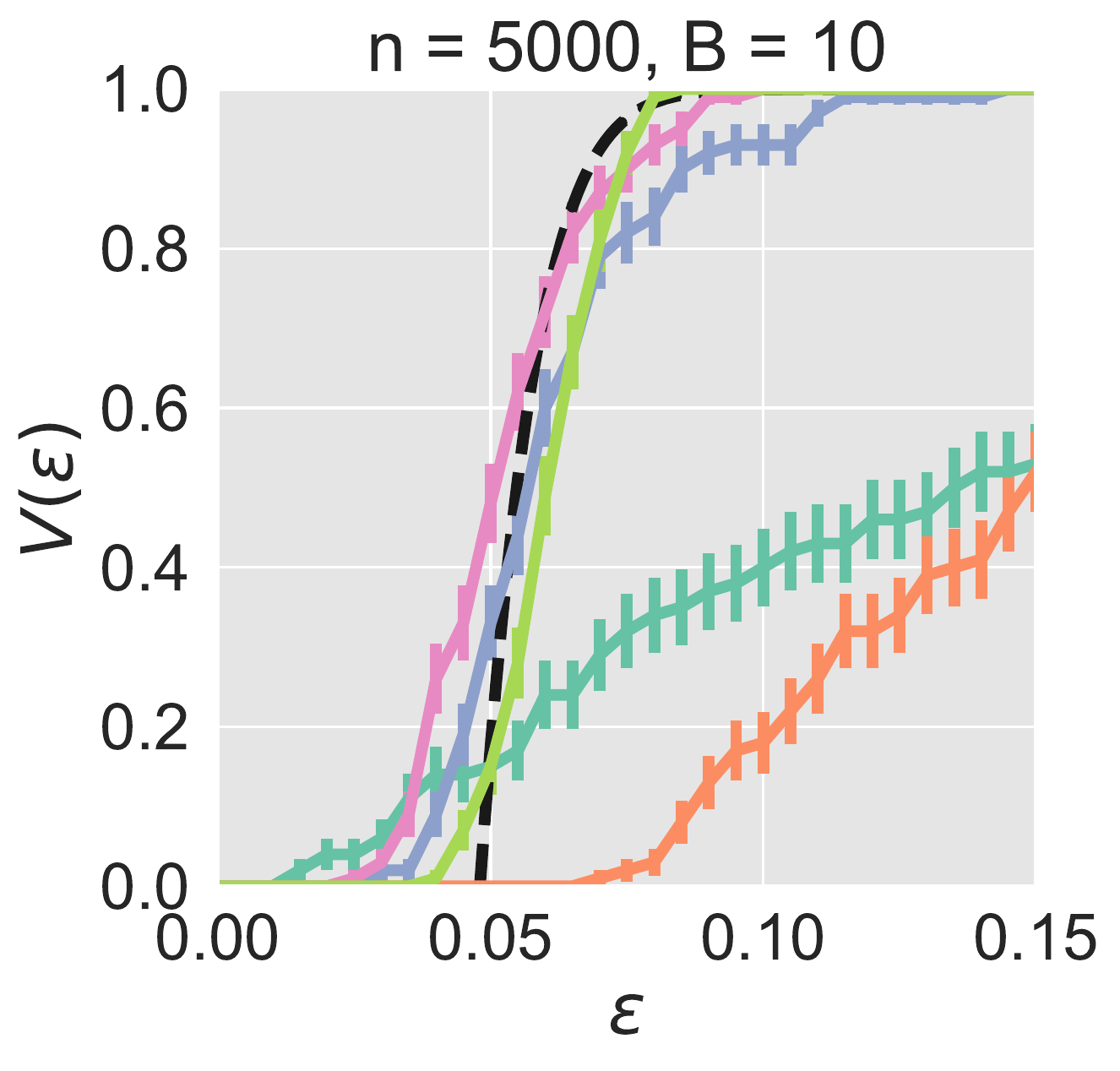}
\includegraphics[width=0.49\linewidth]{conditional_0.7.pdf}
\end{subfigure}
\begin{subfigure}[c]{0.25\linewidth}
\includegraphics[width=\linewidth,trim=0cm -2cm 0cm 0cm,clip]{legend.png}
\end{subfigure}

\caption{Conditional validity plots comparing \ze to other binning methods. \ze and scaling-binning are the best methods for conditional calibration at nearly all values of $n, \varepsilon$. Scaling-binning performs slightly better for small $n$ whereas \ze performs slightly better for large $n$. The performance of \ze is closely explained by its theoretical guarantee. %The performance of \ze and the tightness of its guarantee improves as sample size increases. When $n=1$K, scaling-binning performs the best. 
}
\label{fig:comparison-across-methods-conditional-all}
\end{figure*}

In Section~\ref{sec:simulations}, we compared \ze to other binning methods on the \credit dataset, for $n=3$K and $n=7$K.  Here, we present plots for $n=1$K and $n=5$K (for easier comparison, we also show the plots for $n=3$K and $n=7$K). The marginal validity plots are in Figure~\ref{fig:comparison-across-methods-marginal-all}, and the conditional validity plots are in Figure~\ref{fig:comparison-across-methods-conditional-all}. Apart from additional evidence for the same observations made in Section~\ref{sec:simulations}, we also see some interesting behavior in the low sample case ($n=1$K). First, the Theorem~\ref{thm:umd-binary-randomized} curve does not explain performance as well as the other plots. We tried the Clopper Pearson exact confidence interval \citep{clopper1934use} instead of Hoeffding and obtained nearly identical results (plots not presented). It would be interesting to explore if a tighter  guarantee can be shown for small sample sizes. Second, for $n=1$K, scaling-binning performs better than \ze in both the marginal and conditional validity plots, and is competitive with isotonic regression in the marginal validity plot. This behavior occurs since in the small sample regime, while all other binning methods attempt to re-estimate the biases of the bins using very little data, scaling-binning relies on the statistical efficiency of the learnt $g$ which was trained on 15K training points. A similar phenomenon was observed by \citet{Niculescu2005predicting} when comparing Platt scaling and isotonic regression: Platt scaling performs better at small sample sizes since it relies more on the underlying efficiency of $g$, compared to isotonic regression.

\begin{figure*}[t]
\begin{subfigure}[c]{0.37\linewidth}
\includegraphics[width=\linewidth]{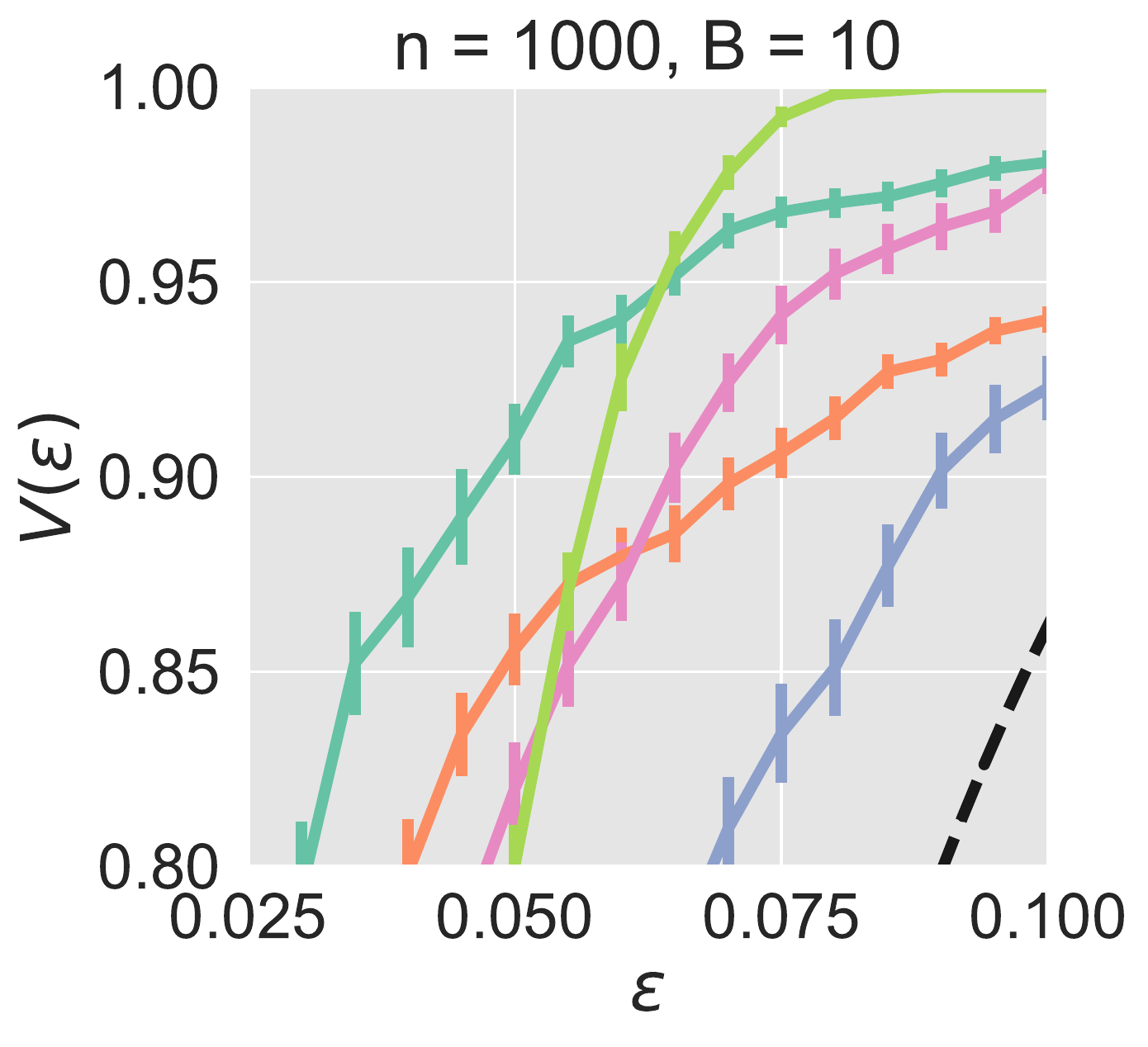}
\caption{Marginal validity plot.}
\end{subfigure}
\begin{subfigure}[c]{0.37\linewidth}
\includegraphics[width=\linewidth]{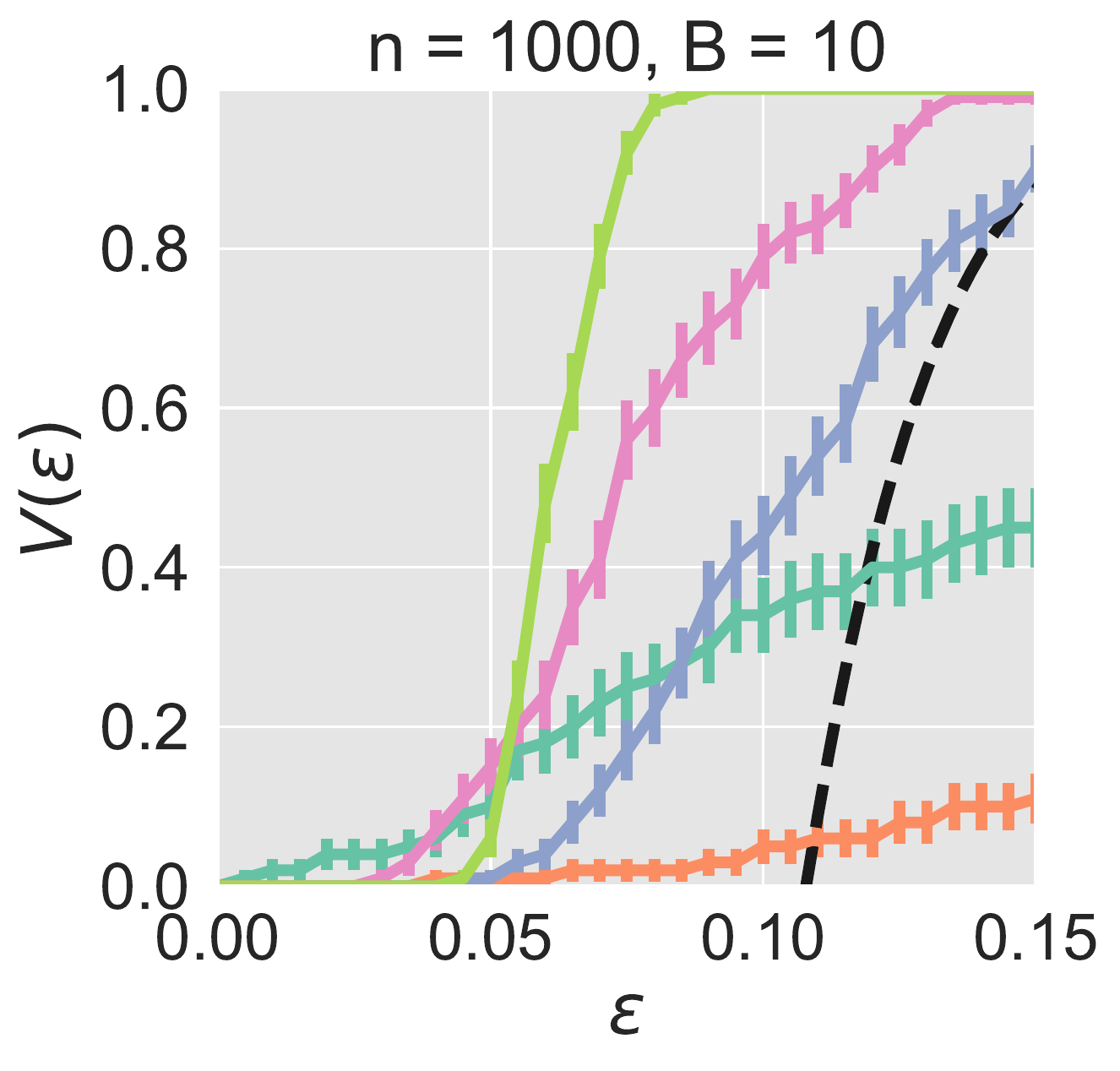}
\caption{Conditional validity plot.}
\end{subfigure}
\begin{subfigure}[c]{0.24\linewidth}
\includegraphics[width=\linewidth,trim=0cm -2cm 0cm 0cm,clip]{legend.png}
\end{subfigure}

\caption{Validity plots comparing \ze to other binning methods with fewer points used for recalibration. Namely, 14K points are used for training logistic regression, 1K for Platt scaling, and 1K for binning. Overall, scaling-binning performs quite well, since it relies on the underlying efficiency of logistic regression more than the other methods. 
}
\label{fig:comparison-across-methods-small}
\end{figure*}

While the experiments considered so far use 10K points for training logistic regression, 5K points for Platt scaling, and between 0.5-10K points for binning, a practically common setting is where most points are used for training the base model, and a small fraction of points are used for recalibration. On recommendation of one of the ICML reviewers, we ran experiments with 14K points for training logistic regression, 1K for Platt scaling, and 1K for binning. The marginal and conditional validity plots for this experiment are displayed in Figure~\ref{fig:comparison-across-methods-small}. We observe that these plots are very similar to the marginal and conditional validity plots in Figures~\ref{fig:comparison-across-methods-marginal-all} and \ref{fig:comparison-across-methods-conditional-all} for $n=1$K, and the same conclusions described in the previous paragraph can be drawn. 

%As far as I know, no `estimator' $R'$ exists such that $\E \abs{R'- c} = \abs{p-c}$. If such an $R'$ existed, we could estimate $\abs{p-c}$ using multiple resamples.

\end{document}